\documentclass[12pt]{article}
\usepackage{graphicx}
\usepackage{axodraw}

\newcommand{\drawsquare}[2]{\hbox{%
\rule{#2pt}{#1pt}\hskip-#2pt
\rule{#1pt}{#2pt}\hskip-#1pt
\rule[#1pt]{#1pt}{#2pt}}\rule[#1pt]{#2pt}{#2pt}\hskip-#2pt
\rule{#2pt}{#1pt}}

\newcommand{\kbox}{\drawsquare{7}{0.6}}
\usepackage{epsfig,amsmath,amssymb,verbatim,mathrsfs,subfigure}

\def\beq{\begin{eqnarray}}
\def\eeq{\end{eqnarray}}
\def\bea{\begin{eqnarray}}
\def\eea{\end{eqnarray}}

\def\xsection#1{\section{#1}}

\def\tev{\, {\rm TeV}}
\def\gev{\, {\rm GeV}}

\def\xfb{\, {\rm fb}}
\newcommand{\gsim}{\lower.7ex\hbox{$\;\stackrel{\textstyle>}{\sim}\;$}}
\newcommand{\lsim}{\lower.7ex\hbox{$\;\stackrel{\textstyle<}{\sim}\;$}}

\def\eq#1{eq.~(\ref{#1})}

\def\Dslash{\gamma^\mu D_\mu}

\def\vector#1#2{\left( \begin{array}{c}#1\\ #2\end{array}\right)}

\begin{document}

\setlength{\baselineskip}{0.22in}


\begin{titlepage}
\noindent
\begin{flushright}
{\small CERN-PH-TH-2009-154} \\
{\small MCTP-09-48}  \\
\end{flushright}
\vspace{-.5cm}

\begin{center}
  \begin{Large}
    \begin{bf}

 {Lectures on Higgs Boson Physics \\ in the Standard Model and Beyond}\vspace{0.2cm}\\
     \end{bf}
  \end{Large}
\end{center}
\vspace{0.2cm}
\begin{center}
\begin{large}
James~D.~Wells
\end{large}
  \vspace{0.3cm}
  \begin{it}

CERN, Theoretical Physics, CH-1211 Geneva 23, Switzerland, and \\
Physics Department, University of Michigan, Ann Arbor, MI 48109

 \vspace{0.1cm}
\end{it}

\end{center}


\begin{abstract}
These lectures focus on the structure of various Higgs boson theories. Topics in the first lectures include: mass generation in chiral theories, spontaneous symmetry breaking, neutrino masses, perturbative unitarity, vacuum stability, vacuum alignment,  flavor changing neutral current solutions with multiple Higgs doublets, analysis of type I theory with $Z_2$ symmetry, and rephasing symmetries. After an  Essay on the Hierarchy Problem, additional topics are covered that more directly relate to naturalness of the electroweak theory. Emphasis is on their connection to Higgs boson physics. Topics in these later lectures include: supersymmetry, supersymmetric Higgs sector in the Runge basis, leading-order radiative corrections of supersymmetric light Higgs boson mass, theories of extra dimensions, and radion mixing with the Higgs boson in warped extra dimensions. And finally, one lecture is devoted to Higgs boson connections to the hidden sector.

\end{abstract}

\vspace{1cm}

\begin{center}
{\it British Universities Summer School in Theoretical Elementary Particle Physics  \\
(BUSSTEPP) Cambridge University 2008 and University of Liverpool 2009 }
\end{center}

\end{titlepage}

\setcounter{page}{2}

\tableofcontents

\vfill\eject

\xsection{The Problem of Mass in Chiral Gauge Theories}

The fermions of the Standard Model and some of the gauge bosons have mass. This is a troublesome statement since gauge invariance appears to allow neither. Let us review the situation for gauge bosons and chiral fermions and introduce the Higgs mechanism that solves it. First, we illustrate the concepts with a massive $U(1)$ theory -- spontaneously broken QED.

\noindent
{\it Gauge Boson Mass}

The lagrangian of QED is
\beq
{\cal L}_{QED}=-\frac{1}{4}F_{\mu\nu}F^{\mu\nu}+\bar \psi (i\Dslash -m)\psi
\eeq
where 
\beq
D_\mu=\partial_\mu +ieA_\mu
\eeq
and $Q=-1$ is the charge of the electron.
This lagrangian respects the $U(1)$ gauge symmetry
\bea
\psi & \to & e^{-i\alpha(x)}\psi \\
A_\mu & \to & A_\mu+\frac{1}{e}\partial_\mu \alpha(x).
\eeq
Since QED is a vector-like theory -- left-handed electrons have the same charge as right-handed electrons --  an explicit mass term for the electron does not violate gauge invariance. 

If we wish to give the photon a mass we may add to the lagrangian the mass term
\beq
{\cal L}_{mass}=\frac{m_A^2}{2}A_\mu A^\mu .
\eeq
However, this term is not gauge invariant since under a transformation $A_\mu A^\mu$ becomes
\beq
A_\mu A^\mu\to A_\mu A^\mu+\frac{2}{e}A^\mu\partial_\mu\alpha+\frac{1}{e^2}\partial_\mu \alpha\partial^\mu\alpha
\eeq
This is not the right way to proceed if we wish to continue respecting the gauge symmetry. 
There is a satisfactory way to give mass to the photon while retaining the gauge symmetry. This is the Higgs mechanism, and the simplest way to implement it is via an elementary complex scalar particle that is charged under the symmetry and has a vacuum expectation value (vev) that is constant throughout all space and time.  This is the Higgs boson field $\Phi$.

Let us suppose that the photon in QED has a mass. To see how the Higgs boson implements the Higgs mechanism in a gauge invariant manner,  we 
introduce the field $\Phi$ with charge $q$  to the lagrangian:
\beq
{\cal L}={\cal L}_{QED}+(D_\mu\Phi)^*(D^\mu \Phi)-V(\Phi)
\eeq
where 
\beq
V(\Phi)=\mu^2|\Phi|^2+\lambda |\Phi|^4
\eeq
where it is assumed that $\lambda>0$ and $\mu^2<0$.  

Since $\Phi$ is a complex field we have the freedom to parametrize it as
\beq
\Phi=\frac{1}{\sqrt{2}}\phi(x) e^{i\xi(x)},
\label{eq:phi parametrization}
\eeq
where $\phi(x)$ and $\xi(x)$ are real scalar fields. The scalar potential with this choice simplifies to
\beq
V(\Phi)\to V(\phi)=\frac{\mu^2}{2}\phi^2+\frac{\lambda}{4}\phi^4.
\eeq
Minimizing the scalar potential one finds
\beq
\left. \frac{dV}{d\phi}\right|_{\phi=\phi_0}=\mu^2\phi_0+\lambda\phi^3_0=0~\Longrightarrow~ 
\phi_0=\sqrt{\frac{-\mu^2}{\lambda}}.
\eeq
This vacuum expectation value of $\phi$ enables us to normalize the $\xi$ field by $\xi/\phi_0$ such that its kinetic term is canonical at leading order of small fluctuation, legitimizing the parametrization of
\eq{eq:phi parametrization}. We can now choose the unitary gauge transformation, $\alpha(x)=-\xi(x)/\phi_0$, to make $\Phi$ real-valued everywhere.
One finds that the complex scalar kinetic terms expand to
\beq
(D_\mu\Phi)^*(D^\mu \Phi)\to \frac{1}{2}(\partial_\mu\phi)^2+\frac{1}{2}e^2q^2\phi^2A_\mu A^\mu
\eeq
At the minimum of the potential $\langle \phi\rangle =\phi_0$, so one can expand the field $\phi$ about its vev, $\phi=\phi_0+h$, and identify the fluctuating degree of freedom $h$ with a propagating real scalar boson. 

The Higgs boson mass and self-interactions are obtained by expanding the lagrangian about $\phi_0$. The result is
\beq
-{\cal L}_{Higgs}=\frac{m_h^2}{2}h^2+\frac{\mu'}{3!}h^3+\frac{\eta}{4!}h^4
\eeq
where
\beq
m^2_h=2\lambda\phi^2_0,~~\mu'=\frac{3m^2_h}{\phi_0},~~\eta=6\lambda=3\frac{m^2_h}{\phi_0^2}.
\eeq
The mass of the Higgs boson is not dictated by gauge couplings here, but rather by its self-interaction coupling $\lambda$ and the vev.

The complex Higgs boson kinetic terms can be expanded to yield
\beq
\Delta {\cal L}=\frac{1}{2}e^2q^2\phi_0^2 A_\mu A^\mu+e^2q^2h A_\mu A^\mu
+\frac{1}{2}e^2q^2h^2A_\mu A^\mu.
\label{vector higgs}
\eeq
The first term is the mass of the photon, $m_A^2=e^2q^2\phi_0^2$. A massive vector boson has a longitudinal degree of freedom, in addition to its two transverse degrees of freedom,  which accounts for the degree of freedom lost by virtue of gauging away $\xi(x)$. The second and third terms of eq.~\ref{vector higgs} set the strength of interaction of a single Higgs boson and two Higgs bosons to a pair of photons:
\bea
hA_\mu A_\nu~{\rm Feynman~rule}~& : & ~ i2e^2q^2\phi_0 g_{\mu\nu} = i2\frac{m_A^2}{\phi_0}\label{hAA}\\
hhA_\mu A_\nu~{\rm Feynman~rule}~& : & ~ i2e^2q^2 g_{\mu\nu}=i2\frac{m_A^2}{\phi_0^2}
\eea
after appropriate symmetry factors are included.

The general principles to retain from this discussion are first that massive gauge bosons can be accomplished in a gauge-invariant way through the Higgs mechanism. The Higgs boson that gets a vev breaks whatever symmetries it is charged under -- the Higgs vev carries charge into the vacuum. And finally, the Higgs boson that gives mass to the gauge boson couples to it proportional to the gauge boson mass.

\noindent
{\it Chiral Fermion Masses}

In quantum field theory a four-component fermion can be written in its chiral basis as
\beq
\psi=\left( \begin{array}{c} \psi_L \\ \psi_R \end{array}\right)
\eeq
where $\psi_{L,R}$ are two-component chiral projection fermions. A mass term in quantum field theory is equivalent to an interaction between the $\psi_L$ and $\psi_R$ components
\beq
m\bar \psi\psi = m\psi_L^\dagger\psi_R+m\psi^\dagger_R \psi_L.
\eeq

In vectorlike QED, the $\psi_L$ and $\psi_R$ components have the same charge and a mass term can simply be written down. However, let us now suppose that in our toy $U(1)$ model, there exists a set of chiral fermions where the $P_L\psi=\psi_L$ chiral projection carries a different gauge charge than the $P_R\psi=\psi_R$ chiral projection.  
In that case, we cannot write down a simple mass term without explicitly breaking the gauge symmetry.

The resolution to this conundrum of masses for chiral fermions resides in the Higgs sector. If the Higgs boson has just the right charge, it can be utilized to give mass to the chiral fermions. For example, if the charges are $Q[\psi_L]=1$, $Q[\psi_R]=1-q$ and $Q[\Phi]=q$ we can form the gauge invariant combination 
\beq
{\cal L}_f=y_\psi\, \psi_L^\dagger \Phi\psi_R+c.c.
\eeq
where $y_f$ is a dimensionless Yukawa coupling.  Now expand the Higgs boson about its vev, $\Psi=(\phi_0+h)/\sqrt{2}$, and we find
\beq
{\cal L}_f=m_\psi\, \psi_L^\dagger\psi_R+ \left(\frac{m_\psi}{\phi_0} \right)h \psi_L^\dagger\psi_R+c.c.
\eeq
where $m_\psi=y_\psi \phi_0/\sqrt{2}$.

We have successfully generated a mass by virtue of the Yukawa interaction with the Higgs boson. That same Yukawa interaction gives rise to an interaction between the physical Higgs boson and the fermions:
\beq
h\bar\psi\psi~{\rm (Feynman~rule)}~:~ i\frac{m_\psi}{\phi_0}.
\eeq
Just as was the case with the gauge bosons, the generation of fermion masses by the Higgs boson leads to an interaction of the physical Higgs bosons with the fermion proportional to the fermion mass.
As we will see in the Standard Model, this rigid connection between mass and interaction is what enables us to anticipate Higgs boson phenomenology with great precision as a function of the unknown Higgs boson mass.

\xsection{Standard Model Electroweak Theory}

The bosonic electroweak lagrangian is an $SU(2)_L\times U(1)_Y$ gauge invariant theory
\beq
{\cal L}_{bos}=|D_\mu \Phi |^2-\mu^2|\Phi |^2-\lambda |\Phi |^4-\frac{1}{4}B_{\mu\nu}B^{\mu\nu}
-\frac{1}{4}W^a_{\mu\nu}W^{a,\mu\nu}
\eeq
where $\Phi$ is an electroweak doublet with Standard Model charges of $({\bf 2},1/2)$ under $SU(2)_L\times U(1)_Y$ ($Y=+1/2$). In our normalization electric charge is $Q=T^3+\frac{Y}{2}$, and the doublet field $\Phi$ can be written as two complex scalar component fields $\phi^+$ and $\phi^0$:
\beq
\Phi=\left( \begin{array}{c} \phi^+ \\ \phi^0\end{array}\right).
\label{eq:higgs em}
\eeq
The covariant derivative and field strength tensors are
\bea
D_\mu \Phi & = &\left( \partial_\mu+ig\frac{\tau^a}{2}W^{a}_{\mu}+ig'\frac{Y}{2}B_\mu\right) \Phi \\
B_{\mu\nu}&=&\partial_\mu B_\nu-\partial_\nu B_\mu \\
W^a_{\mu\nu} & = & \partial_\mu W^a_\nu-\partial_\nu W^a_\mu-g f^{abc}W^b_\mu W^c_\nu
\eea

The minimum of the potential does not occur at $\Phi=0$ if $\mu^2<0$. Instead, one finds that the minimum occurs at a non-zero value of $\Phi$ -- its vacuum expectation value (vev) -- which via a gauge transformation can always be written as
\beq
\langle \Phi\rangle =\frac{1}{\sqrt{2}}\left( \begin{array}{c} 0 \\ v \end{array}\right)~~{\rm where}~~
v\equiv \sqrt{\frac{-\mu^2}{\lambda}}.
\eeq
This vev carries hypercharge and weak charge into the vacuum, and what is left unbroken is electric charge. This result we anticipated in \eq{eq:higgs em} by defining a charge $Q$ in terms of hypercharge and an eigenvalue of the $SU(2)$ generator $T^3$, and then writing the field $\Phi$ in terms of $\phi^0$ and $\phi^+$ of zero and positive $+1$ definite charge. 

Our symmetry breaking pattern is then simply $SU(2)_L\times U(1)_Y\to U(1)_Q$. The original group, $SU(2)_L\times U(1)_Y$, has a total of four generators and $U(1)_Q$ has one generator. Thus, three generators are `broken'. Goldstone's theorem~\cite{Goldstone:1962es} tells us that for every broken generator of a symmetry there must correspond a massless field.  These three massless Goldstone bosons we can call $\phi_{1,2,3}$. We now can rewrite the full Higgs field $\Phi$ as
\beq
\langle \Phi\rangle =\frac{1}{\sqrt{2}}\left( \begin{array}{c} 0 \\ v\end{array}\right)
+\frac{1}{\sqrt{2}} \left( \begin{array}{c} \phi_1+i\phi_2 \\ h+i\phi_3\end{array}\right)
\eeq
The fourth degree of freedom of $\Phi$ is the Standard Model Higgs boson $h$. It is a propagating degree of freedom. The other three states $\phi_{1,2,3}$ can all be absorbed as longitudinal components of three massive vector gauge bosons $Z,W^\pm$ which are defined by
\bea
W^\pm_\mu & = & \frac{1}{\sqrt{2}}\left( W^{(1)}_\mu\mp i W^{(2)}_\mu\right) \\
B_\mu & = & \frac{-g'Z_\mu+gA_\mu}{\sqrt{g^2+g'^2}} \\
W^{(3)}_\mu & = & \frac{g Z_\mu+g' A_\mu}{\sqrt{g^2+g'^2}}.
\eeq
It is convenient to define $\tan\theta_W=g'/g$. By measuring interactions of the gauge bosons with fermions it has been determined experimentally that $g=0.65$ and $g'=0.35$, and therefore
$\sin^2\theta_W=0.23$. 

After performing the redefinitions of the fields above, the kinetic terms for the $W^\pm_\mu, Z_\mu, A_\mu$ will all be canonical. Expanding the Higgs field about the vacuum, the contributions to the lagrangian involving Higgs boson interaction terms are
\bea
{\cal L}_{h\, int}& = &\left[ m_W^2W^+_\mu W^{-,\mu}+\frac{m_Z^2}{2}Z_\mu Z^\mu\right]  \left( 1+\frac{h}{v}\right)^2 \\
& & -\frac{m_h^2}{2}h^2-\frac{\xi}{3!}h^3-\frac{\eta}{4!} h^4
\eea
where
\beq
m^2_W=\frac{1}{4}g^2v^2,~~m_Z^2=\frac{1}{4}(g^2+g'^2)v^2~~\Longrightarrow~~
\frac{m_W^2}{m^2_Z}=1-\sin^2\theta_W
\eeq
\beq
m^2_h=2\lambda v^2,~~\xi=\frac{3m^2_h}{v},~~\eta=6\lambda=\frac{3m^2_h}{v^2}.
\eeq
From our knowledge of the gauge couplings, the value of the vev $v$ can be determined from the masses of the gauge bosons: $v\simeq 246\gev$.

The Feynman rules for Higgs boson interactions are
\bea
hhh & : & -\frac{i3m^2_h}{v} \\
hhhh & : & -i\frac{3m^2_h}{v^2} \\
hW^+_\mu W^-_\nu & : & i2\frac{m^2_W}{v}g^{\mu\nu} \\
hZ_\mu Z_\nu & : & i2\frac{m_Z^2}{v}g_{\mu\nu} \\
hhW^+_\mu W^-_\nu & : & i2\frac{m_W^2}{v^2}g_{\mu\nu} \\
hhZ_\mu Z_\nu & : & i2\frac{m_Z^2}{v^2}g_{\mu\nu}
\eea

Fermion masses are also generated in the Standard Model through the Higgs boson vev, which in turn induces an interaction between the physical Higgs boson and the fermions. Let us start by looking at $b$ quark interactions. The relevant lagrangian for couplings with the Higgs boson is
\beq
\Delta {\cal L}=y_b Q^\dagger_L \Phi b_R+c.c.~~{\rm where}~~Q^\dagger_L=(t^\dagger_L ~ b^\dagger_L)
\eeq
where $y_b$ is  the Yukawa coupling. The Higgs boson, after a suitable gauge transformation, can be written simply as
\beq
\Phi=\frac{1}{\sqrt{2}}\left( \begin{array}{c} 0 \\ v+h \end{array}\right)
\eeq
and the interaction lagrangian can be expanded to 
\bea
\Delta{\cal L} &=&y_bQ^\dagger_L \Phi b_R+c.c.=\frac{y_b}{\sqrt{2}}(t^\dagger_L~b^\dagger_L)
\left( \begin{array}{c} 0 \\ v+h \end{array}\right)b_R+h.c. \\
& = & m_b(b^\dagger_R b_L+b^\dagger_L b_R)\left( 1+\frac{h}{v}\right)=
m_b\, \bar b b\left( 1+\frac{h}{v}\right)
\eea
where $m_b=y_bv/\sqrt{2}$ is the  mass of the $b$ quark. 

The quantum numbers work out perfectly to allow this mass term. See Table~\ref{table:charges} for the quantum numbers of the various fields under the Standard Model symmetries. Under $SU(2)$ the interaction $Q_L^\dagger \Phi b_R$ is  invariant because ${\bf 2}\times {\bf 2}\times {\bf 1}\in {\bf 1}$ contains a singlet. And under $U(1)_Y$ hypercharge the interaction is invariant because $Y_{Q^\dagger_L}+Y_\Phi+Y_{b_R}=-\frac{1}{6}+\frac{1}{2}-\frac{1}{3}$ sums to zero.  Thus, the interaction is invariant under all gauge groups, and we have found a suitable way to give mass to the bottom quark.

\begin{table}
\begin{center}
\begin{tabular}{lccccc}
\hline\hline
Field & $SU(3)$ & $SU(2)_L$ & $T^3$ & $\frac{Y}{2}$ & $Q=T^3+\frac{Y}{2}$ \\
\hline
$g_\mu^a$ (gluons) & {\bf 8} & {\bf 1} & 0 & 0 & 0 \\
$(W_\mu^\pm, W_\mu^0)$ & {\bf 1} & {\bf 3} & $(\pm 1,0)$ &  0 &   $(\pm 1,0)$ \\
$B^0_\mu$  & {\bf 1} & {\bf 1} &  0 & 0 & 0 \\
\hline
$Q_L=\vector{u_L}{d_L}$ & {\bf 3} & {\bf 2} & $\vector{\frac{1}{2}}{-\frac{1}{2}}$ & $\frac{1}{6}$ & $\vector{\frac{2}{3}}{-\frac{1}{3}}$ \\
$u_R$ & {\bf 3} & {\bf 1} & 0 & $\frac{2}{3}$ & $\frac{2}{3}$ \\
$d_R$ & {\bf 3} & {\bf 1}& 0 & $-\frac{1}{3}$ & $-\frac{1}{3}$ \\
$E_L=\vector{\nu_L}{e_L}$ & {\bf 1} & {\bf 2}& $\vector{\frac{1}{2}}{-\frac{1}{2}}$ & $-\frac{1}{2}$ & $\vector{0}{-1}$ \\
$e_R$ & {\bf 1} & {\bf 1}& 0 & $-1$ & $-1$ \\ 
\hline
$\Phi=\vector{\phi^+}{\phi^0}$  & {\bf 1} & {\bf 2} & $\vector{\frac{1}{2}}{-\frac{1}{2}}$ &  $\frac{1}{2}$ & $\vector{1}{0}$ \\
$\Phi^c=\vector{\phi^0}{\phi^-}$  & {\bf 1} & {\bf 2}& $\vector{\frac{1}{2}}{-\frac{1}{2}}$ & $-\frac{1}{2}$ & $\vector{0}{-1}$ \\
\hline
\hline
\end{tabular}
\end{center}
\caption{\em Charges of Standard Model fields. \label{table:charges}}
\end{table}

How does this work for giving mass to the top quark?
Obviously, $Q^\dagger_L \Phi t_R$ is not invariant. However, we have the freedom to create the conjugate representation of $\Phi$ which still transforms as a ${\bf 2}$ under $SU(2)$ but switches sign under hypercharge: $\Phi^c=i\sigma^2 \Phi^*$. This implies that $Y_{\Phi^c}=-\frac{1}{2}$ and
\beq
\Phi^c=\frac{1}{\sqrt{2}}\left( \begin{array}{c} v+h\\ 0 \end{array}\right)
\eeq
when restricted to just the real physical Higgs field expansion about the vev.  Therefore, it becomes clear that $y_t Q_L^\dagger \Phi^c t_R+c.c.$ is now invariant since the $SU(2)$ invariance remains ${\bf 2}\times {\bf 2}\times {\bf 1}\in {\bf 1}$ and $U(1)_Y$ invariance follows from $Y_{Q^\dagger_L}+Y_{\Phi^c}+Y_{t_R}=-\frac{1}{6}-\frac{1}{2}+\frac{2}{3}=0$. Similar to the $b$ quark one obtains an expression for the mass and Higgs boson interaction:
\bea
\Delta{\cal L} &=&y_tQ^\dagger_L \Phi^c t_R+c.c.=\frac{y_t}{\sqrt{2}}(t^\dagger_L~b^\dagger_L)
\left( \begin{array}{c} v+h \\ 0 \end{array}\right)t_R+c.c. \\
& = & m_t(t^\dagger_R t_L+t^\dagger_L t_R)\left( 1+\frac{h}{v}\right)=
m_t\, \bar t t\left( 1+\frac{h}{v}\right)
\eea
where $m_t=y_tv/\sqrt{2}$ is the  mass of the $t$ quark.  

The mass of the charged leptons follows in the same manner, $y_eE^\dagger_L \Phi e_R+c.c.$, and interactions with the Higgs boson result. In all cased the Feynman diagram for Higgs boson interactions with the fermions at leading order is
\beq
h\bar f f ~ : ~ i\frac{m_f}{v}.
\eeq

We see from this discussion several important points. First, the single Higgs boson of the Standard Model can give mass to all Standard Model states, even to the neutrinos as we will see in the next lecture. It did not have to be that way. It could have been that quantum numbers of the fermions did not enable just one Higgs boson to give mass to everything. This is the Higgs boson miracle of the Standard Model.  The second thing to keep in mind is that there is a direct connection between the Higgs boson giving mass to a particle and it interacting with that particle. We have seen that all interactions are directly proportional to a mass factor. This is why Higgs boson phenomenology is completely determined in the Standard Model as long as one assumes, or ultimately knows, the Higgs boson mass itself.

\section{The Special Case of Neutrino Masses\label{sec:neutrinos}}

For many years it was thought that neutrinos might be exactly massless. Although recent experiments have shown that this is not the case, the masses of neutrinos are extraordinarily light compared to other Standard Model fermions. In this section we discuss the basics of neutrino masses~\cite{neutrino reviews}, with emphasis on how the Higgs boson plays a role.

Some physicists define the Standard Model without a right-handed neutrino. Thus, there is no opportunity to write down a Yukawa interaction of the left and right-handed neutrinos with the Higgs boson that gives neutrinos a mass. A higher-dimensional operator is needed,
\beq
{\cal O}_\nu =\frac{\lambda_{ij}}{\Lambda} (E_{iL}^\dagger H^c)^\dagger (E_{jL}^\dagger H^c)
\label{eq:neutrino op}
\eeq
where $E_L=(\nu_L~e_L)$ is the $SU(2)$ doublet of left-handed neutrino and electron. Taking into account the various flavors $i=1,2,3$ 
results in a $3\times 3$ mass matrix for neutrino masses
\beq
(m_{v})_{ij}=\lambda_{ij}\frac{v^2}{\Lambda}.
\label{eq:neutrino mass}
\eeq
$\Lambda$ can be considered the cutoff of the Standard Model effective theory (see lecture~\ref{sec:hierarchy}), and the operator given by \eq{eq:neutrino op} is the only gauge-invariant, Lorentz-invariant operator that one can write down at the next higher dimension ($d=5$) in the theory. Thus, it is a satisfactory approach to neutrino physics, leading to an indication of new physics beyond the Standard Model at the scale $\Lambda$. For this reason, many view the existence of neutrino masses as a signal for physics beyond the Standard Model.

The absolute value of neutrino masses has not been measured but the differences of mass squareds between various neutrino masses have been measured and range from about $10^{-5}$ to $10^{-2}\, {\rm eV}^2$~\cite{neutrino reviews}.  It is reasonable therefore to suppose that the largest neutrino mass in the theory should be around $0.1\, {\rm eV}$.  If we assume that this mass scale is obtained using the natural  value of $\lambda\sim 1$ in \eq{eq:neutrino mass} and a large mass scale $\Lambda$, this sets the scale of the cutoff $\Lambda$ to be
\beq
\Lambda\simeq \frac{(246\gev)^2}{0.1\, {\rm eV}}\simeq 10^{15}\gev
\eeq
This is a very interesting scale, since it is within an order of magnitude of where the three gauge couplings of the Standard Model come closest to meeting, which may be an indication of grand unification. The scale $\Lambda$ could then be connected to this Grand Unification scale.

Another approach to neutrino masses is to assume that there exists a right-handed neutrino $\nu_R$. After all, there is no strong reason to banish this state, especially since there is an adequate right-handed partner state to all the other fermions. Furthermore, if the above considerations are pointing to a grand unified theory, right-handed neutrinos are generally present in acceptable versions, such as $SO(10)$ where all the fermions are in the ${\bf 16}$ representation, including $\nu_R$.  Quantum number considerations indicate that $\nu_R$ is a pure singlet under the Standard Model gauge symmetries, and thus we have a complication in the neutrino mass sector beyond what we encountered for the other fermions of the theory. In particular, we are now able to add a Majorana mass term $\nu_R^T i\sigma^2\nu_R$ that is invariant all by itself without the need of a Higgs boson. The full mass interactions available to the neutrino are now
\beq
{\cal L}_{\nu}=y_{ij} E^\dagger_{iL} \Phi^c \nu_{jR}+\frac{M_{ij}}{2}\nu^T_{iR} i\sigma^2\nu_{jR}+c.c.
\label{eq:neutrino lag}
\eeq
The resulting $6\times 6$ mass matrix in the $\{ \nu_L,\nu^c_R\}$ basis is
\beq
m_\nu=\left( \begin{array}{cc}
0 & m_D \\
m_D^T & M \end{array}\right)
\label{eq:seesaw}
\eeq
where $M$ is the matrix of Majorana masses with values $M_{ij}$ taken straight from \eq{eq:neutrino lag}, and $m_D$ are the neutrino Dirac mass matrices taken from the Yukawa interaction with the Higgs boson
\beq
(m_D)_{ij}=\frac{y_{ij}}{\sqrt{2}}v.
\eeq

Consistent with effective field theory ideas, there is no reason why the Majorana mass matrix entries should be tied to the weak scale. They should be of order the cutoff scale of when the Standard Model is no longer considered complete.  Therefore, it is reasonable and expected to assume that $M_{ij}$ entries are generically much greater than the weak scale. In that limit, the seesaw matrix of \eq{eq:seesaw} has three heavy eigenvalues of  ${\cal O}(M)$, and three light eigenvalues that, to leading order and good approximation, are eigenvalues of the $3\times 3$ matrix
\beq
m^{\rm light}_\nu=-m_D^T M^{-1}m_D\sim y^2 \frac{v^2}{M}
\eeq
which is parametrically of the same form as \eq{eq:neutrino mass}. This is expected since the light eigenvalues can be evaluated from the operators left over after integrating out the heavy right-handed neutrinos in the effective theory. That operator is simply \eq{eq:neutrino op}, where schematically $\Lambda$ can be associated with the scale $M$ and $\lambda$ can be associated with $y^2$.

Neutrino physics is a rich field with many implications for both electroweak symmetry breaking and mass scales well above the weak scale. Thus, it may be especially sensitive to specifics of the theory at the cutoff scale.  Our desire here was to show how neutrinos get mass, consistent with Standard Model gauge symmetries and the principles of effective field theory. For more details about neutrino mass and mixing measurements and phenomenology, I recommend consulting reviews dedicated to that purpose~\cite{neutrino reviews, Kayser PDG}.

\xsection{Experimental Searches for the Standard Model Higgs Boson}

As we emphasized in the last lecture, any particle that gets a mass through the Higgs boson vacuum expectation value will also couple to it proportional to its mass. The phenomenology of the Higgs boson is then completely determined once the mass of the Higgs boson itself is specified. It is important to emphasize that despite this rigidity in the phenomenology predictions, the Higgs boson is a speculative object. There is no direct proof of its existence, although the indirect proof based on compatibility with the data is tantalizing.  

Experimental searches for the Higgs boson have been going on for several decades. It would be impossible to summarize the history, but recent developments, which are the most constraining and relevant, can be given. There are three experimental efforts relevant to this discussion. The first is the search by the LEP2 collaborations at the $e^+e^-$ collider at CERN. The second is the search by the D0 and CDF collaborations at the $p\bar p$ collider at Fermilab. And the third is the precision electroweak analysis that utilizes the results of a great many experiments, including LEP, SLC, Tevatron, etc.

LEP2 ran their $e^+e^-$ collider at energies as high as $\sqrt{s}=209\gev$ center of mass.  The primary search mode for the Higgs boson at this collider was $e^+e^-\to hZ$. The would-be signal is clean and the barrier to discovery is primarily the limitation of the center of mass energy. Kinematically, the maximum Higgs boson mass that could be produced on-shell at the collider in this mode is $m_h=\sqrt{s}-m_Z=118\gev$. The cross-section drops rather precipitously near this threshold so the sensitivity cannot be quite at the kinematic limit. Statistical fluctuations of candidate events can also affect the final lower limit if a signal is not established. Indeed, a signal was not established at LEP2 and the final mass limit arrived at by the collaborations~\cite{Barate:2003sz} taking all into account is
\beq
m_h< 114.4\gev~~{\rm excluded~ at}~95\% ~{\rm CL}~{\rm (LEP2)}.
\eeq

Tevatron has some sensitivity to the Higgs boson mainly through channels such as $gg\to h^{(*)}\to WW$ where the $W$'s can decay leptonically on one side and into jets on the other, along with many other channels such as $q\bar q\to Wh\to l\nu b\bar b$, etc.  The gluon partons that initiate the first set of events are of course plucked out of the $p$ and $\bar p$ hadrons.  Running at $\sqrt{s}=1.96\tev$ with $2.0-3.6\xfb^{-1}$ of luminosity analysed at CDF and $0.9-4.2\xfb^{-1}$ of luminosity analysed at D0, the combined effort~\cite{tevhiggslimit} yields an exclusion of
\beq
160\gev < m_h<170\gev~~{\rm excluded~at}~95\%~{\rm CL~~(D0/CDF)}.
\eeq

Lastly, we discuss the indirect limits on the Standard Model Higgs boson. Many observables are measured so precisely that  a full quantum loop computation is needed to show compatibility of the collected measurements of experiment with the Standard Model theory. The collection of observables include $Z$ decays measured at LEP1 and SLC, the $W$ mass at LEP2 and Tevatron, the top mass at Tevatron, muon decay, $e^+e^-\to {\rm hadrons}$ data at many low-energy machines as input to the determination of $\alpha(m_Z)$, etc.  The Higgs boson shows up in the quantum corrections in various ways. It contributes to the self-energies of the $W$ and $Z$ bosons most especially. Thus, the self-consistency of all the experimental measurements depends on the assumed value of $m_h$.  For pedagogical reviews of the precision electroweak program see for example refs.~\cite{Kennedy:1992tj,Wells:2005vk}.

To carry out a complete precision electroweak analysis is a very complicated subject with many uncertainties and correlations that have to be taken into account simultaneously. A somewhat complete picture of the significant effort required can be found in~\cite{:2005ema,Alcaraz:2006mx,Flacher:2008zq}.  Unfortunately, we do not have time to go through all of the issues. Nevertheless, let us simplify the discussion to illustrate how the limit on the Higgs boson is obtained. Let us approximate the situation by saying that all parameters of the Standard Model lagrangian besides the Higgs boson sector can be represented as $\{p\}$.  The remaining parameter of the Higgs boson sector, as established above, is merely the Higgs boson mass. Everything observable can be predicted by knowing $\{p\}$ and $m_h$.  

The prediction for the $i^{\rm th}$ observable, such as the $Z$ width or $Z\to l^+l^-$ branching ratio, we can write as ${\cal O}^{th}_i(m_h,\{p\})$.  The measurement of the observable is ${\cal O}^{\rm expt}_i$ with uncertainty $\Delta {\cal O}^{\rm expt}_i$.  We want to somehow cycle over all our parameters and find the optimal set that matches best the experimental measurements. The formal way to do this is to construct a $\chi^2$ function which when minimized gives the best fit to the data:
\beq
\chi^2=\sum_i \frac{({\cal O}^{th}_i(m_h,\{p\})-{\cal O}^{\rm expt}_i)^2}{(\Delta {\cal O}^{\rm expt}_i)^2}.
\eeq
Now, with this $\chi^2$ you can ask two questions. The first question is whether the theory matches the data. The answer is affirmative if at the minimum of the $\chi^2$ function its value per degree of freedom is not much larger than 1: $\chi^2_{min}/{\rm d.o.f.}\sim 1$. If it's much less than 1 then the laws of statistics are being violated and there is a systematic bias among the experimentalists to get ``the right answer".   In the Standard Model, the $\chi^2_{min}/{\rm d.o.f.}$ is 17.8/13 (see Table 10.2 on page 133 of~\cite{Alcaraz:2006mx}), which is good enough to establish that the Standard Model is compatible with the data.

\begin{figure}[t]
\begin{center}
\includegraphics[width=0.6\textwidth]{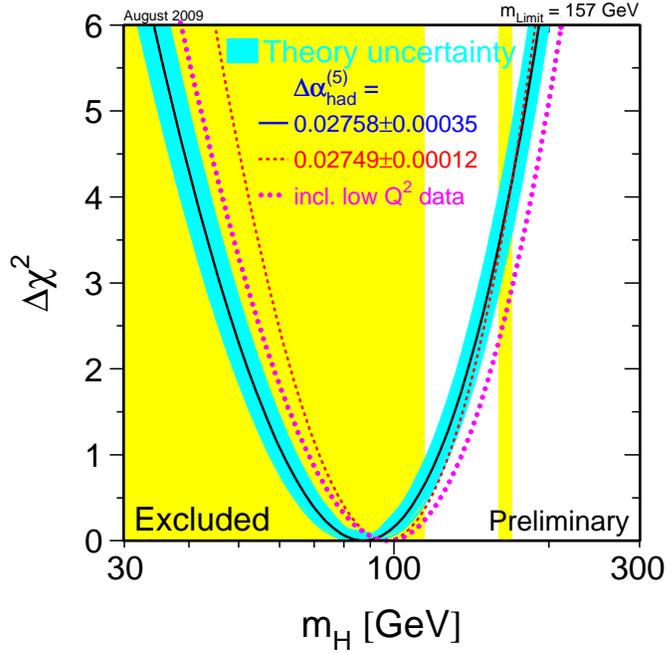}
\vspace{0.3cm}
\caption{$\Delta\chi^2$ fit to electroweak precision observables as a function of the Higgs boson mass in the Standard Model. Detailed description of the plot can be found in~\cite{Alcaraz:2006mx}. This particular plot is the latest from the summer 2009 update~\cite{LEPEWWG 2009}. The 95\% CL interval is for $\Delta\chi^2<(1.96)^2$, leading to $m_h\lsim 200\gev$ upper limit from precision electroweak analysis.
\label{fig:chi2}}
\end{center}
\end{figure}

Once it has been established that the theory is compatible with the data, one can ask a separate interesting question. What is the allowed interval for a particular parameter of the fit? That is decided at the 95\% CL by constructing a 
\beq
\Delta\chi^2(m_h,\{p\})\equiv \chi^2(m_h,\{ p\})-\chi^2_{min}
\eeq
function, and finding the maximum interval range of $m_h$, allowing all variations of $\{ p\}$ needed to minimize $\Delta\chi^2$, such that $\Delta\chi^2<(1.96)^2$~\cite{chi2}. This has been done in the Standard Model, for example in fig.\ 10.5 on page 137 of~\cite{Alcaraz:2006mx}. I reprint here in fig.~\ref{fig:chi2} the $\Delta\chi^2$ determination from the Summer 2009 update~\cite{LEPEWWG 2009} of the LEP Electroweak Working Group. The interval has a lower limit well below the direct limit of $114.4\gev$, but the upper limit of the interval is $\sim 200\gev$. There are various precise numbers given, for various assumptions of how to treat $\alpha(m_Z)$ and what complete collection of observables to include, so I am  being a little conservative and quoting a squiggly number a little above most of the precise numbers listed. See the LEP Electroweak Working group website for the very latest in this evolving 
story~\cite{LEPEWWG}. Thus, I will maintain somewhat loosely that the indirect precision electroweak fits suggest that there is high range of Higgs masses excluded by the data
\beq
m_h\gsim 200 \gev~~{\rm excluded~at}~95\%~{\rm CL~~(Precision~ EW)}
\eeq
This indirect limit from precision electroweak should not be taken as rigidly as the direct limits from LEP2 and Tevatron, since it is relatively easy to form a conspiracy with other new states that allow for a heavier Higgs boson~\cite{Peskin:2001rw}.

Putting it all together, the current expectation is that the Higgs boson must have mass somewhere in one of two regions
\beq
114.4\gev < m_h<160\gev~~~{\rm or}~~~170\gev<m_h\lsim 200\gev.
\eeq
Another slightly more provocative way of saying it is that the Standard Model is incompatible with the data unless the Higgs boson mass falls within this limited range.
Discovery will have to wait for much more data at the Tevatron, or more probably, the LHC. LHC discovery channels and prospects are also evolving, especially as the energy of the machine may go through various unanticipated phases. Excellent places to read about the latest studies in this regard are at~\cite{Aad:2009wy,CMSTDR}.

\xsection{Perturbative Unitarity}

In the previous lecture we summarized the current experimental constraints on the Higgs boson and some of the ways that it can be found and studied at the Large Hadron Collider. In this lecture we would like to stretch and pull the electroweak theory to find regimes where the theory might not make sense any more. After all, the Fermi theory of four-fermion interactions described phenomena well when it was introduced many years ago, but physicists knew that it only had a finite energy range of applicability before the theory became strongly coupled and not useful. Might a similar fate befall the Standard Model electroweak sector?

The first place to poke at the electroweak theory is obviously in high-energy vector boson scattering.  The reason is that a divergence develops at increasing energy in the longitudinal polarization vector of the massive electroweak gauge bosons.  What might this mean to the calculability of our theory? To answer this question let us begin with considering the three polarization vectors of the massive gauge bosons $V=W^\pm,Z^0$ traveling with three-momentum $\vec k$  in the $\hat z$ direction with magnitude $k$: $\vec k=k\hat z$.  The on-shell four vector of this motion is $k^\mu=(E_k;\vec k)=(E_k;0,0,k)$ where $E_k^2=k^2+m_V^2$ is the energy. The three polarization vectors are
\bea
\epsilon^\mu_+(\vec k)&=& \frac{1}{\sqrt{2}}(0;1,i,0)~~{\rm (righthanded~polarized)}\\
\epsilon^\mu_-(\vec k)&=& \frac{1}{\sqrt{2}}(0;1,-i,0)~~{\rm (lefthanded~polarized)}\\
\epsilon^\mu_L(\vec k)&=&\frac{1}{m_V}(k;0,0,E_k)~~{\rm (longitudinally~polarized)}\label{eq:Lpol}
\eea
where the polarization vectors satisfy the required identities
\beq
k_\mu\epsilon^\mu_a(\vec k)=0~~{\rm and}~~\epsilon^\mu_a(\vec k)\epsilon^*_{b\mu}(\vec k)=-\delta_{ab}
\eeq
for all polarizations $a,b=+,-,L$.

Even the most casual inspection of these equations throws up a caution flag: the longitudinal polarization vector diverges without bound for $E_k\gg m_V$. Thus, any computations of cross-sections with external longitudinally polarized massive vector bosons could very well ``go strong" as the center of mass energy increases above some critical energy, signaling the breakdown of the electroweak theory.  Investigation is warranted.  The more external longitudinally polarized vector bosons that are in the process the better, so let us as a thought experiment consider $W_L^+W^-_L\to W_L^+W^-_L$ scattering and ask at what energy it breaks down.

Not surprisingly, this concern was recognized  in the very early days of the electroweak theory. An excellent paper that summarizes the situation in the electroweak theory is by 
Lee, Quigg and Thacker~\cite{Lee:1977eg}.  Among other processes in their comprehensive study, they considered $W^+_LW^-_L\to W^+_LW^-_L$. There are seven tree-level diagrams to compute in the electroweak theory that can be grouped into three classes: \\ \\
{\it Four point interaction (FP)}: $WWWW$ four-point interaction. \\
{\it Gauge exchange (GE):} $s$- and $t$-channel $\gamma$ and $Z$ boson exchange.\\
{\it Higgs exchange (HE):} $s$- and $t$-channel Higgs boson exchange.\\ \\
The amplitudes of any of these classes of diagrams can be written as an exchange in the high center of mass energy limit $s,t\gg m_V^2,m^2_h$:
\beq
{\cal A}={\cal A}^{(2)}s^2+{\cal A}^{(1)}s+{\cal A}^{(0)}
\eeq
Computations reveal that 
\bea
{\cal A}^{(2)}& = &{\cal A}^{(2)}_{FP}+{\cal A}^{(2)}_{GE} \to 0
\\
{\cal A}^{(1)}&=& {\cal A}^{(1)}_{FP}+{\cal A}^{(1)}_{GE}+{\cal A}^{(1)}_{HE}\to 0 \\
{\cal A}^{(0)}&=& {\cal A}^{(0)}_{FP}+{\cal A}^{(0)}_{GE}+{\cal A}^{(0)}_{HE}\to 
-\frac{2m_h^2}{v^2} \label{eq:A0}
\eea
A miracle of cancellations has happened.  The amplitude does not grow without bound as we go to higher and higher energy.  Instead, the amplitude asymptotes to a constant value.

This miracle is equivalent to the miracle of the Goldstone boson equivalence 
theorem~\cite{Chanowitz:1985hj}, which states that amplitudes of longitudinal boson scattering at high energy are equivalent to amplitudes with the Goldstone bosons that ultimately are absorbed as the longitudinal components of the vector bosons up to $m_V^2/s$ corrections:
\beq
{\cal A}(W^+_LW^-_L\to W^+_LW^-_L)={\cal A}(\phi^+\phi^-\to \phi^+\phi^-)
+{\cal O}\left(\frac{m_W^2}{s}\right).
\eeq
The charged Goldstone bosons, $\phi^\pm$ and $\phi^0$, are  the three states in the Standard Model Higgs doublet
\beq
\Phi=\vector{\phi^+}{\frac{1}{\sqrt{2}}(v+h)-i\frac{\phi^0}{\sqrt{2}}}~~{\rm and}~~\Phi^\dagger=\vector{\phi^-}{\frac{1}{\sqrt{2}}(v+h)+i\frac{\phi^0}{\sqrt{2}}}^T.
\eeq
Expanding the Higgs potential 
\beq
V(\Phi)=\lambda \left( \Phi^\dagger\Phi-\frac{v^2}{2}\right)^2
\eeq
about the vev, one finds the relevant interaction lagrangian
\beq
{\cal L}_\phi=-\frac{m_h^2}{v}h\phi^+\phi^--\frac{m^2_h}{2v^2}\phi^+\phi^-\phi^+\phi^-+\cdots
\label{eq:goldstones}
\eeq
where $m_h^2=2\lambda v^2$.

From the interaction lagrangian of eq.~(\ref{eq:goldstones}) one can compute the amplitude
\beq
{\cal A}(\phi^+\phi^-\to \phi^+\phi^-)=-\frac{m^2_h}{v^2}\left( \frac{s}{s-m_h^2}+\frac{t}{t-m^2_h}\right)
\to -\frac{2m_h^2}{v^2}
\eeq
where the far right term comes from taking the limit of $s,t\gg m_h^2$. This result matches what was obtained in eq.~(\ref{eq:A0}). Of course, in this approach we did not necessarily expect a problem.
Scalar bosons do not have ``diverging polarization vectors" to worry about. This is a case where looking at the problem with more appropriate degrees of freedom reveals simply that a problem we thought might exist never can exist.

Despite the successes of the electroweak theory in controlling its scattering of longitudinal vector bosons, a concern remains. The amplitude scales with the as-yet unknown Higgs boson mass. If the Higgs boson mass is too large then the theory is strongly interacting and we cannot compute anymore. This is not a surprise. The amplitude is really just the Higgs boson self-coupling $m^2_h/v^2\sim \lambda$, and any coupling that grows too large will create difficulties in a perturbation theory expansion.  For example, the $W^+W^+\to (WW)_{\rm loop}\to W^+W^-$ one-loop amplitude should scale as $\sim 2\lambda^2/16\pi^2$. This is appropriately sketchy -- the factor of 2 is from two one-loop diagrams for this process, and the factor of $16\pi^2$ is the generic loop factor. Thus, the one-loop contribution would compete with the tree-level amplitude of $W^+W^-\to W^+W^-$ of size $\lambda$ if $\lambda\sim 8\pi^2$. So, naively, we can say that the electroweak Higgs theory breaks down if $m_h\sim \sqrt{2\lambda}v=4\pi v\simeq 3.1\tev$.

As an interlude, we can approach the question of perturbativity from the perspective of the decays of the Higgs boson into longitudinal vector boson states. The partial widths of Higgs boson decays to $W$-bosons and $Z$-bosons are
\bea
\Gamma(h\to WW) & = & \frac{1}{16\pi}\frac{m_h^3}{v^2}\left( 1-4\frac{m_W^2}{m_h^2}+12\frac{m^4_W}{m^4_h}\right) \sqrt{1-4\frac{m_W^2}{m^2_h}}\, ,  \\ 
\Gamma(h\to ZZ) & = & \frac{1}{32\pi}\frac{m_h^3}{v^2}\left( 1-4\frac{m_Z^2}{m_h^2}+12\frac{m^4_Z}{m^4_h}\right) \sqrt{1-4\frac{m_Z^2}{m^2_h}} \, .
\eea
The width grows quite strongly with $m_h$, which is another manifestation of the strong coupling involving longitudinal vector bosons at high energy.  Explicit calculation of the longitudinally polarized final states compared to transversely polarized as a function of $m_h$ shows that the ratio scales as $m_h^4/8m_W^4$ for large $m_h$, which is expected. One measure of perturbativity is to ask at what Higgs boson mass does the tree-level computed width equal the mass.  Of course, when they are equal the tree-level computation is not valid, but it is a well-defined algorithm to understand the scales at which the theory is behaving badly. Using the above equations, which dominate the width\footnote{For example, the only would-be competitor is the final state of top quarks, but the branching fraction into top quarks is never the leading decay of the Higgs boson for any Higgs mass. Furthermore, as the Higgs boson mass increases well above $t\bar t$ threshold, its branching fraction decreases rapidly with respect to the vector boson branching fractions because of the latter's $m_h^3$ scaling.}, we find that $\Gamma_h=m_h$ when $m_h\simeq 1.4\tev$.   Likewise, the width is well above $m_h/2$ for $m_h\gsim 1\tev$, and so we can say that the Higgs boson is not a respectable, narrow width particle if its mass is in the trans-TeV regime\footnote{The computer program HDECAY~\cite{Djouadi:1997yw} gives $\Gamma_h\simeq 650\gev$ for $m_h=1\tev$.}. This is one measure of validity of perturbation theory.

Let us go back to the longitudinal scattering process $W^+_LW^-_L\to W^+_LW^-_L$. The limit of $3.1\tev$ that we derived earlier for the Higgs boson mass based on the validity of a perturbative expansion of longitudinal $W$ scattering was not very rigorous. We can do better by asking ourselves what Higgs boson mass corresponds to a formal violation of unitarity if we compute only at tree level. The path to answering this question starts with expanding the amplitude in terms of 
partial waves~\cite{partial wave}
\beq
{\cal A}=\sum_{\ell=0}^\infty {\cal A}_\ell,~~{\rm where}~~{\cal A}_\ell=16\pi(2\ell +1)P_\ell(\cos\theta)a_\ell .
\eeq
The variable $\ell$ labels the spin-$\ell$ partial wave, and $P_\ell(\cos\theta)$ are Legendre Polynomials with $\theta$ being the angle at which the final state $W^+$ deviates from the $W^+$ incoming direction.  Using orthogonality of the Legendre polynomials, and integrating over $|A|^2$ one finds the cross-section
\beq
\sigma=\sum_\ell \sigma_\ell,~~{\rm where}~~\sigma_\ell=\frac{16\pi}{s}(2\ell +1) |a_\ell |^2.
\label{eq:sigma1}
\eeq
Conservation of probability for elastic scattering requires
\beq
a_\ell =\frac{e^{2i\delta_\ell}-1}{2i}.
\label{eq:al}
\eeq
Varying $2\delta_\ell$ from $0$ to $2\pi$ sweeps out a circle of radius $\frac{1}{2}$ centered at $(0,\frac{1}{2})$ in the complex plane of ${\rm Im}(a_\ell)$ vs.\ ${\rm Re}(a_\ell)$. Nowhere on that circle is it possible to have ${\rm Re}(a_\ell)>1/2$ or ${\rm Re}(a_\ell)<-1/2$, which implies the perturbativity rule that
\beq
|{\rm Re}(a_\ell)|\leq \frac{1}{2}.
\label{eq:Rehalf}
\eeq
Since the theory ultimately is perturbative, violating at tree-level the condition expressed by \eq{eq:Rehalf} is equivalent to saying the tree-level computation is unreliable and our perturbative description of the theory is not valid at high energies. Of course, it does not mean that the tree-level contribution cannot be greater than $\frac{1}{2}$, since there can be cancellations at higher order to bring a tree-level result of $\frac{1}{2}+\epsilon$ down to below $\frac{1}{2}$. Nor should it be considered as the rigorous value to compare with the reliability of a calculation.  However, in a perturbation theory, it is expected that being near the unitarity limit of $a_\ell$ at some fixed low order is an indication that the perturbation expansion may be in trouble.  The criteria is well-defined, but as usual with any discussion of this nature, the physics content of the precise statement is not as precise. But let us take it seriously.

To compute the values of the Higgs boson mass that violates \eq{eq:Rehalf} we must first find the various partial wave expansion coefficients. In the very high energy limit $s\gg m_W^2$ the zeroth partial wave is
\beq
a_0=\frac{1}{16\pi s}\int_{-s}^0{\cal A}dt=-\frac{m^2_h}{16\pi v^2}
\left[ 2+\frac{m^2_h}{s-m^2_h}-\frac{m^2_h}{s}\log\left( \frac{s}{m_h^2}\right)
-\sum_{n=2}^\infty \frac{(-1)^n}{n-1}\left(\frac{m_h^{2}}{s}\right)^n  \right]\nonumber
\eeq
In the  limit that the energy is much greater than the Higgs mass one finds
\beq
a_0\to -\frac{m^2_h}{8\pi v^2}~~{\rm in~the~limit}~~s\gg m_h^2.
\eeq
Applyling \eq{eq:Rehalf} to this result one finds
\beq
\left| -\frac{m^2_h}{8\pi  v^2}\right|\leq \frac{1}{2}~~\Longrightarrow~~
m_h<2v\sqrt{\pi}=870\gev.
\eeq
What this is purported to mean (see caveats below) is that there is no perturbative description of the Standard Model for arbitrarily high energies if the Higgs boson mass is greater than this critical mass of $870\gev$. 

The perturbativity limit of $870\gev$ can be reduced even further down to about $710\gev$ by taking into account more $2\to 2$ scattering amplitudes that depend on the Higgs boson mass, such as $W_LW_L\to Z_LZ_L$,  etc.~\cite{HHG,Djouadi:2005gi,Reina:2005ae}.  There is a matrix of these kinds of amplitudes.  When diagonalizing it, one finds a particular linear combination of incoming states and outgoing states that has the highest $a_0$ partial wave.  After some analysis one finds that the amplitude is such that a Higgs boson mass greater than $710\gev$ violates \eq{eq:Rehalf}.

Be careful how you think about this bound of $710\gev$. The number is computed by a precise {\it definition} -- tree-level partial wave unitarity of two-to-two processes in the electroweak theory -- but the number's {\it physical meaning} is not as precise, for the same reasons that we discussed above. For example, the electroweak theory does not go from a highly convergent well-behaved perturbation theory at $m_h=709\gev$ to a disastrously out of control non-perturbative theory at $m_h=711\gev$. Although it is true that the  pretense that the calculation is under control is self-evidently suspect above the perturbative unitarity limit,  higher-order corrections are still required to make decent  predictions when the Higgs mass is large even if below the perturbative unitarity limit.  Indeed, higher-order corrections may push the amplitude over the perturbative unitarity edge. Likewise, if you dream up new physics that cancels the tree-level graphs and enables a Higgs boson of, say, $900\gev$ to satisfy the unitarity limits at tree-level, that does not mean that at the next order of computation things remain perturbatively under control.  Declarations of perturbative unitarity cannot be made solely upon unprincipled, manufactured cancellations at any finite order in perturbation theory.

\xsection{High Scale Perturbativity and Vacuum Stability}

When considering the full domain of applicability of the electroweak theory, we must ask what the behavior of the couplings is at very high energy. In the Standard Model there are several couplings that are reasonably large at the electroweak scale: the gauge couplings $g_i=\{ 0.41,0.64,1.2\}$, the top Yukawa coupling $y_t=\sqrt{2}m_t/v\simeq 1$, and the Higgs boson self-coupling $\lambda =m^2_h/2v^2$.  When scattering at very high energies there can be large logarithms $\log(E/M_{EW})$ with prefactors of these couplings. Large logarithms are best summed by renormalization group techniques.  Therefore, in this lecture, we will look at the behavior of the large couplings of the Standard Model in the ultraviolet by employing the techniques of renormalization group evolution.  These RG equations are~\cite{Machacek:1984zw,Arason:1991ic}\footnote{In the RG equations I have taken into account that my Higgs self-coupling is defined by $V=\lambda|\Phi^\dagger \Phi|^2+\cdots$ (i.e., $m_h^2=2\lambda v^2$ where $\langle \Phi\rangle=v/\sqrt{2}$), whereas the one of refs.~\cite{Machacek:1984zw,Arason:1991ic} is defined by $V=\frac{1}{2}\lambda|\Phi^\dagger \Phi|^2+\cdots$ (i.e., $m_h^2=\lambda v^2$ where $\langle \Phi\rangle=v/\sqrt{2}$).}
\bea
\frac{dg_1}{dt}&=& \frac{41}{10} \frac{g_1^3}{16\pi^2},~~~
\frac{dg_2}{dt}= -\frac{19}{6} \frac{g_2^3}{16\pi^2},~~~
\frac{dg_3}{dt}= -7\frac{g_3^3}{16\pi^2} \label{eq:RGEs} \label{eq:sm rges} \\
\frac{dy_t}{dt}&=& \frac{y_t}{16\pi^2}\left(-\frac{17}{20} g_1^2- \frac{9}{4} g_2^2 - 8 g_3^2+\frac{9}{2}y_t^2\right)\nonumber \\
\frac{d\lambda}{dt}&=& \frac{1}{16\pi^2}\left( 24\lambda^2
-\lambda \left( \frac{9}{5}g_1^2+9g_2^2 +12y_t^2\right)
+\frac{9}{8}\left( \frac{3}{25}g_1^4+\frac{2}{5}g^2_1g^2_2+g_2^4\right) 
-6y_t^4\right) \nonumber
\eea
where $t\equiv \ln (Q/Q_0)$, and
\bea
g_1&=& \sqrt{\frac{5}{3}}\frac{\sqrt{4\pi\alpha(m_Z)}}{\cos\theta_W}\simeq 0.46~~{\rm (GUT~normalized)} \\
g_2 & = & \frac{\sqrt{4\pi\alpha(m_Z)}}{\sin\theta_W}\simeq 0.65 \\
g_3 & = &\sqrt{4\pi \alpha_3(m_Z)}\simeq 1.2 \\
\lambda & = &\frac{m_h^2}{2v^2}~~{\rm where}~v=246\gev .
\eeq

We will look for two breakdowns of the theory in our analysis: perturbative validity and vacuum stability in the UV. We begin with perturbative validity~\cite{Cabibbo:1979ay}. If any of the couplings gets very strong at some high scale we will posit that the Standard Model theory is no longer a good description above that scale. The $\lambda$ coupling is uncertain since we do not know the Higgs boson mass, but if it is large it will continue to grow into the ultraviolate due to the $+24\lambda^2$ term in its RG equation. When the  coupling goes strong by some definition, e.g.\ $\lambda>4\pi$ or $\lambda>\sqrt{4\pi}$, at some scale, it is also the case that it will quickly develop a divergence, or Landau pole, at a scale very soon above that. We of course are discussing computations that are being performed at fixed order in perturbation theory and are not technically valid in the strong coupling regime. Nevertheless, we have no reason to suspect that a strongly coupled theory at one loop would suddenly be very well behaved at higher order, so it is a good approximation to continue with the analysis.  Thus, we will conflate the two definitions into one and call this scale $Q_{LP}$ -- the scale at which $\lambda$ diverges into a Landau Pole.  If our theory encounters a scale $Q_{LP}$ in its RG evolution into the UV, we will say that it is perturbatively valid for all $E<Q_{LP}$, but unknown or not valid for $E>Q_{LP}$.

\begin{figure}[t]
\begin{center}
\includegraphics[width=0.7\textwidth]{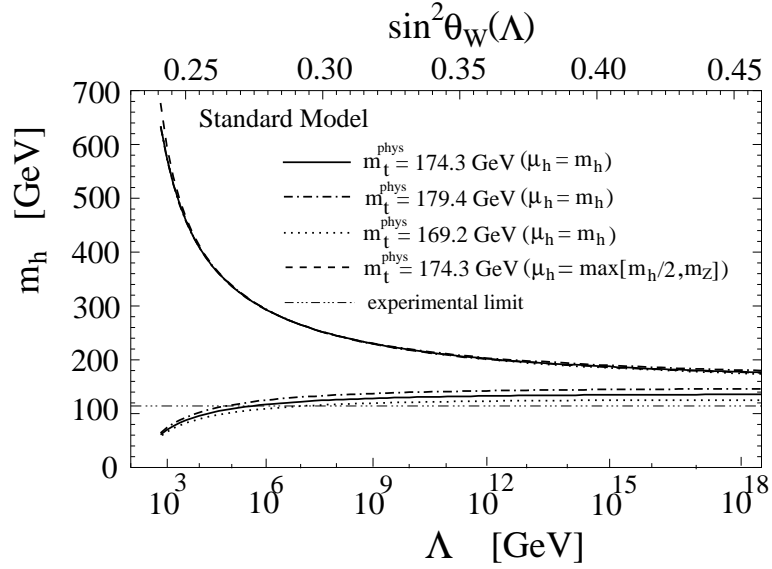}
\vspace{0.3cm}
\caption{The upper curve is to be interpreted as the Higgs mass $m_h$ that leads to a Landau pole in the Higgs self coupling at the scale $\Lambda\, (=Q_{LP})$. The lower curve is to be interpreted as the Higgs mass $m_h$ that leads to the Higgs self coupling turning negative at the scale $\Lambda\, (=Q_{NG})$. Plot taken from ref.~\cite{Tobe:2002zj}.
\label{fig:nightmare}}
\end{center}
\end{figure}

Inspection of the RG equations~(\ref{eq:RGEs}) shows us that in the limit of large Higgs mass the term that dominates the RG equation of the Higgs self-coupling is the $+24\lambda^2$ term.  The RG equation in this limit simplifies to 
\beq
\frac{d\lambda}{dt}\simeq \frac{3\lambda^2}{2\pi^2}
\eeq
which is easy to solve
\beq
\lambda(Q)=\frac{\lambda(Q_0)}{1-\frac{3\lambda(Q_0)}{2\pi^2}\ln(Q/Q_0)}.
\eeq
The Landau pole occurs where the denominator goes to zero, which enables us to solve for the Landau Pole $Q_{LP}$ scale in terms of $\lambda(Q_0)$ at the scale $Q_0$. Choosing $Q_0=m_h$ one finds
\beq
1-\frac{3\lambda(m_h)}{2\pi^2}\ln\left(\frac{Q_{LP}}{m_h}\right)=0~~\Longrightarrow~~
Q_{LP}=m_h\exp \left( \frac{4\pi^2v^2}{3m_h^2}\right)
\eeq
A few example values are if $m_h=200\gev$ $(300\gev)$ the Landau pole scale would be
$Q_{LP}\simeq 9\times 10^9\gev$ $(2\times 10^6\gev)$.  In fig.~\ref{fig:nightmare} the upper curve traces out the ordered pair values of $(m_h,Q_{LP})$ where $\Lambda$-axis of the plot should be interpreted as $Q_{LP}$ for this upper curve.  The plot was made more carefully~\cite{Tobe:2002zj} than the simple approximation above, with all couplings included in the RG evolution at two loops.  Uncertainties in the top quark mass are not terribly important for this particularly calculation. What we see is that if the Higgs boson mass is less than $180\gev$ the theory is perturbative all the way up to the Planck scale of $\sim 10^{18}\gev$.

The second breakdown of the theory may be vacuum instability~\cite{Cabibbo:1979ay,Sher:1988mj,Altarelli:1994rb,Tobe:2002zj}.
Motivated by the various terms in the RG equation for the Higgs boson self-coupling, and the uncertainty in $\lambda$ due to present ignorance of the Higgs mass, we can envision a scenario for a light Higgs boson where the $-6y_t^4$ term dominates in the RG equation for $\lambda$. If that is the case then at some scale $\lambda$ turns negative. Let us call this scale $Q_{NG}$ where NG is short for ``$\lambda$ goes negative." This is a disaster for the Higgs potential, since $\lambda \phi^4$ is the only important term in the potential when the field values of $\phi$ are very large. Thus, it is to be expected that new physics should come in at scales roughly near $Q_{NG}$ to lift the potential up, whatever the appropriate degrees of freedom are, and keep our vacuum stable for at least the lifetime of the universe.

In fig.~\ref{fig:nightmare} the lower curve traces the ordered pair $(m_h,Q_{NG})$ solution. The $x$-axis label $\Lambda$ should be interpreted as $Q_{NG}$ for the lower curve. For example, if the Higgs mass were to have been found at about $60\gev$ it would have implied that our vacuum is unstable, unless new physics came in at about the TeV scale to lift the potential. If the Higgs mass is greater than about $130\gev$ then there is no vacuum stability concern all the way up to the Planck scale. 

As a final comment to this lecture, some people have labeled the Higgs boson mass range of $130\gev \lsim m_h\lsim 180\gev$ the ``nightmare scenario", since by the arguments above there would be no firm computational reason to declare with certainty that new physics must be present below the Planck scale. Anxiety increases when one realizes that the precision electroweak analysis and direct limits of the Higgs boson that we discussed in an earlier lecture are forcing us into that Higgs mass window independently.   To me this concern is not so severe, since I interpret the data as suggesting that the Higgs sector coupling $\lambda$ must match to a reasonable perurbative coupling at some high scale, perhaps through a few steps but nevertheless the physics from here to the Planck scale is perturbative. This gives slight preference for the Higgs mass to be in the ``nightmare" range without the terror. Minimal supersymmetry is an example theory of this category.

\xsection{Adding Another Higgs Boson: Electroweak Symmetry Breaking\label{add higgs}}

We have analyzed the Standard Model Higgs boson and find an appealing framework for giving mass to vector bosons and the fermions. However, it is natural to ask ourselves about the possibility of having more than one Higgs boson, just as we have more than one electron (i.e., muon and tau leptons), more than one down quark  (i.e., strange and bottom quarks), and more than one up quark (i.e., charm and top quarks). 
Let us take that option seriously in this section and discuss some of the issues relevant for electroweak symmetry breaking.

As we will see in the next section, adding another Higgs boson can be very dangerous from the perspective of flavor changing neutral currents (FCNC) if we allow it to couple arbitrarily to Standard Model fermions. For this reason, let us suppose that the new Higgs boson $\Phi_2$ does not couple to the Standard Model fermions, although its Standard Model quantum numbers $({\bf 2},1/2)$ are the same as the first Higgs boson $\Phi_1$. This can be ensured by making it odd under a $Z_2$ symmetry, while the other Higgs boson is even:
\bea
\Phi_1 & \to & \Phi_1 \\
\Phi_2 & \to & -\Phi_2.
\eea
The fermions of the Standard Model can be odd or even $f\to \pm f$ with impunity, and the gauge bosons are even. 

The most general Higgs potential that we can write down with these symmetries is
\bea
V(\Phi_1,\Phi_2) & = & \mu^2_1|\Phi_1|^2+\mu^2|\Phi_2|^2+\lambda_1|\Phi_1|^4+\lambda_2|\Phi_2|^4+\lambda_3|\Phi_1|^2|\Phi_2|^2 \nonumber \\
& & +\lambda_4 (\Phi_2^\dagger\Phi_1)(\Phi^\dagger_1\Phi_2)+
\left[ \frac{\lambda_5}{2}(\Phi_1^\dagger\Phi_2)^2+c.c.\right].
\label{eq:type I potential}
\eea
Without loss of generality all the $\lambda_i$ couplings are real. Hermiticity demands it for all $\lambda_i$ except $\lambda_5$, which can be rotated to real by $\Phi_2$ absorbing its phase. 

The potential must be bounded from below in all field directions. One can test for dangerous runaway directions by parametrizing field excursions such as $(\Phi_1,\Phi_2)\to (a,a)$ where $a$ can be arbitrarily large in value.  Here are a few field directions to consider and their corresponding unbounded from below (UFB) constraints:
\beq
\begin{array}{cc}
(\Phi_1,\Phi_2)~{\rm direction} & {\rm UFB~constraint} \\
\hline
(a,0)& \lambda_1>0 \\
(0,a)& \lambda_2>0 \\
(a,a)& \sum_i\lambda_i>0 \\
(\lambda_2^{1/4}a,\lambda_1^{1/4}a) & \lambda_3+\lambda_4+\lambda_5+2\sqrt{\lambda_1\lambda_2}>0
\end{array}
\eeq
The third constraint is never as powerful as the fourth constraint and  is superfluous to write down. 

The most general vacuum expectations values for the two $\Phi_{1,2}$ Higgs fields can be expressed 
as (see, e.g., \cite{DiazCruz:1992uw})
\beq
\Phi_1=\vector{0}{v_1},~~{\rm and}~~\Phi_2=\vector{u_2}{v_2e^{i\xi}}.
\eeq
A non-zero $u_2$ would indicate the full breaking of $SU(2)\times U(1)_Y$, and in particular the photon would obtain mass. Let us carry forward for now with this general vacuum structure to investigate the consequences.

For the potential to be stable we must be at a minimum, which is to be determined by setting $dV/d\phi_i=0$ for all real fields $\phi_i$ defined in
\beq
\Phi_1=\vector{\phi_1+i\phi_2}{\phi_3+i\phi_4},~~{\rm and}~~\Phi_2=\vector{\phi_5+i\phi_6}{\phi_7+i\phi_8}.
\eeq
The minimization conditions~\cite{DiazCruz:1992uw} derived from each of these derivatives are
\beq
\begin{array}{cl}
\phi_1: & (\lambda_4+\lambda_5)u_2v_1v_2\cos\xi=0 \\
\phi_2: & (\lambda_4-\lambda_5)u_2v_1v_2\sin\xi =0 \\
\phi_3: & v_1[\mu^2_1+2\lambda_1v^2_1+\lambda_4v^2_2+\lambda_5v^2_2\cos 2\xi +\lambda_3(u^2_2+v^2_2)]=0 \\
\phi_4: & \lambda_5v_1v_2^2\sin 2\xi =0 \\
\phi_5: & u_2[\mu^2_2+\lambda_3v^2_1+2\lambda_2(u^2_2+v^2_2)]=0\\
\phi_6: & 0=0 \\
\phi_7: & v_2\cos\xi [\mu^2_2+(\lambda_3+\lambda_4+\lambda_5)v^2_1+2\lambda_2(u_2^2+v^2_2)]=0 \\
\phi_8: & v_2\sin\xi[\mu^2_2+(\lambda_3+\lambda_4-\lambda_5)v^2_1+2\lambda_2(u_2^2+v^2_2)]=0 
\end{array}
\eeq
Inspection of these equations tells us that except for possibly at special points where $\lambda_4\pm \lambda_5=0$ there is no hope in satisfying the minimization conditions if all three vevs $v_1,v_2,u_2$ are nonzero. 

If we allow both $(v_1,u_2)$ to be non-zero and $v_2=0$, we are in the situation of full breaking of $SU(2)\times U(1)_Y$.  When analyzing the spectrum one finds four massless scalars, which are Goldstone bosons to be eaten by the four generators of $SU(2)\times U(1)_Y$ to form longitudinal components of massive $W^\pm$, $Z$ and $A$.  There remain four physical scalar states in the spectrum
\bea
m^2_{\tilde \phi_1} & = & (\lambda_4+\lambda_5)(u^2_2+v^2_1) \\
m^2_{\tilde\phi_2}& = & (\lambda_4-\lambda_5)(u_2^2+v^2_1) \\
m^2_{\tilde\phi_3}& \simeq & (4\lambda_2-\lambda_3^2/\lambda_1)u^2_2\\
m^2_{\tilde\phi_4}&\simeq & 4\lambda_1v^2_1
\eeq
where the last two mass eigenstates are derived under the assumption that $u_2\ll v_2$. In order for this theory to have a stable minimum all $m^2_{\tilde\phi_i}$ must be positive, which puts the condition on the couplings that
\bea
\lambda_1>0,~~\lambda_4+\lambda_5 >0,~~\lambda_4-\lambda_5>0,~~4\lambda_1\lambda_2-\lambda^2_3>0
\eea
which is easily satisfied over large parts of parameter space. Thus, a random dart throw in the space of couplings of a general two Higgs doublet model can ``just as often" give a massive photon as a massless photon. It is for this reason that some people are turned off~\cite{Veltman:1997nm} by the general two Higgs doublet model compared to the Standard Model Higgs doublet theory that guarantees the photon does not get mass.

But let us carry on. We are more interested in the case where the symmetry breaking is proper  $SU(2)\times U(1)_Y\to U(1)_{em}$.  Thus, we take the other case where $v_1$ and $v_2$ are non-zero. The minimization conditions  become
\beq
\begin{array}{cl}
\phi_1: & 0=0 \\
\phi_2: & 0 =0 \\
\phi_3: & v_1[\mu^2_1+2\lambda_1v^2_1+\lambda_3 v^2_2+\lambda_4v^2_2+\lambda_5v^2_2\cos 2\xi ]=0 \\
\phi_4: & \lambda_5v_1v_2^2\sin 2\xi =0 \\
\phi_5: & 0=0\\
\phi_6: & 0=0 \\
\phi_7: & v_2\cos\xi [\mu^2_2+(\lambda_3+\lambda_4+\lambda_5)v^2_1+2\lambda_2v^2_2]=0 \\
\phi_8: & v_2\sin\xi[\mu^2_2+(\lambda_3+\lambda_4-\lambda_5)v^2_1+2\lambda_2 v^2_2]=0 
\end{array}
\eeq
Choosing $\mu^2_1$ and $\mu^2_2$  appropriately to zero out the conditions $\phi_3$ and $\phi_7$, we are left with only two non-trivial conditions yet to be satisfied:
\beq
\begin{array}{cl}
\phi_4: & \lambda_5v_1v^2_2\sin 2\xi=0 \\
\phi_8: & \lambda_5 v^2_1v_2\sin\xi=0 
\end{array}
\eeq
We will see shortly that we need $\lambda_5\neq 0$ so that leaves us with the requirement that $\xi=0$ or $\pi$.  We  choose $\xi=0$ for our convention, since the opposite sign $(\xi=\pi)$  can be reabsorbed by a field rephasing of $\Phi_1$ or $\Phi_2$, take your pick, which simultaneously flips the sign in front of the $\lambda_5$ term of the potential. 

We need to check if this solution is stable. To do that we require that the second derivative of the potential, i.e.\ the mass matrix, be positive definite.  The $8\times 8$ mass matrix in the $\{ \phi_1,\phi_2,\ldots,\phi_8\}$ basis is ${\cal M}^2_{\phi_i\phi_j}=$
\begin{tiny}
\beq
\left( \begin{array}{cccc|cccc}
-(\lambda_4+\lambda_5)v^2_2 & 0 & 0&0&(\lambda_4+\lambda_5)v_1v_2 & 0&0&0\\
0&-(\lambda_4+\lambda_5)v^2_2&0&0&0&(\lambda_4+\lambda_5)v_1v_2&0&0\\
0&0&4\lambda_1v^2&0&0&0&2(\lambda_3+\lambda_4+\lambda_5)v_1v_2 &0\\
0&0&0&-2\lambda_5v^2_2&0&0&0&2\lambda_5v_1v_2\\ \hline
(\lambda_4+\lambda_5)v_1v_2&0&0&0&-(\lambda_4+\lambda_5)v^2_1&0&0&0\\
0&(\lambda_4+\lambda_5)v_1v_2&0&0&0&-(\lambda_4+\lambda_5)v^2_1&0&0\\
0&0&2(\lambda_3+\lambda_4+\lambda_5)v_1v_2 & 0&0&0&4\lambda_2v^2_2&0\\
0&0&0&2\lambda_5v_1v_2&0&0&0&-2\lambda_5v^2_1 
\end{array}\right)\nonumber
\eeq
\end{tiny}
The large number of zeros in this matrix enables us to find very quickly what the eigenvalues are by solving four $2\times 2$ matrices. These matrices arise from $\phi_k\phi_{k+4}$ mixing for $k=1,2,3,4$. To begin with, we look at the $\phi_1\phi_5$ and $\phi_2\phi_6$ mixings, which have the same $2\times 2$ mass matrix:
\beq
{\cal M}^2_{\phi_1\phi_5}={\cal M}^2_{\phi_2\phi_6}=
\left( \begin{array}{cc}
 -(\lambda_4+\lambda_5)v^2_2 & (\lambda_4+\lambda_5)v_1v_2 \\
 (\lambda_4+\lambda_5)v_1v_2 & -(\lambda_4+\lambda_5)v^2_1
 \end{array}\right)
 \eeq
 which leads to four eigenstates
 \beq
 m^2_{G^\pm} & = & 0~~{\rm (charged~Goldstone~bosons)} \\
 m^2_{H^\pm}& =& -(\lambda_4+\lambda_5)(v^2_1+v^2_2)~~{\rm (charged~Higgs~bosons)}.
\eeq
Now let us look at $\phi_4\phi_8$ mixing:
\beq
{\cal M}^2_{\phi_4\phi_8}=
\left( \begin{array}{cc}
 -2\lambda_5v^2_2 & 2\lambda_5 v_1v_2  \\
2\lambda_5 v_1v_2 & -2\lambda_5 v^2_1
 \end{array}\right).
 \eeq
This leads to two eigenstates
\bea
m^2_{G^0} & = & 0 ~~{\rm (neutral~Goldstone~bosons)}  \\
m^2_{A^0}&=& -2\lambda_5(v^2_1+v^2_2)~~{\rm (neutral~pseudoscalar~boson)}.
\eea
Finally, there is $\phi_3\phi_7$ mixing:
\beq
{\cal M}^2_{\phi_3\phi_7}=
\left( \begin{array}{cc}
4\lambda_1v^2_1 & 2(\lambda_3+\lambda_4+\lambda_5)v_1v_2 \\
2(\lambda_3+\lambda_4+\lambda_5)v_1v_2 & 4\lambda_2v^2_2
\end{array}\right)
\eeq
This is the $2\times 2$ mass matrix for the two physical neutral scalar Higgs bosons of the theory, $h^0$ and $H^0$. The sum of the eigenvalues is the trace of the matrix
\beq
m^2_{h^0}+m^2_{H^0}=4\lambda_1v^2_1+4\lambda_2v^2_2.
\eeq
The mixing angle to rotate from $\{\phi_3,\phi_7\}$ basis to $\{h^0,H^0\}$ basis is usually called $\alpha$, which is defined by convention to satisfy
\beq
\vector{H^0}{h^0}=
\left( \begin{array}{cc}
\cos\alpha & \sin\alpha \\
-\sin\alpha & \cos\alpha
\end{array}\right)\vector{\phi_3}{\phi_7}.
\eeq
The solutions are obtained by simple eigenvalue, eigenvector analysis of the $2\times 2$ matrix, and one obtains
\bea
\sin 2\alpha & = & \frac{\eta v_1v_2}{\sqrt{(\lambda_1v_1^2-\lambda_2v_2^2)^2+4\eta^2 v^2_1v^2_2}} \\
\cos 2\alpha & = & \frac{\lambda_1v_1^2-\lambda_2v_2^2}{\sqrt{(\lambda_1v_1^2-\lambda_2v_2^2)^2+4\eta^2 v^2_1v^2_2}}
\eea
and
\bea
m^2_{H^0} & = &2\lambda_1v^2_1+2\lambda_2v^2_2+|2\lambda_1v^2_1-2\lambda_2v^2_2|
\sqrt{1+4\left(\frac{\eta v_1v_2}{\lambda_1v_1^2-\lambda_2v_2^2}\right)^2} \\
m^2_{h^0} & = & 2\lambda_1v^2_1+2\lambda_2v^2_2-|2\lambda_1v^2_1-2\lambda_2v^2_2|
\sqrt{1+4\left(\frac{\eta v_1v_2}{\lambda_1v_1^2-\lambda_2v_2^2}\right)^2}
\eea
where for simplicity I have defined 
\beq
\eta\equiv \frac{1}{2}(\lambda_3+\lambda_4+\lambda_5).
\eeq

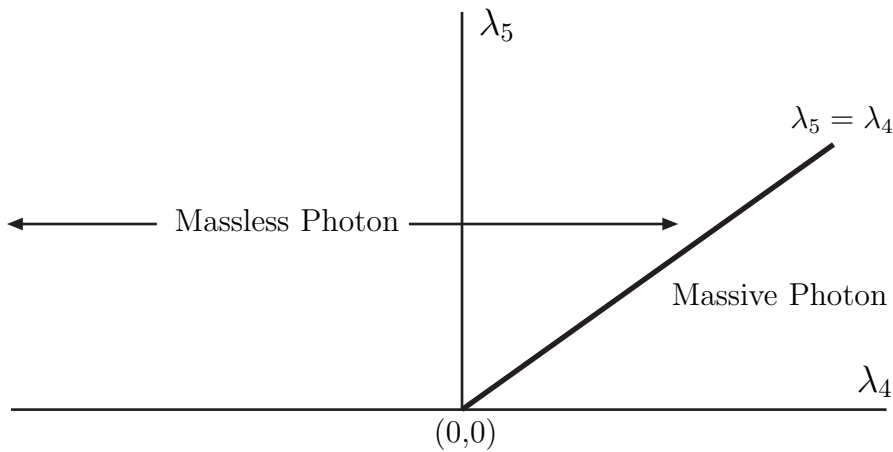
\begin{figure}[t]
\begin{center}
\begin{picture}(350,260)(0,90)
\SetWidth{1.0}
\Line(10,100)(340,100)
\Line(180,100)(180,250)
\SetWidth{2.0}
\Line(180,100)(320,200)
\Text(330,105)[lb]{\large $\lambda_4$}
\Text(187,240)[lb]{\large $\lambda_5$}
\Text(305,205)[lb]{$\lambda_5=\lambda_4$}
\Text(260,140)[lb]{Massive Photon}
\Text(72,168)[lb]{Massless Photon}
\SetWidth{1.0}
\LongArrow(160,170)(260,170)
\LongArrow(65,170)(10,170)
\Text(170,85)[lb]{(0,0)}
\end{picture}
\end{center}
\caption{Parameter space for massive versus massless photon in the type I two Higgs doublet model of \eq{eq:type I potential}. The partition of the parameter space between massless and massive photon for $\lambda_5<0$ is obtained by reflecting this plot around the $\lambda_5=0$ axis.}
\label{fig:lambdas}
\end{figure}

From our solution we have learned several things. We have computed  three massless states that correspond to the Goldstone bosons of $SU(2)\times U(1)_Y\to U(1)_{em}$ symmetry breaking. These states become the longitudinal components of $W^\pm$ and $Z^0$. We also require that the mass matrix be positive definite, which puts important constraints on the parameters of the theory. For example, from the charged Higgs and pseudo-scalar Higgs boson masses we know that
\beq
\lambda_4+\lambda_5<0,~~{\rm and}~~\lambda_5<0
\eeq
is required.  Note, the sign condition on $\lambda_4+\lambda_5$ is exactly opposite to that of the case where $u_2\neq 0$ and the photon gets a mass. An interesting question is whether the ``photon mass region" we specified earlier, through conditions only on $\lambda_4$ and $\lambda_5$, could have been consistent with $u_2=0$ and $v_2\neq 0$ vacua if other parameters were adjusted. The results here answer that question as a definitive no. If $\lambda_4+\lambda_5>0$ it is impossible to have a vacuum with $v_1,v_2\neq 0$, and the only option remaining is the massive photon vacuum of $u_2\neq 0$.   Fig.~\ref{fig:lambdas} plots the parameter space in the $\lambda_4$ vs.\ $\lambda_5$ plane that corresponds to massive photon and massless photon cases, in agreement with~\cite{DiazCruz:1992uw}.

Finally, it appears that there is a large region of $\lambda_5>0$ and $\lambda_4-\lambda_5<0$ that has no solution. However, the vacuum structure is symmetric about either sign of $\lambda_5$ because of the rephasing that we can do to change $v_2<0$ ($\xi=\pi$ case) to $v_2\to -v_2$ and $\lambda_5\to -\lambda_5$. Thus, the partition of the parameter space between massless and massive photon for $\lambda_5<0$ is obtained by reflecting fig.~\ref{fig:lambdas} around the $\lambda_5=0$ axis.

Let us point out another interest result in this analysis. The physical pseudoscalar mass scales to zero $m_{A^0}^2\to 0$ as  $\lambda_5\to 0$.  A massless electroweak pseudoscalar is not allowed by the data, and so if we want to avoid this problem we have to be sure that our theory allows a $\lambda_5$ coupling. The reason behind this result is the usual reason behind such scaling behavior. There is a symmetry that protects the pseudoscalar mass as $\lambda_5\to 0$.  If it were not for the $\lambda_5$ term in our potential then it would be possible to rephase one of the fields, say $\Phi_1$, arbitrarily with respect to the other field. When the field gets a vev, this global symmetry is spontaneously broken, thereby leading to a Goldstone boson. In that limit, $A^0$ is identified as a Goldstone boson of $\Phi_1$ rephasing symmetry.  To give $A^0$ mass we need  explicit breaking somewhere in the theory. The $\lambda_5$ term provides that for us since $(\Phi_1^\dagger\Phi_2)^2$ does not allow arbitrary rephasing of $\Phi_1$ with respect to $\Phi_2$.  For small coupling $\lambda_5$ the pseudoscalar mass is much smaller than the characteristic scale of the problems, $v^2_1+v^2_2$, and we might call it a pseudo-Nambu Goldstone boson. If $\lambda_5\sim 1$ then we can drop all such  modifiers since we can no longer make the pretense that the rephasing symmetry is approximately or nearly valid.

This kind of argument occurs time and time again in particle physics. The most famous early application is in pion physics. Why is the pion so small in mass with respect to the proton? The reason is that massless quarks respect an $SU(2)_L\times SU(2)_R$ chiral flavour symmetry -- left and right handed quarks can be separately rephased in $SU(2)$ space with respect to each other. Chiral symmetry breaking of QCD spontaneously breaks this to the vector subgroup $SU(2)_L\times SU(2)_R\to SU(2)_V$ leading to three massless Goldstone bosons $m^2_{\pi^{\pm,0}}=0$.  But we know pions are not exactly massless. They get their mass by virtue of the quarks' elementary masses $m_q q^\dagger_Lq_R+c.c.$, i.e.\ from tiny explicit breaking of the full chiral flavor symmetry that does not allow $q_L$ to be rephased completely independently from $q_R$. Thus, pions ultimately do get mass $m_\pi^2\propto m_q m_{\rm proton}$. Since $m_q\ll m_{\rm proton}$ we conclude that $m^2_{\pi}\ll m_{\rm proton}^2$. In this analogy, $\lambda_5$ is like the quark masses, and $v_{1,2}^2$ is like $\Lambda_{QCD}$ or the proton mass.

You might be concerned that in supersymmetric theories there is no such coupling $\lambda_5$. But do not fear. In that theory there is no $Z_2$ symmetry between the two Higgs doublets and we are allowed the all important $B_\mu \Phi_u\Phi_d$ interaction that explicit breaks any attempted independent rephasing. Thus, we already know that the pseudoscalar mass in supersymmetry must be $m^2_{A^0}\propto B_\mu$, as we will see later in lecture~\ref{susy higgs lecture}.

\xsection{Adding Another Higgs Boson: Flavor Changing Neutral Currents\label{sec:FCNC}}

In the previous section we assumed that the second Higgs boson possessed an odd charge under a global $Z_2$ discrete symmetry in order to forbid its coupling to fermions. We did this using the argument that generically one expects large flavor changing neutral current (FCNC) problems otherwise. In this lecture we discuss this challenge to a multi-Higgs doublet theory, and conclude the discussion with a simple theorem that expresses how a large class of theories with multiple Higgs bosons can avoid the bane of tree-level FCNC.

Our theory is the general two-Higgs doublet model with fields $\Phi_1$ and $\Phi_2$, just as in the previous lecture, but without any extra $Z_2$ symmetry. We will not focus on the electroweak symmetry breaking aspects of this model since the techniques of the previous lecture immediately apply. Rather, we worry about the Higgs boson interactions with the fermions. From the perspective of both EWSB and flavor physics, there is a Higgs field basis that is particularly interesting. Let me call it the ``Runge\footnote{Runge is my 3${}^{\rm rd}$\,great-grandfather of physics and was the world's expert on vectors. He wrote a famous book on the subject {\it  (Die Vektoranalysis des dreidimensionalen Raumes},\,1919), which led to his name being attached, somewhat undeservedly, to the Runge-Lenz vector of classical mechanics. In that tradition, and also because of its winsome euphony, I call it the ``Runge basis."} basis".
It is defined to be the basis in which one Higgs field carries the full vev, $\Phi_{vev}$, and the other Higgs field is perpendicular to it, $\Phi_\perp$:
\bea
\Phi_{vev}&=&\frac{v_1}{v}\Phi_1+\frac{v_2}{v}\Phi_2=\cos\omega\,\Phi_1+\sin\omega\,\Phi_2 \\
\Phi_\perp&=&-\frac{v_2}{v}\Phi_1+\frac{v_1}{v}\Phi_2=-\sin\omega\,\Phi_1+\cos\omega\,\Phi_2
\eea
where $v^2\equiv v^2_1+v^2_2$, $\cos\omega=v_1/v$ and $\sin\omega=v_2/v$.  The angle $\omega$ is usually denoted by $\beta$ in the literature, but to minimize confusions I want to only use $\beta$ in the supersymmetric two-Higgs doublet model later.

The Runge basis is very helpful from the EWSB point of view also, since we know the Goldstone bosons must be contained entirely within the field that gets the vev $\Phi_{vev}$. Thus, in this basis we can identify
\bea
\Phi_{vev}=\vector{G^\pm}{\frac{1}{\sqrt{2}}(v+\varphi'_1+iG^0)},~~{\rm and}~~
\Phi_{\perp}=\vector{H^\pm}{\frac{1}{\sqrt{2}}(\varphi'_2+iA^0)}
\eea
where $G^{\pm,0}$ are the Goldstone bosons, $H^\pm$ the physical charged Higgs bosons, $A^0$ the pseudoscalar Higgs boson, and $\varphi_{1,2}$ the physical scalar Higgs bosons.  

The Runge basis is helpful from the perspective of flavor physics since all the fermions must get mass only through $\Phi_{vev}$. Thus, after a suitable rotation of the fermion weak eigenstates into mass eigenstates, the couplings to $\varphi'_1$ must be diagonal,
\bea
\varphi'_1 \bar f_i f_j~~& : &~~ i \frac{m^f_{i}}{v}\delta_{ij}~~{\rm (Feynman~rule)},
\eeq
where $f$ indicates one of three species fermions: up-type quarks ($f_i=\{u,c,t\}$), down-type quarks ($f_i=\{d,s,b\}$) or leptons ($f_i=\{e,\mu,\tau\}$).
The couplings to $\varphi'_2$ and $A^0$ can be anything and are in general not diagonal:
\bea
\varphi'_2 \bar f_i f_j~~& : &~~ i \xi^f_{ij}~~{\rm (Feynman~rule)} \\
A^0\bar f_if_j ~~&:&~~i\xi^f_{ij}\gamma_5~~{\rm (Feynman~rule)},
\eea
and we will take $\xi_{ij}^f=\xi_{ji}^{f*}$~\cite{Atwood:1996vj}.

The CP even mass-eigenstates are a linear combination of $\varphi'_1$ and $\varphi'_2$ through the mixing angle $\alpha'$
\beq
\vector{H^0}{h^0}=\left( \begin{array}{cc} 
\cos\alpha' & \sin\alpha' \\
-\sin\alpha' & \cos\alpha' \end{array}\right)
\vector{\varphi'_1}{\varphi'_2}
\eeq
which leads to final expressions for the Feynman rules of the mass eigenstate scalars with the fermions:
\bea
H\bar f_if_j~~& : &~~i\cos\alpha'\, \frac{m^f_{i}}{v}\delta_{ij}+i\sin\alpha' \xi_{ij}^f\\
h\bar f_if_j~~&:&~~-i\sin\alpha'\,  \frac{m^f_{i}}{v}\delta_{ij} +i\cos\alpha'\, \xi^f_{ij}.
\eeq

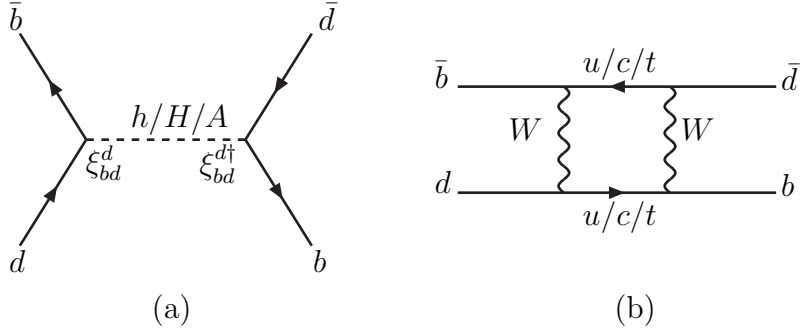
\begin{figure}[t]
\begin{center}
\begin{picture}(300,150)(0,30)
\SetWidth{1.0}
\DashLine(30,100)(90,100){3}
\ArrowLine(5,60)(30,100)
\ArrowLine(30,100)(5,140)
\ArrowLine(115,140)(90,100)
\ArrowLine(90,100)(115,60)
\ArrowLine(290,120)(170,120)
\ArrowLine(170,80)(290,80)
\Photon(210,80)(210,120){2}{4}
\Photon(250,80)(250,120){2}{4}
\Text(1,142)[lb]{$\bar b$}
\Text(1,50)[lb]{$d$}
\Text(118,142)[lb]{$\bar d$}
\Text(116,50)[lb]{$b$}
\Text(47,102)[lb]{$h/H/A$}
\Text(162,120)[lb]{$\bar b$}
\Text(162,80)[lb]{$d$}
\Text(293,119)[lb]{$\bar d$}
\Text(293,79)[lb]{$b$}
\Text(190,100)[lb]{$W$}
\Text(255,100)[lb]{$W$}
\Text(218,123)[lb]{$u/c/t$}
\Text(218,65)[lb]{$u/c/t$}
\Text(30,85)[lb]{$\xi^d_{bd}$}
\Text(73,84)[lb]{$\xi^{d\dagger}_{bd}$}
\Text(55,30)[lb]{(a)}
\Text(230,30)[lb]{(b)}
\end{picture}
\end{center}
\caption{Flavor changing neutral current contributions to $B_d^0-\bar B_d^0$ mixing from (a)  Higgs exchange diagrams in arbitrary two-Higgs doublet model, and (b) Standard Model gauge contributions. Note that the Standard Model diagrams are one-loop whereas the competing Higgs exchange is tree-level. Experiment is consistent with Standard Model results, which implies severe constraints on the Higgs flavor-changing neutral current couplings $\xi_{ij}^f\ll 1$.}
\label{fig:fcnc}
\end{figure}

The existence of arbitrary flavor couplings $\xi^f_{ij}$ in this extended Higgs sector is dangerous to flavor physics. Stringent flavor observables within this context are mass splittings in $F^0-\overline{F}^0$ mixing~\cite{Atwood:1996vj}.  Within the Standard Model, these mass splittings are accomplished via box diagrams with $W^\pm$ and quarks in the loop (see Fig.~\ref{fig:fcnc}). Experimental results are nicely consistent with these being the dominant source of FCNC. With generic $\xi^f_{ij}$ couplings in the two-Higgs doublet model, the mass splitting predictions can be significantly higher since they can occur through tree-level interactions of the sort $\bar q q'\to {\rm higgs}^*\to\bar q' q$.

Let us quantify the extent of this challenge to flavor physics compatibility. The Standard Model diagram and the Higgs exchange diagrams are given in Fig.~\ref{fig:fcnc}. The mass-splitting of  $F^0-\overline{F}^0$  resulting from the exchange of neutral Higgs bosons can be parametrized as
\beq
M_F\Delta M_F=B_F(\xi^f_{ij})^2\left( \cos^2\alpha'\frac{S_F}{m^2_{h^0}}
+\sin^2\alpha'\frac{S_F}{m^2_{H^0}}-\frac{P_F}{m^2_{A^0}}\right)
\label{eq:mfdmf}
\eeq
where
\bea
S_F &=& \frac{B_F f_F^2 M_F^2}{6}\left(1+\frac{ M_F^2}{(m^f_i +m^f_j)^2}\right)\\
P_F &=& -\frac{B_F f_F^2 M_F^2}{6}\left(1+\frac{11 M_F^2}{(m^f_i +m^f_j)^2}\right)
\eea
and $ij=ds, db$ and $uc$ for $K^0,B^0_d$ and $D^0$ respectively. $B_F$ constants are recalibration factors for having used the vacuum insertion approximation, and $f_F$ are the decay constants. See Table~\ref{table:mesons} for their values.  The formulae of eq.~(\ref{eq:mfdmf}) with these constants should give computations to accuracy within factors of ${\cal O}(1)$.
The  experimental values for $\Delta M_F$ are given in  Table~\ref{table:mesons}.
\begin{table}[t]
\centering
\begin{tabular}{cccc}
Meson (quarks) & $B_F$ & $f_F$ (GeV) & $\Delta M_F^{\rm expt}$ (GeV) \\
\hline
$K^0\,(\bar sd)$  & 0.79 & 0.159&  $(3.476\pm 0.006)\times 10^{-15} $ \\                                              
$B_d^0\, (\bar bd)$  & 1.28 & 0.216 &  $(3.337\pm0.033)\times 10^{-13} $  \\
$D^0\,(\bar cu)$  & 0.82 & 0.165 & $(0.95\pm0.37)\times 10^{-14} $  \\ 
\hline
\end{tabular}
\caption{Data associated with the neutral mesons $K^0$, $B^0_d$ and $D^0$. Values are obtained from~\cite{Lunghi:2007ak,Lubicz:2007yu}.}
\label{table:mesons}
\end{table}

It is straightforward to use the results of the theory computation above and the experimental limits of Table~\ref{table:mesons} to place limits on how large $\xi_{ij}$ can be~\cite{Atwood:1996vj,Gupta:soon}. These limits depend on many factors, including $\alpha'$ and the three different Higgs boson masses. For simplicity, let us assume that all Higgs mass are equal to the common value $m_{\rm Higgs}$ and define 
\beq
\tilde\xi^f_{ij}\equiv \xi^f_{ij}\,\left( \frac{120\gev}{m_{\rm Higgs}}\right)^2.
\eeq

The Standard Model prediction for $\Delta M_{B^0_d}$, for example, is $\Delta M_{B^0_d}^{SM}=(4.5\pm1.0)\times 10^{-13}$ GeV ~\cite{Lunghi:2007ak}.  This gives a sense for the computational uncertainties involved. Therefore, let us be simple-minded and conservative here to illustrate the important point that $\xi_{ij}^f$'s need to be small. Let's require that the Higgs boson contributions for the $B^0_d$ mass splitting is bounded by $\Delta M_{B^0_d}<10^{-12}\gev$.  This translates into a bound of $\tilde\xi_{db}^d  \lsim  10^{-4}$, which is quite small. Similar results follow for the other flavor-changing $\xi_{ij}^f$ couplings, and it is hard to imagine that random choices for the entries would satisfy the constraints.

There is a general class of solutions to this problem while admitting the existence of extra Higgs bosons in the spectrum. Tree-level FCNCs do not arise if Higgs boson interactions with the fermions take the form
\begin{eqnarray}
\Delta {\cal L}_{f}=y^d_{ij}\bar Q_i\,  F_u(\{ \Phi_k\})\, d_{jR} + y^u_{ij}\bar Q_i\, F_d(\{ \Phi_k\}) \, u_{jR} +
y^e_{ij}\bar L_i\, F_e(\{ \Phi_k\})\, e_{jR} + c.c.
\label{no fcnc} 
\end{eqnarray}
where $F_{u,d,e}(\{\Phi_k\})$ are functions of Higgs fields $\{ \Phi_k\}$, constrained only by the requirements that they are independent of the fermionic flavor indices $i,j$ and that $F_u$ transform like an $SU(2)_L$ doublet with hypercharge $-1/2$, and $F_d$ and $F_e$ transform like $SU(2)_L$ doublets with hypercharge $1/2$.

The generalized form of \eq{no fcnc} subsumes many ideas already present in the literature. For example, the Standard Model Higgs sector is $F_u=H^c_{SM}$ and $F_d=F_e=H_{SM}$. The Type II~\cite{Higgs type models} two-Higgs doublet model~\cite{HHG} is $F_u=H_u$ and $F_d=F_e=H_d$. The type I two-Higgs doublet model~\cite{HHG} is $F_u=\Phi_1$ and $F_d=F_e=\Phi^*_1$ with an additional $\Phi_2$ that does not couple to fermions. The leptophilic Higgs model of ref.~\cite{Su:2009fz} is $F_u=\phi^*_q$, $F_d=\phi_q$ and $F_e=\phi_l$. 

There are an infinite variety of models that can satisfy \eq{no fcnc}. However, principles are expected to be at work to fall into this general class if there are more than one Higgs boson. In the case of supersymmetry, as we will see, the type II structure follows from holomorphy of the superpotential. In the case of type I theories, it is usually assumed that the second Higgs boson has, for example,  a discrete $Z_2$ symmetry $\Phi_2=-\Phi_2$ that forbids its direct coupling to fermions whereas $\Phi_1$ does not. This was the example that we initially studied in lecture~\ref{add higgs} with the promise that it was a good illustration that did not violate flavor constraints. This approach has been nicely illustrated recently in the model of~\cite{Ambroso:2008kb}, where a selection rule was identified that forbids the second Higgs doublet from coupling to fermions.  

The main summary point here is that  extra  Higgs bosons are likely to violate FCNC constraints due to tree-level mediated interactions unless a principle is invoked that ensures compliance with the condition of \eq{no fcnc}. The most straightforward principles that can do this are restrictive global symmetries or, in supersymmetric theories, holomorphy of the superpotential. Of course, concerns about FCNCs do not completely vanish when the tree-level contributions are eradicated. One-loop contributions generally are always present. Most famously, the charged Higgs boson and top quark loop contribution to $b\to s\gamma$ inclusive rare decays competes with the Standard Model $W$-boson loop and top quark loop-induced decay, yielding important constraints on the masses and couplings of any multi-Higgs boson doublet model~\cite{Misiak:2006zs}.

\xsection{Essay on the Hierarchy Problem\label{sec:hierarchy}}

The Standard Model with its postulated Higgs boson is an unsatisfactory theory for many reasons. There are several direct data-driven reasons why it is incomplete. The Standard Model has no explanation for the baryon asymmetry of the Universe. For some reason there are many more protons than anti-protons, and if the Universe is cooling from some primordial hot state with particles in thermal equilibrium that is unexpected. Some mechanism that goes beyond the Standard Model dynamics must be at play. Similarly, there is plenty of astrophysical evidence for dark matter in the Universe. This dark matter helps to explain structure formation, cosmic microwave background radiation, galactic rotation curves, etc. The problem is the Standard Model has no candidate explanation, and new physics must be invoked.  

There are many other reasons to consider physics beyond the Standard Model. The three gauge forces could be unified and the matter unified within representations of a grand unified symmetry. The many different parameters of the flavor sector are hard to swallow without envisaging deeper principles that organize them. Furthermore, the integration of the Standard Model with quantum gravity is not obvious, and many people think a deeper structure, such as that built from strings and branes, is needed for their coexistence. 

So, there are many reasons to believe that there is physics beyond the Standard Model. But the issue that is front and center for us now, relevant to Higgs boson physics and electroweak explorations at the Large Hadron Collider, is the Hierarchy Problem. The Hierarchy Problem is often expressed as a question: Why is the weak scale ($\sim 10^2\gev$) so much lighter than the Planck scale ($\sim 10^{18}\gev$)? It is  a bit uninspirting when phrased this way, since it begs the question of why we should be concerned at all about a big difference in scales. Blue whales are much bigger than nanoarchaeum equitans but we do not believe nature must reveal a dramatic new concept for us to understand it.

A knowing-just-enough-to-be-dangerous naive way to look at the Standard Model is that it is the ``Theory of Particles", valid up to some out-of-reach scale where gravity might go strong, or some other violence is occurring that we do not care about. It is a renormalizable theory. I can compute everything at multiple quantum loop order, set counter terms, cancel infinities that are fake since they do not show up in observables, and then make predictions for observables that experiment agrees with. Quadratic divergences of the Higgs boson self-energy, which so many people make a fuss about, are not even there if I use dimensional regularization. The theory is happy, healthy, stable, and in no need of any fixes.  New physics {\it near the electroweak scale} can still be justified~\cite{Wells:2003tf,split susy} after dismissing naturalness as impossibly imprecise to understand at this stage, or as merely a purely philosophical problem\footnote{One definition of philosophy could be the study of incompatible views held by smart, fully informed people of high integrity. In addition to such lofty debates as Free Will vs.\ Determinism, it includes deciding on the utility of various Naturalness criteria in theory building.}, but the urgency is certainly diminished for it being {\it at the electroweak scale}.

This viewpoint  that the Standard Model is complete can be challenged right at the outset. It is simply not the ``Theory of Particles" -- it does breakdown. It is an effective theory, even if one thinks there is a way to argue it being valid to some very remote high scale where gravity goes strong, such as $M_{Pl}$. As an effective theory, all operators should have their dimensionality set by the cutoff of the theory~\cite{Polchinski:1992ed}. 
If operator ${\cal O}^{(d)}$ has dimension $d$ then its coefficient is $c\Lambda^{4-d}$, where $\Lambda$ is the cutoff of the theory and $c$ is expected to be $\sim 1$ in value. Irrelevant operators with $d>4$ cause no harm. Same goes for $d=4$ marginal operators. The Standard Model is almost exclusively a theory of $d=4$ marginal operators with its kinetic terms, gauge interaction terms, and Yukawa interaction terms.  What is potentially problematic is the existence of any $d<4$ relevant operators. In that case, the coefficients should be large, set by the cutoff of the theory.

Does the Standard Model have any gauge-invariant, Lorentz-invariant relevant $d<4$ operators to worry about? Yes, two of them. The right-handed neutrino Majorana mass interaction terms $\nu_R^Ti\sigma^2\nu_R$, which is $d=3$, and  the Higgs boson mass operator  $|H|^2$, which is $d=2$.  The expectations of effective field theories is that the scale of the coefficients of these operators should be set by high-scale cutoffs of the theory and disconnected from any other surviving mass scale in the infrared. As we saw in lecture~\ref{sec:neutrinos} this expectation is nicely met in the neutrino case, where we have actually measured the masses and see a self-consistent picture for large Majorana masses for the right-handed neutrinos, which serve as cutoff scale coefficients. These coefficients are tied to lepton number violation, for example, and not electroweak symmetry breaking, and therefore have naturally large values above the weak scale.

It did not have to be that way with neutrino physics. It could have been that the neutrino sector was shown experimentally to have independent left and right-handed components and the masses were of order the weak scale. This would have been in violation of effective field theory expectations, unless new symmetries tied to the weak scale were discovered to protect the right-handed neutrino from getting a large Majorana mass.  The fact that that the neutrino sector conforms with effective field theory expectations should be viewed as contributing evidence for these concepts.

In contrast to the neutrino operator, the $d=2$ Higgs mass operator in the Standard Model is unwelcome if its coefficient is not set to the weak scale. From our effective field theory expectations, the Lagrangian operator  should be
\beq
\Delta {\cal L}_{rel}=c\Lambda^2 |H|^2
\label{eq:eff op}
\eeq
This is a potential disaster for the theory, since from our previous work on the Higgs potential we stated that the Higgs mass must be  $-\mu^2\sim v^2$, where $v\simeq 246\gev$ is the Higgs boson vacuum expectation value needed to reproduce the $W$ and $Z$ masses. If we assume the Standard Model to be a valid theory to very high energies $E\gg v$, that implies the cutoff of the Standard Model effective theory is $\Lambda\gg v$, which implies the coefficient of $|H|^2$ is $|\mu^2|=\Lambda^2\gg v^2$, which is in contradiction to $v\simeq \sqrt{-\mu^2}$.  The effective theory would then need the coefficient $c$ in \eq{eq:eff op} to be finetuned to an extraordinarily small and unnatural~\cite{Giudice:2008bi} value $c\sim v^2/\Lambda^2$ to make all the scales work out properly. The concern about how this can be so is the Hierarchy Problem.

The discussion is a bit abstract, but it bears fruit with direct computations. As one example out of an infinite number that would demonstrate the Hierarchy Problem, consider the possible existence of other scalar fields $\phi_i$ at higher energies. The assumption is that if there is a Higgs boson in the theory, then there is every reason to believe that there can be other scalars. They can have mass at the weak scale, intermediate scale, Planck scale, whereever.  Let us suppose that we put one $\phi$ at the cutoff scale $\Lambda$ of the theory. The operator $|\phi|^2|H|^2$ immediately gives a quantum correction to the Higgs mass operator coefficient of $\sim \Lambda^2/16\pi^2$. Although the $1/16\pi^2$ can help a little, if $\Lambda \gg 4\pi v$ there is serious problem, and the weak scale cannot exist naturally with such a hierarchy.  For this reason, it is often assumed that naturalness of the Higgs boson sector of the Standard Model effective theory requires new physics to show up at some scale below $\Lambda\sim 4\pi v\sim {\rm few}\tev$.

There are many different approaches to solving the Hierarchy Problem. They can be put into three categories. The first category suggests that  there is new physics at the TeV scale and the cutoff $\Lambda$ in \eq{eq:eff op} is in the neighborhood of the weak scale. Supersymmetry~\cite{Martin:1997ns}, little Higgs~\cite{little higgs}, conformal theories~\cite{conformal}, and extra dimensions~\cite{xdim reviews} can be employed in this approach. For example, supersymmetry accomplishes the task by a softly broken symmetry, where $\Lambda$ is the supersymmetry breaking mass scale. All quadratic divergences to the Higgs boson mass operator cancel up to supersymmetry breaking terms. Extra dimensions accomplishes it by banishing all mass scales accessible to the Higgs boson above the TeV scale.  The second category suggests that  fundamental scalars are banished from the theory that could form invariant $|\varphi|^2$ operators. For example, this is the approach of Technicolor~\cite{Lane:2009ct} and top-quark condensate theories~\cite{top condensate} that try to reproduce the symmetry breaking of a Higgs boson with the condensate of a fermion bilinear operator. Higgsless theories and their variants are also in this category~\cite{higgsless}. And finally, the third category of solutions to the Hierarchy Problem suggests that large statistics of finetuned solutions dominate over the fewer number of non-tuned solutions in the landscape, leading to a higher probability of our Universe landing in a highly tuned solution ($c\ll 1$). Thus, guided by concerns over the cosmological constant problem\footnote{Although not directly related to external particle physics interactions, the cosmological constant can be considered as the coefficient of yet another gauge-invariant, Lorentz-invariant operator -- the operator being merely a constant: $-{\cal L}_{cc}=\Lambda_{cc}^4$.  The tiny value of this coefficient, $\Lambda^4_{cc}\simeq (10^{-3}\, {\rm eV})^4$, is well below any conceivable theory expectation. It is the elephant in the room for effective field theories. However, it is an unexpressed article of faith among most particle physicists that the solution to the Cosmological Constant Problem lies in the details of mysterious quantum gravity, and that the new concepts buried in that unknown solution do not materially affect the natural solution to the Hierarchy Problem. Landscapists question that assumption.}, it has been suggested that this statistical, stringy naturalness  over the landscape may take precedence over normal naturalness envisioned from effective field theories~\cite{Douglas:2006es}. This is controversial with conflicting claims over unrealistic theories; nevertheless, it is an interesting idea that might one day be impactful. 

In the following sections, I will address the issue from the first perspective. I will start by positing  supersymmetry and investigate its implications. These implications are numerous, but I will focus on very direct issues of the Higgs boson masses and interactions and not so much the phenomenological implications of superpartners, which are required to stabilize the cutoff at the electroweak scale.  I will then discuss the implications for the Higgs boson sector of Randall-Sundrum warped extra dimensions. There is a new state, the radion, whose properties are very similar to the Higgs boson. When it mixes with the Higgs boson there is no pure Higgs state nor radion state, but an admixture which has implications for collider phenomenology.  Finally, I will end the lectures with comments about how the $|H|^2$ operator does not only cause us headaches with the Hierarchy Problem, but also can be a window to new hidden sector physics not accessible through any other means except through mixing with the Higgs boson mass relevant operator.

\xsection{Higgs Sector of Minimal Supersymmetry\label{susy higgs lecture}}

There are two good reasons why supersymmetry requires two Higgs doublets.  The first reason is anomaly cancellation. The superpartner of the Higgs boson is a fermion, which contributes to triangle gauge anomalies. Since Standard Model fermions already cancel the gauge anomalies by themselves, the addition of another fermion charged under $SU(2)_L\times U(1)_Y$ introduces an uncompensated contribution. A second fermion that is the vector complement of the first cancels the anomalies. For this reason, the minimal supersymmetric Standard Model (MSSM) introduces a $H_u$ Higgs doublet and its vector complement $H_d$.

The second reason why two Higgs doublets are required is the inability otherwise to give mass to all the fermions. A supersymmetric theory is defined by the particle content, gauge symmetries, and its superpotential.  The superpotential is a dimension-three potential constructed out of the chiral superfields of the theory but not their conjugates.  A chiral multiplet $\hat \Phi$ consists of a scalar $\Phi$ and a fermion $\psi$: $\hat\Phi=(\Phi,\psi)$. Holomorphy of the superpotential says that if the scalar component of the chiral multiplet $\hat H_u$ gives mass to the up-quarks then it cannot give mass to the down quarks since $\hat H^*_u$ is not allowed in the superpotential.  

By  convention the two Higgs doublets are $H_u$ with hypercharge $+1/2$ and $H_d$ with hypercharge $-1/2$:
\beq
H_u=\vector{\varphi_u^+}{\varphi^0_u}~~~{\rm and}~~~
H_d=\vector{\varphi_d^0}{\varphi_d^-}
\eeq
where $\varphi$ fields are complex.

Starting with our two Higgs doublets, we construct the  general gauge-invariant superpotential of the MSSM 
\beq
W=y_u \hat Q\cdot\hat H_u \hat u^c-y_d\hat Q\cdot\hat H_d \hat d^c+y_e\hat L\cdot \hat H_d\hat e^c
-\mu \hat H_u\cdot \hat H_d
\eeq
where the $\cdot$ symbol is by definition the $SU(2)$ contraction $A\cdot B=\epsilon_{ij} A_iB_j$, where $i,j$ are $SU(2)$ indices and $\epsilon_{12}=-\epsilon_{21}=1$ and $\epsilon_{11}=\epsilon_{22}=0$. We have also assumed $R$-parity conservation to disallow baryon number and lepton number violating interactions. Supersymmetry invariance of the lagrangian is automatically achieved when the lagrangian is derived by applying a few easy rules to the superpotential and the fields:
\beq
{\cal L}_W &=&-\left[ \frac{1}{2}\left( \frac{\partial^2W}{\partial\hat \Phi_i\partial\hat\Phi_j}\right)_{\hat \Phi=\Phi}\psi_i\psi_j+c.c.\right] -F_i^*F_i-\frac{1}{2}D^a D^a-\frac{1}{2}D'^2
\label{eq:susylag}
\eeq
where
\bea
F^*_i&=&\left. \frac{\partial W}{\partial \hat\Phi_i}\right|_{\hat\Phi=\Phi} \\
D^a & = & g\sum_i (\Phi^*_i)_m T^a_{mn} (\Phi_i)_n \\
D' & = & g' \sum_i \frac{Y_i}{2}\Phi^*_i\Phi_i 
\eea
where the index $i$ is for particle species, and the indices $m,n$ run over the $SU(2)$ components of the representation under which $\Phi_i$ is charged, and $T^a_{mn}$ are the $SU(2)$ generators.

Applying these rules to our theory we find that the Higgs boson interactions with fermions are governed by
\beq
{\cal L}_y= y_u Q^\dagger\cdot H^*_u u_R-y_d  Q^\dagger\cdot H^*_d d_R-y_e L^\dagger\cdot H^*_d e_R + c.c.
\eeq
which are the same kinds of interactions that we are used to.  The fermion masses are then simply 
\beq
m_{u}=y_u \langle H^0_u\rangle, ~~~m_d=y_d\langle H^0_d\rangle,~~{\rm and}~~
m_e=y_e \langle H^0_d\rangle .
\eeq
We will come back to these mass terms later.

The focus now turns to the Higgs boson sector. The supersymmetric lagrangian terms that involve only Higgs boson interactions are from the $F$-terms and $D$-terms provided in \eq{eq:susylag}. Supersymmetry of course is not an exact symmetry and must be broken by some mechanism. Softly broken supersymmetry, i.e., supersymmetry breaking that does not reintroduce quadratic sensitivities to the cutoff, gives two types of terms beyond ${\cal L}_W$  of relevance for our discussion. First, each chiral superfield gets a ``soft mass term" for its scalar component. In other words,  if the theory contains the chiral superfield $\hat \Phi$, the lagrangian contains a supersymmetry breaking mass term $m_\Phi^2|\Phi|^2$.
The second kind of supersymmetry breaking term of importance here are the so-called ``$A$-terms" and ``$B$-terms". The rule for identifying these terms is to take each operator of the superpotential and evaluate the chiral fields at their scalar field values.  A coefficient of one dimension higher than the superpotential coefficient is needed, and that is identified with a supersymmetry breaking mass. If the superpotential operator is dimension three then the resulting supersymmetry breaking interaction in the lagrangian is called an $A$-term by convention, and if the superpotential operator is dimension two it is called a $B$-term:
\bea
W=\lambda \hat \Phi_1\hat\Phi_2\hat\Phi_3 &\Longrightarrow & {\cal L}_{soft}=A_\lambda \Phi_1\Phi_2\Phi_3 +c.c. \\
W=m \hat \Phi_1\hat\Phi_4& \Longrightarrow &{\cal L}_{soft}=B_m \Phi_1\Phi_4+c.c.
\eea
where $\lambda$ and $m$ have dimensions of zero and one respectively, whereas $A_\lambda$ and $B_m$ have dimensions of one and two respectively.
In the Higgs sector of the Standard Model there is only a bilinear term that can be utilized to make a soft supersymmetry breaking interaction of this type. It is the $\mu$ term in the lagrangian, which leads to a $B_\mu$ term in the lagrangian
\beq
W=-\mu  \hat H_u\cdot \hat H_d~\Longrightarrow~ {\cal L}=-B_\mu H_u\cdot H_d+c.c.
\eeq

Putting all the terms together, the complete Higgs potential of the MSSM is
\beq
V&=&(|\mu|^2+m^2_{H_u})|H_u^2|+(|\mu|^2+m^2_{H_d})|H_d|^2 +(B_\mu H_u\cdot H_d+c.c.) \nonumber\\
& & +\frac{1}{8}(g^2+g'^2)(|H_u|^2-|H_d|^2)^2+\frac{1}{2}g^2|H_u^\dagger H_d|^2.
\label{eq:higgs potential}
\eeq
I'll assume in these lectures that all the parameters of the Higgs potential, the $\mu$ term and the $B_\mu$ term, as well as the gaugino masses and Yukawa-like $A$-terms are all simultaneously real. For this reason, I will subsequently write $|\mu|^2$ as $\mu^2$.  If all the parameters could not be rotated to real simultaneously  we would have complex parameters to deal with, which only complicates the issues I would like to highlight here. However, we can make two comments about it. If large CP violating phases were present in the supersymmetric theory beyond the normal CKM phase, the theory encounters a significant challenge to make compatible with measured CP violation limits from such well-constrained observables as the electric dipole moments of the electron and neutron.   Assuming this challenge is overcome, the CP violating effects could be an important aspect of multi-Higgs boson phenomenology~\cite{Pilaftsis:1999qt,Carena:2002bb}.

We assume that the neutral components of each Higgs boson get a real and positive vev, consistent with the expectations we developed in lecture~\ref{add higgs}.  By convention I'll define
\beq
\langle H_u^0\rangle=\frac{v_u}{\sqrt{2}},~~
\langle H_d^0\rangle=\frac{v_d}{\sqrt{2}},~~
\tan\beta=\frac{v_u}{v_d},~~
{\rm and}~v^2=v_u^2+v_d^2\simeq (246\gev)^2.
\eeq
Given these definitions, we could proceed straight  with the Higgs potential of \eq{eq:higgs potential} as our starting point to directly analyze the vacuum structure and particle spectrum of our theory.  However, we will find it convenient again to first transform our Higgs potential into the Runge basis and then analyze, as we did in lecture~\ref{add higgs}.

Recall that in the Runge basis we identify one Higgs boson as the state that gets all the vev, and the other Higgs boson(s) orthogonal to it. There is one subtlety here that we did not experience in lecture~\ref{add higgs}. In lecture~\ref{add higgs} the Higgs boson doublets all had the same quantum numbers. The convention of the MSSM is to define $H_u$ and $H_d$ with opposite hypercharge. This of course was totally unnecessary to do, but alas we must pay our respects to history.  The Higgs boson is also charged under $SU(2)$ of course, but the great thing about $SU(2)$, unlike any other $SU(N)$ with $N>2$, is that the conjugate representation is equivalent to the original representation.  Thus, we are free to construct a field from $H_d$ that has the same quantum numbers as $H_u$. That field is $H^c_d=i\sigma^2H_d^*$.  Now we can define the Runge basis by adding an apple ($H_u$) with an apple ($H_d^c$) rather than an apple ($H_u$) with an orange ($H_d$):
\beq
\Phi_{vev}&=& \frac{v_d}{v}H_d^c+ \frac{v_u}{v}H_u \\
\Phi_\perp &=& \frac{v_u}{v}H_d^c-\frac{v_d}{v}H_u
\eeq
or, since, $\sin\beta=v_u/v$ and $\cos\beta=v_d/v$ we can write
\beq
\vector{\Phi_{vev}}{\Phi_\perp}=\left( \begin{array}{cc} \cos\beta & \sin\beta \\ -\sin\beta & \cos\beta
\end{array}\right)\vector{H^c_d}{H_u}.
\eeq

Because this is a two Higgs doublet model and $\Phi_{vev}$ contains all the vev for electroweak symmetry breaking we know immediately where the Goldstone bosons ($G^{\pm,0}$) and heavy Higgs bosons ($H^\pm,A^0$) reside just as in lecture~\ref{add higgs}. We show this as the components of the new Higgs fields:
\beq
\Phi_{vev}=\frac{1}{\sqrt{2}}\vector{\sqrt{2}G^+}{v+h'+iG^0}~~{\rm and}~~
\Phi_{\perp}=\frac{1}{\sqrt{2}}\vector{\sqrt{2}H^+}{H'+iA^0}.
\eeq
The $h',H'$ states are the neutral scalar Higgs bosons. Unlike the case for the other fields in this basis, there is no reason why $h'$ and $H'$ need be mass eigenstates. Despite this, the Runge basis is very convenient to analyze these states as well, as we will see below.

Let us go ahead and rewrite \eq{eq:higgs potential} in the Runge basis:
\beq
V&=& (\mu^2+s_\beta^2m^2_{H_u}+c_\beta^2m^2_{H_d}-s_{2\beta}B_\mu)|\Phi_{vev}|^2
+(\mu^2+c_\beta^2 m^2_{H_u}+s_\beta^2 m_{H_d}^2+s_{2\beta}B_\mu)|\Phi_\perp|^2 \nonumber \\
& & +\left[c_\beta s_\beta (m_{H_u}^2-m_{H_d}^2)-c_{2\beta}B_\mu\right]
(\Phi^\dagger_{vev}\Phi_\perp+c.c.) \label{eq:higgs runge} \\
&& +\frac{1}{8}(g^2+g'^2)\left[c_{2\beta}(|\Phi_\perp|^2-|\Phi_{vev}|^2)+s_{2\beta}(\Phi_{vev}^\dagger
\Phi_\perp +c.c.)\right]^2 +\frac{1}{2}g^2|\Phi_{vev}\cdot \Phi_\perp|^2 \nonumber
\eeq
Let us analyse the conditions for electroweak symmetry breaking and particle spectroscopy using this potential.  To do so, we define eight real scalar fields that are the components of $\Phi_{vev}$ and $\Phi_\perp$:
\beq
\Phi_{vev}=\frac{1}{\sqrt{2}}\vector{\phi_1+i\phi_2}{\phi_3+i\phi_4}~~~{\rm and}~~~
\Phi_\perp=\frac{1}{\sqrt{2}}\vector{\phi_5+i\phi_6}{\phi_7+i\phi_8}.
\eeq

The potential is at its minimum if all $dV/d\phi_i=0$ when evaluated at $\phi_i=0$ for all $\phi_i$ except $\phi_3=v$.  Six of the eight conditions are trivially satisfied ($0=0$), leaving us with two for consideration:
\bea
\phi_3: & & \frac{1}{8}(g^2+g'^2)v^2c^2_{2\beta}+\mu^2+m_{H_d}^2c^2_\beta+m^2_{H_u}s^2_\beta-B_\mu s_{2\beta}=0 \\
\phi_7: && -\frac{1}{8}(g^2+g'^2)v^2c_{2\beta}s_{2\beta} +\frac{1}{2}(m_{H_u}^2-m_{H_d}^2)s_{2\beta} -B_\mu c_{2\beta} =0
\eea
Taking two different linear combinations of these equations puts them in a form that is more familiar:
\bea
\frac{2}{v}\left[ (\phi_3:)c_{2\beta}-(\phi_7:)s_{2\beta}\right] &\Longrightarrow &
\mu^2+\frac{m_Z^2}{2}=\frac{m^2_{H_d}-\tan^2\beta\, m^2_{H_u}}{\tan^2\beta -1} \label{eq:ewsb1} \\
\frac{1}{v}\left[(\phi_e:)s_{2\beta}+(\phi_y:)c_{2\beta}\right] &\Longrightarrow &
\frac{B_\mu}{\sin 2\beta}=\mu^2+\frac{1}{2}(m^2_{H_d}+m^2_{H_u}) \label{eq:ewsb2}
\eeq
The reason why these conditions are frequently more useful than the $(\phi_3:)$ and $(\phi_7:)$ conditions is because model builders often feel more confident about setting scalar mass boundary conditions, such as $m_{H_u}^2$ and $m^2_{H_d}$. These masses are also correlated well with the other sparticle masses, such as the top squark, selectron, etc. On the other hand, a theoretical understanding of the origin of $\mu$ and $B_\mu$ is more elusive. For example, $\mu$ is not a soft breaking mass, but rather shows up in the superpotential itself. Although explanations exist for why the $\mu$ scale should be close to the soft supersymmetry breaking mass scale and the electroweak scale, none are overwhelmingly compelling. This is an ongoing active area of research but for a generic review of the situation see~\cite{Polonsky:1999qd}. Thus,
oftentimes we wish merely to know what $\mu$ and $B_\mu$ need to be in order to have electroweak symmetry breaking work out for a given set of soft supersymmetry breaking parameters -- theory can then adjust to that ``experimental need." But we run too far afield in discussing the famous $\mu$ and $B_\mu$ problem of supersymmetry.

The upshot of our symmetry breaking conditions is that we have two equations, eqs.~(\ref{eq:ewsb1}) and~(\ref{eq:ewsb2}), that set four parameters ($\mu^2,B_\mu,m_{H_d}^2,m_{H_u}^2$).   If we choose to set $\mu^2$ and $B_\mu$ in terms of $m_{H_d}^2$ and $m_{H_u}^2$ from these two minimization conditions, we can proceed to compute the full mass matrix from ${\cal M}^2_{ij}=d^2V/d\phi_id\phi_j$ evaluated at the minimum. This matrix will depend on all the various remaining parameters $g,g',v,\tan\beta,m_{H_d}^2,m_{H_u}^2$.  If you do compute the matrix with these parameters you will find that the terms do not  depend on $m_{H_d}^2$ and $m_{H_u}^2$ in independent combinations, but rather only on the difference $m_{H_d}^2-m_{H_u}^2$.  Now, if we define for convenience
\beq
\tilde m^2=-\frac{1}{4}(g^2+g'^2)v^2+(m^2_{H_u}-m^2_{H_d})\sec 2\beta
\eeq
the full $8\times 8$ mass matrix in the $\{\phi_1,\phi_2,\ldots,\phi_8\}$ basis becomes
\begin{small}
\beq
{\cal M}^2_{ij}=\left( \begin{array}{cccc|cccc}
0&0&0&0&0&0&0&0 \\
0&0&0&0&0&0&0&0 \\
0&0& m^2_Z c^2_{2\beta} &0&0&0&-m^2_Zs_{2\beta}c_{2\beta}&0 \\
0&0&0&0&0&0&0&0 \\
\hline
0&0&0&0&\tilde m^2+m^2_W&0&0&0 \\
0&0&0&0&0&\tilde m^2+m^2_W&0&0 \\
0&0&-m^2_Zs_{2\beta}c_{2\beta}&0&0&0&\tilde m^2+m^2_Zs^2_{2\beta}&0 \\
0&0&0&0&0&0&0& \tilde m^2
\end{array}\right) \label{eq:runge matrix}
\eeq
\end{small}

From the mass matrix of \eq{eq:runge matrix} we can readily identify that, as promised, the zero mass modes of $\phi_1$, $\phi_2$, and $\phi_4$ are the Goldstone bosons of the theory. The massive modes are
\bea
m^2_{A^0} &= & \tilde m^2 \\
m^2_{H^\pm}&=& \tilde m^2+m_W^2,
\eea
Note that using eqs.~\ref{eq:ewsb1} and~\ref{eq:ewsb2} we can rewrite $\tilde m^2=2B_\mu/\sin 2\beta$, thus proving the claim we made at the end of lecture~\ref{add higgs} that $m_A^2\propto B_\mu$ in supersymmetry.

The mass matrix for the $\{h',H'\}$ system comes from mixing among $\phi_3\phi_7$,
\beq
{\cal M}^2_{\rm Higgs}=\left( \begin{array}{cc}
m^2_Z c^2_{2\beta} & -m_Z^2 s_{2\beta}c_{2\beta} \\
-m_Z^2 s_{2\beta}c_{2\beta} & \tilde m^2+m_Z^2 s^2_{2\beta}
\end{array}\right).
\label{eq:higgs runge matrix}
\eeq
This matrix can be easily diagonalized and mass eigenvalues and  eigenvectors identified.

As promised the scalar Higgs bosons are quite revealing in the  Runge basis, even though they are not mass eigenstates.  In contrast to other bases, the only place where a supersymmetry breaking parameter shows up is in the $22$ component of the mass matrix. This parameter is very likely to be above the weak scale, perhaps significantly above the weak scale, and is only bounded by naturalness concerns.  It is a theorem of positive definite matrices that the smallest eigenvalue is always smaller than the smallest diagonal component. Therefore, we have proven at this order of computation quite directly that
\beq
m^2_h\leq m^2_Z\cos^2 2\beta.
\eeq
which is a famous result of supersymmetry. Furthermore, we can see that the bound is saturated as the supersymmetry breaking mass decouples from the eigenvalue equation, i.e.\ $\tilde m^2\to \infty$.

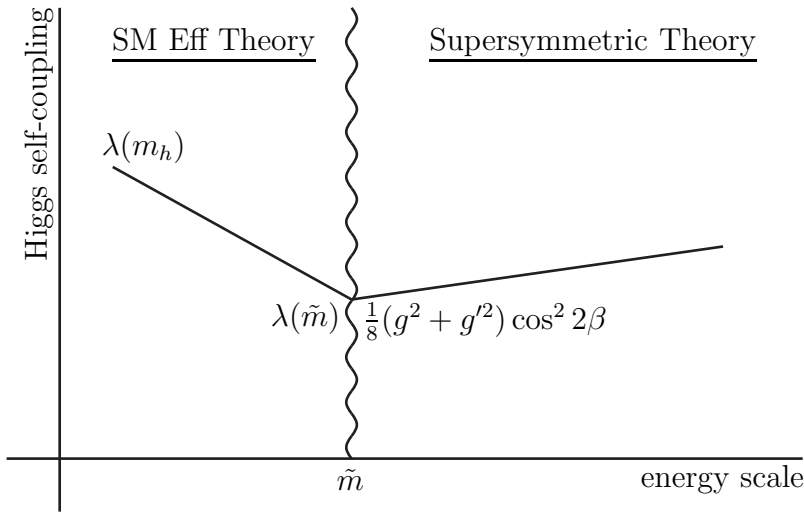
\begin{figure}[t]
\begin{center}
\begin{picture}(300,200)(0,0)
\SetWidth{1.0}
\Line(0,20)(300,20)
\Line(20,0)(20,190)
\Photon(130,20)(130,190){2}{10}
\Line(130,80)(270,100)
\Line(40,130)(130,80)
\Text(160,170)[lb]{\underline{Supersymmetric Theory}}
\Text(40,170)[lb]{\underline{SM Eff Theory}}
\Text(125,8)[lb]{$\tilde m$}
\Text(100,67)[lb]{$\lambda(\tilde m)$}
\Text(36,132)[lb]{$\lambda(m_h)$}
\Text(135,65)[lb]{$\frac{1}{8}(g^2+g'^2)\cos^2 2\beta$}
\Text(240,7)[lb]{energy scale}
\rText(3,95)[lb][l]{Higgs self-coupling}
\end{picture}
\end{center}
\caption{The light Higgs boson self-coupling transitions from $\frac{1}{8}(g^2+g'^2)c^2_{2\beta}$ in the supersymmetric theory to $\lambda$ in the Standard Model effective theory, with the matching scale being the superpartner mass $\tilde m$. Renormalization group flow of $\lambda$ from $Q=\tilde m$ down to $Q=m_h$ in the Standard Model effective theory introduces the famous $\Delta m_h^2\sim \frac{3m_t^4}{\pi^2 v^2}\log\left( \frac{\tilde m}{m_t}\right)$ leading-log correction as explained in the text.}
\label{fig:susy higgs rad cor}
\end{figure}

This formulation in the Runge basis also aids in our ability to immediately recognize what the leading radiative correction is to the mass of the lightest Higgs boson.  Let us suppose that all superpartner masses are at the scale $\tilde m$, including the stop mass, pseudoscalar Higgs boson mass, etc. Just above that scale we have a supersymmetric theory, and just below that scale we have the Standard Model effective theory.  We need to match the $\lambda$ self-coupling of the Standard Model Higgs boson with a self-coupling of the Higgs boson that remains light in supersymmetry. We computed the  $33$ component of the mass matrix in \eq{eq:runge matrix}, which is exactly that light supersymmetric Higgs boson in the limit of large $\tilde m\gg m_Z$. To be more precise, we actually computed
\beq
m^2_{h}=2\left[\frac{1}{8}(g^2+g'^2)c^2_{2\beta}\right] v^2 ~~{\rm (SUSY~Theory:~}Q>\tilde m)
\eeq
which should be compared with the Standard Model effective theory computation of 
\beq
m^2_h=2\lambda v^2 ~~{\rm (SM~Theory:~}Q<\tilde m)
\label{eq:SM Higgs mass}
\eeq
Thus, the matching of the UV supersymmetric theory with the IR Standard Model theory is done at the scale $Q=\tilde m$ where
\beq
\lambda(\tilde m)=\frac{1}{8}(g^2+g'^2)c^2_{2\beta}
\eeq
as depicted in fig.~\ref{fig:susy higgs rad cor}.
To get the true value of the Higgs mass we need to renormalize $\lambda(Q)$ down to the Higgs boson mass scale. If that scale is not too far from $\tilde m$, the linear approximation to solving the RG equation for $\lambda$ can be employed
\beq
\lambda(Q)=\lambda(\tilde m)-\beta_\lambda \ln\left( \frac{\tilde m}{ Q}\right).
\eeq
The leading term of the $\beta_\lambda$ function is (see \eq{eq:sm rges})
\beq
\beta_\lambda = -\frac{3y_t^4}{8\pi^2}+\cdots
\eeq
Recognizing that $y_t=\sqrt{2}m_t/v$, and that the top quark  contribution to the $\beta_\lambda$ function shuts off for $Q<m_t$ (i.e., $\beta_\lambda\simeq 0$ is the approximation for $Q<m_t$), one finds that
\bea
\lambda(m_h)& = &\lambda(\tilde m)+\frac{3m_t^4}{2\pi^2 v^4}\ln\left( \frac{\tilde m_t}{m_t}\right) \nonumber \\
& = & \frac{1}{8}(g^2+g'^2)c^2_{2\beta}+\frac{3m_t^4}{2\pi^2 v^4}\ln\left( \frac{\tilde m_t}{ m_t}\right) 
\eea
where I've gone ahead and written $\tilde m=\tilde m_t$ suggestively, since all superpartner masses are at $\tilde m$. Substituting this radiatively corrected value of $\lambda$ in the effective theory into \eq{eq:SM Higgs mass} we find
\bea
m^2_h & = & m^2_Z\cos^2 2\beta+\frac{3m^2_t}{\pi^2 v^2} \ln\left( \frac{\tilde m_t}{m_t}\right) \nonumber \\
&=& m^2_Z\cos^2 2\beta+\frac{3g^2}{4\pi^2}\frac{m^4_t}{m^2_W}\ln\left( \frac{\tilde m_t}{m_t}\right) 
\label{eq:leading rad}
\eea
which is the famous leading log formula for the Higgs boson mass of 
supersymmetry~\cite{Haber:1990aw,Okada:1990vk,Ellis:1990nz}. A more complete description of the loop-corrected Higgs mass can be found, e.g., in~\cite{Carena:2002es}. 

\begin{figure}[t]
\begin{center}
\includegraphics[width=0.7\textwidth]{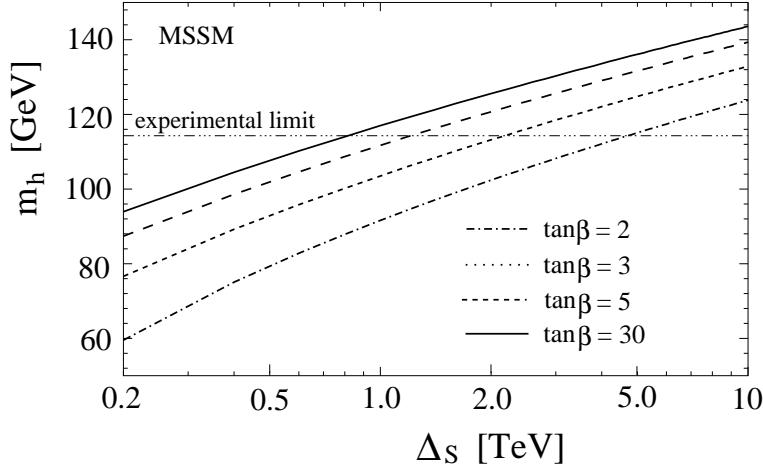}
\vspace{0.3cm}
\caption{Lightest Higgs mass as a function of $\Delta_S$ for various $\tan\beta$ values using \eq{eq:leading rad} and identifying $\Delta_S=\tilde m$. Plot taken from ref.~\cite{Tobe:2002zj}.
\label{fig:higgs mass}}
\end{center}
\end{figure}

There are several interesting features to note from~\eq{eq:leading rad}. First, the superpartner mass only enters the expressions for the lightest Higgs mass logarithmically. Since a logarithm increases very slowly with its argument, the increase in Higgs mass prediction as a function of $\tilde m_t$ is weak. 
Fig.~\ref{fig:higgs mass} shows the lightest Higgs mass prediction as a function of $\Delta_S$, which is the same as $\tilde m_t$ in \eq{eq:leading rad}.  Depending on the value of $\tan\beta$ the superpartner mass scale needs to be greater than $700\gev$ ($\tan\beta=50$) or $5\tev$ ($\tan\beta=2$) in order to evade the Higgs mass bound of $114\gev$, which is an applicable bound in this limit.  Such high masses strain our credulity, since they are the masses that get thrown into the electroweak Higgs potential which ultimately must pop out the $Z$ mass after the cranks are turned. It should be noted that the Higgs mass limit can be increased somewhat by inclusion of stop left-right mixing terms, and higher-order corrections (see, e.g., \cite{Carena:2002es,Martin:2002wn}). Nevertheless, some physicists still worry that the Higgs mass bound of $114\gev$ makes the MSSM unacceptably finetuned -- a little hierarchy problem. I do not share the viewpoint that going beyond the MSSM is necessary because of the Higgs mass prediction, but I am in agreement that seeking further models of supersymmetry that are less finetuned is a worthy exercise.

We now wish to compute the couplings of the CP-even mass eigenstates to the Standard Model fermions and gauge bosons.  In particular, we wish to know the couplings of the lightest Higgs boson since it is the state that stays light when all other supersymmetry masses get heavy.  The scalar Higgs boson mass matrix in the Runge basis~(\ref{eq:higgs runge matrix}) is not diagonal, so we have to do another rotation to get to the mass basis. Diagonalizing \eq{eq:higgs runge matrix} is accomplished by definition through a rotation of angle $\delta$:
\beq
\vector{h'}{H'}=\left(\begin{array}{cc}
\cos\delta & \sin\delta \\ -\sin\delta & \cos\delta \end{array}\right)
\vector{h}{H}
\eeq
where $h',H'$ are Runge states  and $h,H$ are mass eigenstates.
Expanding about $m_Z^2/m_A^2$ one finds
\beq
\delta = -c_{2\beta}s_{2\beta}\frac{m_Z^2}{m_A^2}-c_{2\beta}s_{2\beta}(c^2_{2\beta}-s^2_{2\beta})\frac{m_Z^4}{m_A^4}+\ldots
\label{eq:small delta}
\eeq
which is a small and positive angle.
Applying this rotation along with the rotation that takes $H^c_d,H_u$ into the $h',H'$ basis, one finds
\beq
\vector{{\rm Re}(H_d^{c,0})}{{\rm Re}(H^0_u)}=\frac{1}{\sqrt{2}}\left( \begin{array}{cc}
\cos(\beta-\delta) & -\sin(\beta-\delta) \\ \sin(\beta-\delta) & \cos(\beta-\delta) \end{array}\right)
\vector{h}{H}
\eeq

The couplings of the mass eigenstates to the vector bosons compared the Standard Model Higgs boson couplings are 
\beq
\frac{hZZ}{h_{SM}ZZ}&=& \frac{\langle h\rangle}{\langle h_{SM}\rangle}
= \frac{\cos(\beta-\delta)\langle H^{c,0}_d\rangle+\sin(\beta-\delta) \langle H^0_u\rangle}{v}=
\cos\delta  \\ 
\frac{HZZ}{h_{SM}ZZ}&=&\frac{\langle H\rangle}{\langle h_{SM}\rangle}
=\frac{-\sin(\beta-\delta)\langle H^{c,0}_d\rangle+\cos(\beta-\delta) \langle H^0_u\rangle}{v}=
\sin\delta 
\eeq
In the decoupling limit these couplings become to leading order in $m_Z^2/m_A^2$
\bea
\cos\delta &=& 1-c^2_{2\beta}s^2_{2\beta}\frac{m_Z^4}{m_A^4}+\ldots \\
\sin\delta & =& -c_{2\beta}s_{2\beta}\frac{m_Z^2}{m_A^2}+\ldots
\eea
Note, the light Higgs boson mass eigenstate couplings to Standard Model vector bosons decouple very rapidly, as $\sim 1/m^2_A$, to the Standard Model value. Deviations of $h$ from Standard Model expectations will be small unless $m_A$ is reasonably close to $m_Z$.

Let us now look at the couplings of the Higgs bosons to fermions. In supersymmetry, the masses of the $b$-quark and $t$-quark, which I will use as representatives of ``down-type" fermions and ``up-type" fermions, are given by
\bea
m_b& = &y_b\frac{v_d}{\sqrt{2}}=y_b\frac{v}{\sqrt{2}}\cos\beta ~\longrightarrow~ y_b=\sqrt{2}\, \frac{m_b}{v}\frac{1}{\cos\beta}\\
m_t &=& y_t\frac{v_t}{\sqrt{2}}=y_t\frac{v}{\sqrt{2}}\sin\beta~\longrightarrow~y_t=\sqrt{2}\, \frac{m_t}{v}\frac{1}{\sin\beta}
\eea
which is to be compared to the Standard Model Yukawa couplings of $y_b^{SM}=\sqrt{2}m_b/v$ and $y_t^{SM}=\sqrt{2}m_t/v$. Now, the interactions of the quarks with Higgs boson eigenstates will introduce various mixing angles. We compute them here:
\bea
y_bQ^\dagger H_d^* b_R &\to & \frac{y_b^{SM}}{\cos\beta}Q\left[ \cos(\beta-\delta) h-\sin(\beta-\delta)H\right] b_R \\
y_tQ^\dagger H_u^* t_R &\to & \frac{y_t^{SM}}{\sin\beta}Q\left[ \sin(\beta-\delta)h+\cos(\beta-\delta)H\right] t_R
\eea
and therefore, 
\bea
\frac{hbb}{h_{SM}bb}&=& \frac{\cos(\beta-\delta)}{\cos\beta}\to 1+\delta \tan\beta +\cdots \label{eq:b}\\
\frac{Hbb}{h_{SM}bb}&=&-\frac{\sin(\beta-\delta)}{\cos\beta} \to -\tan\beta+\delta+\cdots\\
\frac{htt}{h_{SM}tt}&=&\frac{\sin(\beta-\alpha)}{\sin\beta} \to 1-\frac{\delta}{\tan\beta}+\cdots \\ 
\frac{Htt}{h_{SM}tt}&=&\frac{\cos(\beta-\delta)}{\sin\beta}\to \frac{1}{\tan\beta}+\delta +\cdots
\eeq
where the last expression of each line is the expansion in terms of small $\delta$ (see \eq{eq:small delta}).

The expansion of the couplings of the scalar Higgs boson mass eigenstates to Standard Model gauge bosons and fermions  by the angle $\delta$, which takes us from the Runge basis to the mass basis, shows instantly  the implications of the decoupling limit when $\delta\sim m_Z^2/m_A^2$ is small.  In particular, for large $m_A^2$ the lightest Higgs boson $h$ interacts with the Standard Model in precisely the same way as the Standard Model Higgs boson. Thus, it may be difficult to distinguish the two theories from light Higgs boson observables alone. 

The heavy Higgs boson decouples rapidly from the vector bosons, but does not decouple from the fermions. This is expected. As the scale of supersymmetry breaking increases, a complete $SU(2)$ multiplet must decouple from the spectrum with it. Indeed, that is what happens here as the $H'$ state in the Runge basis becomes more and more identified with the decoupled mass eigenstate $H$. On the other hand, there is no need from any such consideration that the fermions must decouple in that limit from the $H$ state. Indeed, they do not. However, it is to be noted that if $\tan\beta\to \infty$ the heavy state does not couple to up-type quarks since all the vev is in ${\rm Re}(H_u^0)$ which then is increasingly  identified with the mass eigenstate $h$ while ${\rm Re}(H^{c,0}_d)$ becomes increasingly identified with $H$, and therefore no coupling to top quarks. Similar reasoning tells us that in the opposite limit of $\tan\beta\to 0$, the coupling of $H$ to down quarks decouples as well.  Because of these decoupling features, searches for heavy Higgs bosons of supersymmetry  mostly focus on decays into fermions rather than decays into vector bosons. 

I have presented the Higgs sector in a different manner than is usually done. I have first gone to the Runge basis and then rotated by an angle $\delta$ to the mass eigenstate basis. The advantages of this are largely pedagogical in that one can instantly see that the Goldstones are all contained in $\Phi_{vev}$, the heavy charged Higgs and pseudoscalar are in $\Phi_\perp$, the scalar Higgs are a mixture of the components of $\Phi_{vev}$ and $\Phi_\perp$ but as $m_Z^2/m_A^2$ becomes small, i.e., the decoupling limit, the light Higgs boson $h$ asymptotes to being the component of $\Phi_{vev}$ and the heavy Higgs  boson $H$ a component of $\Phi_\perp$. Leading order radiative corrections are easily understood by the clean separation in this basis of the heavy states in the full supersymmetric theory and the matching of its remaining light state $h$ to the Higgs boson of the Standard Model effective theory below the scale of supersymmetry breaking.  Additional radiative corrections can be quite important, especially with respect to the coupling of the light Higgs boson to the $b$ quarks. This interaction can vary widely with large $\tan\beta$, as demonstrated by the term $\delta \tan\beta$ in \eq{eq:b} which can be large if $\tan\beta$ is large.  The special importance of a more complete analysis of radiative corrections for the Higgs boson couplings with large $\tan\beta$ has been emphasized in refs.~\cite{Loinaz:1998ph,Carena:1998gk,Mrenna:2000qh,Heinemeyer:2000fa,Haber:2000kq,Gunion:2002zf}.

\xsection{Radion Mixing with the Higgs Boson in Warped Compact Space}

A completely different approach to solving the hierarchy problem is to explain the weakness of gravity not by an extraordinarily high mass scale, but by having it propagate in extra dimensional 
space~\cite{ArkaniHamed:1998rs,Antoniadis:1998ig}. In some approaches gravity is weak because its field lines dissipate in higher dimensions. The fundamental gravity scale $M_D$ is of order the weak scale, leading to no troubles with the hierarchy problem, and Planck's constant is a `fake' derived scale
\beq
M_{\rm Pl}^2=M_D^{2+\delta}R^\delta
\eeq
where $R$ is the radius that characterizes the size of the $\delta$-dimensional compact space. Much interesting phenomenology follows from this supposition (see, e.g., refs.~\cite{ArkaniHamed:1998nn,Giudice:1998ck,Hewett:2002hv,Kribs:2006mq}).
 
A different approach is to warp the extra dimension(s) and have all high mass scales be warped to weak-scale masses in our $3+1$-dimensional worldview. This is the approach Randall and Sundrum took~\cite{Randall:1999ee}. Let me review the gravitational set up first and then move to implications for the Higgs boson.

The representative idea is that we live in a 5 dimensional spacetime, where the 5th dimension is a compactified $S^1/Z_2$ orbifold. In other words, think of the 5th dimension as a circle with radius $r_c$, but identify points at $\phi$ and $-\phi$ as equivalent.   This is sometimes called compactifying on a line segment, since all coordinates in the physical space can be represented on a line $0<\phi<\pi$ with coordinate $\phi$.  Branes are placed at the orbifold fixed points $\phi=0$ (`hidden brane' or `Planck brane' location) and $\phi=\pi$ (`visible brane' or `SM brane' location).  

The Randall-Sundrum hypothesized action is
\beq
S=S^{(5)}+S^{(4)}_0+S^{(4)}_c
\eeq
where
\bea
S^{(5)} &=& \int d^4x \int_{-\pi}^{\pi} d\phi\sqrt{G}(-\Lambda+2M^3R) \\
S^{(4)}_{0}&=&\int d^4x \sqrt{-g_0}(-V_0+{\cal L}_0) \\
S^{(4)}_{c}&=&\int d^4x\sqrt{-g_c}(-V_c+{\cal L}_c).
\eeq
The 5D metric is $G_{MN}$ where $M,N=\mu,\phi$ with $\mu=0,1,2,3$. The 4D induced metrics $g_0$ and $g_c$ are obtained by
\bea
g_{0,\mu\nu}(x^\mu)&=&G_{\mu\nu}(x^\mu,\phi=0) \\
g_{c,\mu\nu}(x^\mu)&=&G_{\mu\nu}(x^\mu,\phi=\pi).
\eea
The 5D metric ansatz is 
\beq
ds^2=W(r_c\phi)g_{\mu\nu}dx^\mu dx^\nu-r_c^2d\phi^2,~~{\rm where}~~W(r_c\phi)=e^{-2kr_c|\phi|}.
\eeq
This metric describes an AdS space for the 5th dimension with AdS curvature parameter $k$.  It is assumed that $k\sim M_{\rm Pl}$.

Self-consistency of Einstein's equations and respecting 4D Poincar\'e invariance requires 
\beq
V_0=-V_c=24M^3k~~{\rm and}~~\Lambda=-24M^3k^2
\eeq
This is derived in slightly different ways in~\cite{Randall:1999ee}  and~\cite{Csaki:2004ay}.

The graviton $h_{\mu\nu}$ is identified as the massless fluctuation of $g_{\mu\nu}$ about the metric signature $\eta_{\mu\nu}$: $g_{\mu\nu}=\eta_{\mu\nu}+h_{\mu\nu}$. Thus, we need to expand $S^{(5)}$, look for the term that is proportional to $\sqrt{-g}R_4$, where $R_4$ is the 4D Ricci scalar, and integrate over the extra dimensional coordinate to identify Planck's constant: 
\bea
\int d^4x\int_{-\pi}^{\pi} d\phi 2M^3W(r_c\phi)\sqrt{-g}R_4,
\eea
which implies that
\bea
M_{\rm Pl}^2=M^3r_c\int_{-\pi}^{\pi} d\phi W(r_c\phi)=\frac{M^3}{k}\left[ 1-W(r_c\pi)\right].
\eea
This equation implies that  $M\sim k\sim M_{\rm Pl}$ is reasonable to satisfy the gravitational part of the action.

So far we have merely summarized the gravitational picture in the Randall-Sundrum paper.  There is more. The warp factor $W(r_c\phi)$ can serve to squash all would-be Planck scale masses to weak scale masses. Randall-Sundrum showed, for example, that what looks to be naturally a Planck scale vev $v_0\sim M_{Pl}$ for the Higgs boson on the visible brane is actually warped down to the weak scale by the warp  factor:
\beq
\langle H^\dagger H\rangle =W(r_c\pi)v^2_0=e^{-2kr_c\pi}v^2_0~~{\rm implying}~~kr_c\sim 12.
\eeq
Thus the compactification radius need only be a factor of about 12 larger than $k^{-1}$ in order to warp $M_{Pl}$ down to $e^{-kr_c\pi}M_{Pl}\sim m_W$,  which is not much of a finetuning from that perspective at least. This warping factor suppression is the basic principle that Randall-Sundrum identified as a potential reason behind the hierarchy of the gravitational scale to the electroweak scale.

Besides making the Higgs boson a more palatable object, what does this setup have to do with the Higgs boson? In the setup the visible sector brane was placed at the position $\phi=\pi$. However, there is no built-in mechanism to keep it fixed there. The relative separations of the two branes $\phi$ at the position $x^\mu$ describes a field $T(x^\mu)$. This field is often called the `radion' since it is the dynamical object quantifying radius separation of branes in  the compact space.   Let us sketch how the radion interacts with the normal particles of the Standard Model.

Our first step is to update the metric to take into account the dynamical nature of the compactification radius $r_c$. We do this by promoting $r_c$ to the field $r_c\to T(x)$.  The metric tensor is then defined by
\beq
ds^2=W(T(x)\phi)g_{\mu\nu}(x)dx^\mu dx^\nu-T^2(x)d\phi^2.
\label{eq:RS metric}
\eeq
Without a principle to stabilize the brane separation, the mass of the radion is zero and it acts as a propagating scalar gravitational degree of freedom, leading to an unacceptable scalar-tensor gravity theory. Somehow stabilization must occur.
The dynamical mechanism that is to stabilize the radion requires $\langle T(x)\rangle=r_c$. A method to do this was identified by Goldberger and Wise~\cite{Goldberger:1999uk}, which we will not elaborate on. Substituting the metric of \eq{eq:RS metric} into the original action, and integrating over the extra-dimensional coordinate gives us
\beq
S_{\varphi}=\int d^4x\sqrt{-g}\left[ \frac{2M^3}{k}\left( 1-\frac{\varphi^2}{\Lambda_\varphi^2}e^{-2k\pi r_c}\right) R_4
+\frac{1}{2}\partial_\mu \varphi\partial^\mu \varphi-V(\varphi)+\left( 1-\frac{\varphi}{\Lambda_\varphi}\right) T^\mu_\mu\right] \nonumber
\eeq
where $\varphi(x)$ is the canonically normalized radion field, $V(\varphi)$ is the potential that stabilizes the radion field, and $\Lambda_\varphi$ is its vacuum expectation value
\beq
\varphi(x)\equiv \Lambda_\varphi e^{-k\pi(T(x)-r_c)},~~{\rm and}~~
\Lambda_\varphi \equiv \langle \varphi\rangle =e^{-k\pi r_c}\sqrt{\frac{24M^3}{k}}.
\eeq

The interaction of the radion with the trace of the energy momentum tensor is directly analogous to the graviton interacting with the energy momentum tensor in normal $4D$ gravity.  In that case the interaction is obtained by expanding $\sqrt{-g}{\cal L}$ around the metric signature $g_{\mu\nu}=\eta_{\mu\nu}+h_{\mu\nu}$ at leading order:
\bea
\sqrt{-g}{\cal L}& = &\sqrt{1-\eta^{\mu\nu}h_{\mu\nu}+{\cal O}(h^2)}\, \left({\cal L}(\eta_{\mu\nu})+
 h_{\mu\nu} \left. \frac{\delta {\cal L}}{\delta g_{\mu\nu}}\right|_{g^{\mu\nu}=\eta^{\mu\nu}} +{\cal O}(h^2)\right) \nonumber \\
& = & {\cal L}(\eta_{\mu\nu})+\frac{1}{2}h_{\mu\nu}T^{\mu\nu}
 +{\cal O}(h^2)
\eea
where $T^{\mu\nu}$ is the energy momentum tensor 
\beq
T^{\mu\nu}={2}\frac{\delta S}{\delta h_{\mu\nu}}= \left[
-\eta^{\mu\nu}{\cal L} +2 \frac{\delta {\cal L}}{\delta g_{\mu\nu}}\right]_{g^{\mu\nu}=\eta^{\mu\nu}}.
\eeq
which can be computed straightforwardly from the Standard Model lagrangian.

There is one subtlety regarding the energy-momentum tensor that we must take into account. The Higgs boson is a special field in that at dimension two a gauge invariant, Lorentz invariant operator can be formed, $H^\dagger H$. Because of that we can add to the full action another term that is dimension four~\cite{Giudice:2000av}:
\bea
S_\xi=-\xi\int d^4x\sqrt{-g}RH^\dagger H
\label{eq:Sxi}
\eea
In the weak-field limit we know that
\beq
\sqrt{-g}R=(\eta\partial_\lambda\partial^\lambda-\partial^\mu\partial^\nu)h_{\mu\nu}(x)
\eeq
which when substituted into eq.~(\ref{eq:Sxi}) gives a contribution to the energy momentum tensor of
\beq
 T^{\mu\nu}_\xi=2\xi (\eta\partial_\lambda\partial^\lambda-\partial^\mu\partial^\nu)(H^\dagger H).
\eeq
Including this term, the trace of the energy momentum tensor is then
\beq
T^\mu_\mu=\xi v\kbox h+(6\xi -1) \partial_\mu h\partial^\mu h +6\xi h \kbox h 
+2m_h^2h^2 + m_{ij}\bar\psi_i \psi_j -M_V^2 V_{A\mu}V_A^\mu 
\eeq
where we have only retained terms up to dimension two. 

After shifting $\varphi$ from its vev and expanding the lagrangian, the terms bilinear in Higgs and radion fields are
\beq
{\cal L} =-\frac{1}{2}  \varphi
\left( \kbox +m_\varphi^2 \right) \varphi
-\frac{1}{2} h \left( \kbox +m_h^2 \right) h -\frac{6\xi 
v}{\Lambda_\varphi} \varphi \kbox h
\eeq
The kinetic terms can be made canonical by the transformation
\bea
h\to {h'}\sec\omega ~~{\rm and}~~\phi\to \phi'- h' \tan\omega
\eea
where $\sin\omega=6\xi v/\Lambda_\varphi$.  After this transformation the mass terms mix. The resulting mass matrix can be diagonalized through an orthogonal rotation characterized by the angle $\alpha$:
\beq
\vector{h'}{\phi'}=\left(\begin{array}{cc} \cos\alpha & \sin\alpha \\ -\sin\alpha & \cos\alpha \end{array}\right)
\vector{s_1}{s_2}
\eeq
where $s_{1,2}$ are mass eigenstates, and $\cos\alpha\to 1$ and $\sin\alpha\to 0$ as  $\xi\to 0$.
The final expression for the mixing is
\beq
\vector{h}{\phi}=\left(\begin{array}{cc}
\sec\omega \cos\alpha&\sec\omega \sin\alpha\\
-\sin\alpha -\tan\omega \cos\alpha& \cos\alpha+\tan\omega \sin\alpha
\end{array}\right) \vector{s_1}{s_2}.
\eeq

What we have here is a case of mixing between the Higgs boson and the radion such that no eigenstate is purely one or the other. Furthermore, the couplings of Standard Model states to the radion are very similar to the couplings to the Higgs boson except suppressed by a factor of $v/\Lambda_\varphi$. The interaction lagrangian of the Higgs boson and radion to pairs of Standard Model states demonstrates this well.  Let us go to the original states $h$ and $\varphi$ for simplicity, and the mixing angles can be applied easily later to bring one to the mass eigenstate basis. 

The interaction of the fields 
$\varphi$
and $h$ with fermions and massive gauge bosons is 
\beq
{\cal L} = -\frac{1}{v} \left( m_{ij}\bar\psi_i
\psi_j - M_V^2 V_{A\mu}V_A^\mu \right) \left[ h+ 
\frac{v}{\Lambda_\varphi}  \varphi
\right] .
\eeq
The interactions with massless gluon fields is
\beq
\left[ {\varphi\over \Lambda_\varphi} 
b_3- \frac{1}{2}\left ({\varphi\over
\Lambda_\varphi}+{h\over v}\right ) F_{1/2}(\tau_t)\right]
\frac{\alpha_s}{8\pi}G_{\mu\nu}G^{\mu\nu},
\label{hgg}
\eeq
where $\tau_t=4m^2_t/q^2$, and $F_{1/2}$ is~\cite{HHG}
\bea
F_{1/2}(\tau) & = & -2\tau [1+(1-\tau)f(\tau)]~~~{\rm with} 
\eea
\beq
f(\tau)  =  \left\{ \begin{array}{cc}
      \left[ \sin^{-1}\left( 1/\sqrt{\tau}\right)\right]^2, &
               {\rm if}~\tau\ge 1, \\
      -\frac{1}{4}\left[ \ln (\eta_+/\eta_-)-i\pi\right]^2, &
               {\rm if}~\tau < 1,
                      \end{array}  \right.
\eeq
and
\beq
\eta_{\pm} = 1 \pm \sqrt{1-\tau} .
\eeq
 We identify $q^2$ with the Higgs mass squared when we calculate
the on-shell decays to $gg$.
The first term in eq.~(\ref{hgg}), 
where $b_3=7$ is the QCD $\beta$-function coefficient
in the Standard Model, is generated by the conformal breaking QCD trace anomaly. The second term originates
from one-loop diagrams involving virtual top quarks. The form factor
$F_{1/2}(\tau_t)\to -4/3$ for $\tau\to\infty$,
and $F_{1/2}\to 0$ for $\tau\to 0$.  Thus, for $m_t^2\gg q^2$, the coupling to $\varphi$ becomes
proportional to the $\beta$-function for 5-flavors $b_3+2/3$,
consistent with a smooth decoupling of the top quark.
The coupling to two massless photons is similar in form
\beq
\left \{{\varphi\over \Lambda_\varphi} \left (b_{2}+b_Y\right)
-\left ({\varphi\over
\Lambda_\varphi}+{h\over v}\right )\left (F_1(\tau_W)+
\frac{4}{3}F_{1/2}(\tau_t)\right) \right  \}
\frac{\alpha_{EM}}{8\pi}F_{\mu\nu}F^{\mu\nu},
\label{hgaga}
\eeq
where $F_1(\tau_W)$ is a form factor from the loop with virtual $W$'s,
\beq
F_1(\tau) &= & 2+3\tau+3\tau(2-\tau)f(\tau),
\eeq
and $b_2=19/6$ and $b_Y=-41/6$ are the Standard Model $SU(2)\times U(1)_Y$ $\beta$-funtion
coefficients. For $m_t^2/q^2\to \infty$ 
the coupling to $\varphi$ reduces to the
QED $\beta$-function with $e,\mu,\tau$ and $u,d,s,c,b$
\beq
b_2+b_Y-F_1(\infty)-\frac{4}{3}F_{1/2}(\infty)=-\frac{80}{9}=b_{\rm QED},
\eeq 
which is what we expect for smooth decoupling of heavy top quark and heavy $W$ boson.

The phenomenology of the mixed Radion-Higgs system is quite interesting. The basic signatures are the same as the Standard Model Higgs, with $gg\to s_i\to WW,ZZ,\bar bb,\gamma\gamma$. There are several  important additional qualitative features to emphasize however. First, the QCD trace anomaly contribution to the radion coupling to gluons can significantly enhance the production cross-section of the Higgs bosons through $gg$ fusion. Likewise, the QED trace anomaly contribution can significantly enhance the decay branching fraction to two photons. Both of these effects would be good news for Higgs boson discovery. The other important qualitative feature to emphasize is the prospect that the heavier eigenstate $s_2$ can decay into two  lighter eigenstates $s_1$. Of course, if it is to take place it must be kinematically allowed. Furthermore, the $s_2s_1^2$ interaction is model dependent, as it is partial controlled by the details of the stabilizing mechanism, or in other words, the potential $V(\varphi)$.  It can be added as a free parameter in the model and discovery capabilities or limits estimated. Many more details of collider phenomenology can be found in refs.~\cite{Giudice:2000av,Csaki:2000zn,Hewett:2002nk,Gunion:2003px,Dominici:2009pq,Azatov:2008vm}.

\xsection{Higgs Bosons of a Hidden World}

There are many things we do not know. To name a few, we do not know what generates the baryon asymmetry of the universe, what makes up the dark matter, the substance of the dark energy, the origin of multiple gauge symmetries, the duplication of family symmetries, the hierarchy of fermion masses, how quantum mechanics and general relativity peacefully coexist, and the reason why the weak scale is so much smaller than the Planck scale.  We see that the vast majority of these particle physics problems we are trying to solve have to do with our bodies. The particles we care about are the particles that make up our bodies, or are copies of particles that make up our bodies, or directly interact with the particles that make up our bodies in order to keep bound states, allow decays and transitions, or give mass to them. 

Cosmologists have been fairly good at not restricting inquiries to the problems of our bodies, but not as much in particle physics. We have theories of flavor, theories of electroweak symmetry breaking, MQCD, even string phenomenology, much of it really with an eye toward {\it understanding us.}  Even when particle physicists do venture into cosmology, it is often through the accidental recognition that a {\it theory of us} also gives for free a {\it theory of dark matter}, with the next-to-lightest superpartner of supersymmetry being a good example.

At the 536th anniversary of Copernicus's birth, let's remember that we are not the center of the universe. As Hamlet said to Horatio, ``There are more things in heaven and earth, Horatio, than are dreamt of in your philosophy."  I think most of us, when asked, would say we are sure that the Standard Model fields are not the sum total of all fields in the universe. There is no reason to suspect that to be the case. Let us imagine new states that have nothing to do with solving any problem of our bodies. We just add them and ask if there is a way to detect them in experiment.

As Hamlet knows, there is much more that could be said and done than we will ever know. So, let us start modestly. Let us imagine that in addition to the Standard Model there is a real scalar field $\phi$ that is not charged under any gauge group symmetry of the Standard Model and is odd under a $Z_2$ discrete symmetry.  How can this particle interact with the Standard Model at the renormalizable level?  There is only one way, and it is through the Higgs boson. The lagrangian terms that involve the $\phi$ field are simply
\beq
{\cal L}_\phi= \frac{1}{2}(\partial_\mu \phi)^2-\frac{1}{2}m^2_\phi\phi^2-\frac{\lambda_\phi}{4}\phi^4-y\phi^2H^\dagger H.
\eeq
This simple little model~\cite{Burgess:2000yq,Davoudiasl:2004be} 
if true, would have absolutely colossal effects on our world view and LHC phenomenology. First, the $Z_2$ symmetry makes it stable. Relic abundance can be calculated as a function of the various parameters of the theory including most importantly $m_\phi$, $m_h$ and $y$.  There are accessible regions of parameter space where relic abundance is the value needed to be the CDM of the universe (for latest consideration of this case, see ref.~\cite{Barger:2007im}).  Since $\phi$ does not get a vacuum expectation value, the Higgs boson of the Standard Model cannot mix with $\phi$; however, $H$ can decay into a pair of $\phi$ particles leading to an invisible decay width of the Higgs boson:
\beq
\Gamma(h\to\phi\phi)=\frac{y^2 v^2}{8\pi m_h}\sqrt{1-\frac{4m_\phi^2}{m^2_h}}.
\eeq

Recall that if the Higgs boson mass is less than about $130\gev$, the dominant Standard Model decay width is into $b\bar b$ which is governed by the bottom quark Yukawa coupling. This Yukawa coupling is tiny, $y_b\simeq \sqrt{2}m_b(m_b)/v\simeq 1/40$.  Thus, any value of $y\geq 1/40$, which is quite probable even with relic abundance constraints included~\cite{Burgess:2000yq}, will compete with the dominant Standard Model mode and the invisible decay width could overwhelm the branching fraction of the Higgs boson, qualitatively changing how one would search for this important state.  Several studies have been performed in the theory 
community~\cite{Joshipura:1992ua,Choudhury:1993hv,Gunion:1993jf,Frederiksen:1994me,Binoth:1996au,Martin:1999qf,Godbole:2003it,Davoudiasl:2004aj}
 and the experimental community~\cite{Aad:2009wy} to determine how well an invisibly decaying Higgs boson can be found. There is mostly good news in these studies, in that lower Higgs masses can be found rather well even if decaying invisibly, whereas higher Higgs mass states perhaps not quite as well. But that is probably ok, since as the Higgs mass increases well above the vector boson mass scale, its width into longitudinal $W_L^+W_L^-,Z^0_LZ_L^0$ increases like $\Gamma_h\sim m_h^3/m_Z^2$.  Such strength of decay is unlikely to be buried by an invisible $H\to \phi\phi$ decays, and so a dominant invisible width would not be expected for Higgs bosons above the $2m_Z$ scale.

This introduction of a Standard Model singlet Higgs boson was done in problem solving mode. The problem is explaining dark matter and the solution arrived at is a $\phi$ real scalar with a $Z_2$ symmetry. However, let us interpret this in a different way. Let us interpret the introduction of this new field as one of many possibilities that just is, and not worry about justifying its existence.  If we think in those terms, we start throwing around states in our minds at will and at random. The vast majority do not couple to the Standard Model at a renormalizable way, and so seeing them at a collider in the near term would not be expected.

So let us be a little more clever about selecting the subclass that can be seen amongst the infinite possibilities.  Our best bet is to identify what hidden world states can couple to the Standard Model states at the renormalizable level.  By hidden worlds I mean states that are not part of the Standard Model and are not charged under any Standard Model gauge symmetry. 

The only opportunity that enables us to connect hidden world states with Standard Model states at the renormalizable level (i.e., dimension four level) is if we can find gauge invariant Standard Model operators with dimension less than four.  Not counting the right-handed neutrino Majorana mass, there are only three such operators: the hypercharge field strength tensor, the Higgs boson mass-squared operator, and the Higgs-neutrino operator
\beq
B_{\mu\nu}, ~~~H^\dagger H,~~{\rm and}~~HL
\eeq
which are a Lorentz vector, scalar and spinor respectively. I will ignore the $HL$ operator since the field of neutrino physics, and sterile neutrinos, etc., deals with interactions with this operator. Instead I will focus on the bosonic sector. As we noted earlier in lecture~\ref{sec:hierarchy}, the Higgs boson operator is the only bosonic one with dimension less than four that is both gauge invariant and Lorentz invariant. In my mind, this is one of the most important realizations that one can make about the Standard Model and about the Higgs boson in particular. As a consequence, {\it hidden worlds have a much better chance of communicating with the Higgs boson than any other particle in the Standard Model.}  This is one reason why I think the Higgs boson is particularly susceptible to deviations from expected phenomenology at the LHC.  Furthermore, the accidental narrowness of the data-preferred light Higgs boson with mass $m_h\lsim 2m_W$ makes the state even more ripe for bullying by a hidden sector.

Let us continue forward with the simplest possible example of a hidden sector that couples to both the $B_{\mu\nu}$  tensor and $H^\dagger H$. This theory is an Abelian hidden sector Higgs model, whose particle spectrum is a $U(1)'$ vector boson $X_\mu$ and a complex scalar field $\Phi$, which gets a vev and breaks the $U(1)'$ symmetry, giving mass to the $X$ boson.
The only coupling of this new gauge sector to the Standard Model is through kinetic mixing with
the hypercharge gauge boson $B_\mu$.
The kinetic energy terms of the $U(1)$ gauge groups are
\beq
{\cal L}^{KE}_{BX} = -\frac{1}{4} \hat{B}_{\mu\nu} \hat{B}^{\mu\nu}-\frac{1}{4} \hat{X}_{\mu\nu} \hat{X}^{\mu\nu} + \frac{\chi}{2} \hat{X}_{\mu\nu} \hat{B}^{\mu\nu}.
\label{eq:KEBX}
\eeq
I do not have a strong opinion about the size of $\chi$. In an effective field theory it could be anything, but in many approaches from top down it is often the case that $\chi$ is a one-loop suppressed quantity derived $\chi\sim g^2/16\pi^2$ from some high-scale states that were vectorlike with respect to the low-scale gauge symmetries and were integrated out, but nevertheless perhaps had some small mass splittings. 

We introduce a new Higgs boson $\Phi$ in addition to the usual Standard Model Higgs boson
$H$.
Under $SU(2)_L \times U(1)_Y \times U(1)_X$ we take the representations
$H: (2, 1/2, 0)$ and $\Phi: (1, 0, q_X)$, with $q_X$ arbitrary.
The Higgs sector Lagrangian is
\bea 
{\cal L}_{\Phi} &=& |D_\mu H|^2
+ |D_\mu \Phi |^2 
 + m^2_{\Phi}|\Phi|^2 + m^2_{H}|H|^2 \nonumber\\
& & - \lambda|H|^4 - \rho|\Phi|^4 - \kappa
|H|^2|\Phi|^2, \label{Lphi.EQ} 
\eea
so that $U(1)_X$ is broken spontaneously by $\left< \Phi \right> = \xi/\sqrt{2}$,
and electroweak symmetry is broken spontaneously as usual by
$\left< H\right> = (0,v/\sqrt{2})$. 

One can diagonalize the kinetic terms by redefining $\hat{X}_\mu ,
\hat{Y}_\mu \rightarrow X_\mu , Y_\mu$ with
\beq
\left( \begin{array}{c} X_\mu \\ Y_\mu \end{array} \right) = \left(
\begin{array}{cc} \sqrt{1-\chi^2} & 0 \\ -\chi & 1 \end{array} \right)
\left( \begin{array}{c} \hat{X}_\mu \\ \hat{Y}_\mu \end{array}
\right). \label{eq:kinetic rotate}
\eeq
The covariant derivative is then\beq D_\mu =
\partial_\mu + i (g_X Q_X + g^\prime \eta Q_Y) X_\mu + i g^\prime
Q_Y B_\mu + i g T^3 W^3_\mu \  . \label{DMUGAU.EQ} \eeq
where $\eta \equiv \chi / \sqrt{1-\chi^2}$. Note, the kinetic lagrangian of \eq{eq:KEBX} is symmetric under interchange of $\hat B \leftrightarrow \hat X$, but that symmetric nature is not transparent in \eq{eq:kinetic rotate}. The choice made in \eq{eq:kinetic rotate} is convenient for our purposes, however, since only the $X_\mu$ field will couple to the exotic Higgs boson $\Phi$, and so will decouple from the other states simply (i.e., without having to analyze the full gauge boson mixing matrix) as $\xi\gg v$.

After a $GL(2,R)$ rotation to diagonalize the kinetic terms followed by an $O(3)$ rotation
to diagonalize the $3 \times 3$ neutral gauge boson mass matrix, we can write the
mass eigenstates as (with $s_x\equiv \sin{\theta_x}$,  $c_x\equiv \cos{\theta_x}$)
\bea
\begin{pmatrix} B \\ W^3 \\ X  \end{pmatrix} =
\begin{pmatrix}
c_W & -s_W c_\alpha  & s_W s_\alpha \\
s_W & c_W c_\alpha & -c_W s_\alpha \\
0 & s_\alpha & c_\alpha
\end{pmatrix}
\begin{pmatrix} A \\ Z \\ Z'  \end{pmatrix} \ ,
\label{GAU2MAS.EQ}
\eea
where the
usual weak mixing angle and the new gauge boson mixing angle  are
\beq
s_W \equiv \frac{g^\prime}{\sqrt{g^2 + {g^\prime}^2}} \ ; \quad
\tan{\left( 2\theta_\alpha \right)} = \frac{-2 s_W \eta}{1 - s_W^2\eta^2 - \Delta_Z} \ ,
\label{tan2al.EQ}
\eeq
with $\Delta_Z =
M_{X}^2/M_{Z_0}^2$, $M_{X}^2 = \xi^2 g_X^2 q_X^2$, $M_{Z_0}^2 =
(g^2 + {g^\prime}^2) v^2 / 4$. $M_{Z_0}$ and $M_X$ are masses before mixing. The photon is massless (i.e., $M_A = 0 $), and the two heavier gauge boson
mass eigenvalues are 
\begin{eqnarray} 
M_{Z, Z^\prime} = \frac{M_{Z_0}^2}{2}
\left[\left(1+s_W^2 \eta^2 + \Delta_Z \right) 
\pm \sqrt{\left( 1 - s_W^2 \eta^2 - \Delta_Z \right)^2
+ 4 s_W^2 \eta^2 } \right] ,
\end{eqnarray}
valid for $\Delta_Z < (1-s_W^2 \eta^2)$ ($Z \leftrightarrow Z^\prime$ otherwise). Since we assume that $\eta \ll 1$, mass
eigenvalues are taken as $M_Z \approx M_{Z_0}=91.19$ GeV and 
$M_{Z'} \approx M_X$. More regarding kinetic mixing, precision electroweak constraints, collider constraints and other considerations can be found many places~\cite{kinetic mixing}.

The two real physical Higgs bosons $H$ and $\Phi$ mix after symmetry breaking,
and the mass eigenstates $h, H$ are
\begin{displaymath} \left( \begin{array}{c} H \\ \Phi
\end{array} \right) = \left( \begin{array}{cc} c_h & s_h \\ -s_h &
c_h \end{array} \right) \left( \begin{array}{c} h \\ H \end{array}
\right).
\end{displaymath}
Mixing angle and mass eigenvalues are 
\begin{eqnarray}
\tan{(2\theta_h)} &=& \frac{\kappa v \xi}{\rho \xi^2 - \lambda v^2}
\
\\ M_{h,H}^2 = \left( \lambda v^2 + \rho \xi^2 \right)
            &\mp& \sqrt{ (\lambda v^2 - \rho \xi^2)^2 + \kappa^2 v^2 \xi^2} \ .
\end{eqnarray}

In summary, the model has been completely specified above. The effect of this exotic condensing Higgs sector on LHC phenomenology is to introduce two extra physical states $Z'$ and $H$.  $Z'$ is an extra gauge boson mass eigenstate that interacts with the Standard Model fields because of gauge-invariant, renormalizable kinetic mixing with hypercharge, and $H$ is an extra Higgs boson that interacts with the Standard Model fields  because of renormalizable modulus-squared mixing with the Standard Model Higgs boson.

The Feynman rules are obtained from a straightforward expansion of the above lagrangian in terms of mass eigenstates.  The collider implications of this theory have been detailed 
elsewhere~\cite{Schabinger:2005ei,Bowen:2007ia,Wells:2008xg}, but let me list a few of them:
\begin{itemize}
\item Such theories can lead to the universal suppression in all channels  of the light Higgs boson production rates. Mixing the Higgs boson with a ``sterile" hidden sector state may mean only that the hidden sector steals electroweak coupling away from the lighter Standard Model Higgs, generating two mass eigenstates: a SM-like eigenstate with all couplings reduced by $c_h^2$ and a heavier hidden-like eigenstate with all couplings to the Standard Model states the same as the Standard Model Higgs but with suppression factor $s^2_h$. 
\item The heavier hidden Higgs boson, when it picks up a little electroweak coupling via its mixing with the Standard Model Higgs boson, can then be produced at a collider. Its mass might be significantly above the weak scale, even in the trans-TeV mass region. A trans-TeV Standard Model Higgs boson is a nonsensical proposition, since it is far from perterbative, but a trans-TeV Higgs-like boson with large mixing angle suppression factor can be a sensible scalar state to define and look for. In contrast to the decoupled heavy Higgs bosons of supersymmetry, these heavy hidden-like Higgs bosons will decay primarily into Standard Model vector bosons.
\item The heavy Higgs higgs boson can decay into the lighter Higgs boson if it is mixed with it, leading to interesting double Higgs production and decay phenomenology, such as $gg\to H\to hh\to \gamma\gamma \bar b b$.
\item If the hidden sector has additional light states, the lighter Higgs boson mass eigenstate can decay into them leading to an invisible decay of the Higgs boson. Since the Standard Model Higgs boson is accidentally very narrow in the mass region of $2m_b\lsim m_h\lsim 2m_W$, even a small mixing with such a hidden sector can overwhelm the branching fraction into an invisible width.
\item The simultaneously presence of kinetic mixing of $U(1)_X$ with hypercharge and the mixing of the hidden sector Higgs boson that breaks $U(1)_X$ with the Standard Model Higgs boson can lead to dramatic four lepton decays of the light Higgs boson. This is accomplished by decays such as $h\stackrel{mix}{\longrightarrow}h_{hid}\to XX\stackrel{mix}{\longrightarrow}ZZ\to 4\ell$.  There are essentially no limits on the $X$ boson if the kinetic mixing is a loop factor~\cite{Kumar:2006gm}, and yet this decay can be nearly 100\% of the light Higgs branching fraction if kinematically accessible~\cite{Gopalakrishna:2008dv}.
\end{itemize}
Thus we see the phenomenology can range from dramatic in the early  stages (e.g., $h\to 4\ell$) to quite subtle requiring high luminosity to discover at a later stage (e.g., weakly mixed trans-TeV Higgs boson). The varieties of possibilities with a Higgs boson connecting to a hidden sector are limitless, ranging from collider physics to cosmological implications~\cite{Binoth:1996au,Strassler:2006im,Strassler:2006ri,Bhattacharyya:2007pb,MarchRussell:2008yu,Espinosa:2008kw,Ahlers:2008qc,Gopalakrishna:2009yz}.

\bigskip\bigskip
\noindent
{\sc Acknowledgements:}
I wish to thank Jonathan Evans at Cambridge and Ian Jack at Liverpool for their kind hospitality during the weeks I was in residence delivering these lectures. Thanks must also be extended to them, my fellow lecturers, students, and tutors with whom I had the joy to discuss the material of these lectures and much more.  Regarding the subject matter of these lectures, I have benefitted from discussions with many colleagues, including G. Giudice, C. Grojean, J. Gunion, H. Haber, G. Kane, S. Martin, A. Pierce, R. Rattazzi.







\begin{thebibliography}{9}


\bibitem{Goldstone:1962es}
  J.~Goldstone, A.~Salam and S.~Weinberg,
  Phys.\ Rev.\  {\bf 127}, 965 (1962).

  \bibitem{neutrino reviews}
  For dedicated neutrino physics reviews, see, e.g., Y. Grossman, hep-ph/0305245; A. De Gouvea, hep-ph/0411274; R.N. Mohapatra, hep-ph/0412050; G. Altarelli, arXiv:0711.0161.

\bibitem{Kayser PDG}
For a summary of neutrino masses and mixing constraints, see B. Kayser, ``Neutrino mass, mixing, and flavor change" in ref.~\cite{Amsler:2008zzb}.  
  
\bibitem{Amsler:2008zzb}
  C.~Amsler {\it et al.}  [Particle Data Group],
  Phys.\ Lett.\  B {\bf 667}, 1 (2008).

  
\bibitem{Barate:2003sz}
  R.~Barate {\it et al.}  [LEP Collaborations],
  ``Search for the standard model Higgs boson at LEP,''
  Phys.\ Lett.\  B {\bf 565}, 61 (2003)
  [arXiv:hep-ex/0306033].

 \bibitem{tevhiggslimit}
 TEVNPH Working Group (CDF and D0 Collaborations), ``Combined CDF and D0 Upper Limits on Standard Model Higgs-boson Production with up to $4.2\xfb^{-1}$ of Data," Fermilab-pub-09-060-E, CDF Note 9713, D0 Note 5889 (13 March 2009).
 
\bibitem{Kennedy:1992tj}
  D.~C.~Kennedy,
  ``Renormalization of electroweak gauge interactions,''
TASI Lectures 1991 (Boulder, CO).

\bibitem{Wells:2005vk}
  J.~D.~Wells,
  ``TASI lecture notes: Introduction to precision electroweak analysis,''
  arXiv:hep-ph/0512342.
  
\bibitem{:2005ema}
    LEP and SLD Collaborations,
  ``Precision electroweak measurements on the $Z$ resonance,''
  Phys.\ Rept.\  {\bf 427}, 257 (2006)
  [arXiv:hep-ex/0509008].

\bibitem{Alcaraz:2006mx}
  J.~Alcaraz {\it et al.}  [LEP and SLD Collaborations],
  ``A Combination of preliminary electroweak measurements and constraints on
  the standard model,''
  arXiv:hep-ex/0612034.
  
\bibitem{Flacher:2008zq}
  H.~Flacher, M.~Goebel, J.~Haller, A.~Hocker, K.~Moenig and J.~Stelzer,
  Eur.\ Phys.\ J.\  C {\bf 60}, 543 (2009)
  [arXiv:0811.0009 [hep-ph]].
  
  \bibitem{chi2}
See Table 32.2 of  G. Cowan, ``Statistics" in ref.~\cite{Amsler:2008zzb}.


\bibitem{LEPEWWG 2009}
LEP Electroweak Working Group, Preliminary Results from Summer 2009 Update,
http://lepewwg.web/cern.ch/LEPEWWG/plots/summer2009/ (accessed 18 September 2009).


\bibitem{LEPEWWG}
LEP Electroweak Working Group web page is found at http://lepewwg.web.cern.ch/LEPEWWG/


\bibitem{Peskin:2001rw}
  M.~E.~Peskin and J.~D.~Wells,
  Phys.\ Rev.\  D {\bf 64}, 093003 (2001)
  [arXiv:hep-ph/0101342].


\bibitem{Aad:2009wy}
  G.~Aad {\it et al.}  [The ATLAS Collaboration],
  ``Expected Performance of the ATLAS Experiment - Detector, Trigger and
  Physics,''
  arXiv:0901.0512 [hep-ex].

\bibitem{CMSTDR}
A. De Roeck et al. [The CMS Collaboration], 
``CMS Physics Technical Design Report Volume II: Physics Performance,"
CERN/LHCC 2006-021 (26 June 2006).

\bibitem{Lee:1977eg}
  B.~W.~Lee, C.~Quigg and H.~B.~Thacker,
  Phys.\ Rev.\  D {\bf 16}, 1519 (1977).

\bibitem{Chanowitz:1985hj}
  M.~S.~Chanowitz and M.~K.~Gaillard,
  Nucl.\ Phys.\  B {\bf 261}, 379 (1985).

\bibitem{Djouadi:1997yw}
  A.~Djouadi, J.~Kalinowski and M.~Spira,
  Comput.\ Phys.\ Commun.\  {\bf 108}, 56 (1998)
  [arXiv:hep-ph/9704448].


\bibitem{partial wave}
See, e.g., J.D. Jackson, D.R. Tovey, ``Kinematics" in~\cite{Amsler:2008zzb}; sec.\ 4.7 of
D.H. Perkins, {\it Introduction to High Energy Physics, 2nd \& 3rd ed.}, Addison-Wesley, 1982, 1987; and,  sec.\ 3.7 of S. Weinberg, {\it The Quantum Theory of Fields, vol.I}, Cambridge University Press, 1995.


\bibitem{HHG}
J.F. Gunion, H.E. Haber, G. Kane, S. Dawson, {\it The Higgs Hunter's Guide,} Addison-Wesley, 1990.

\bibitem{Djouadi:2005gi}
  A.~Djouadi,
  Phys.\ Rept.\  {\bf 457}, 1 (2008)
  [arXiv:hep-ph/0503172].


\bibitem{Reina:2005ae}
  L.~Reina,
  arXiv:hep-ph/0512377.


\bibitem{Machacek:1984zw}
  M.~E.~Machacek and M.~T.~Vaughn,
  Nucl.\ Phys.\  B {\bf 249}, 70 (1985).

\bibitem{Arason:1991ic}
  H.~Arason et al.,
  Phys.\ Rev.\  D {\bf 46}, 3945 (1992).

\bibitem{Cabibbo:1979ay}
  N.~Cabibbo, L.~Maiani, G.~Parisi and R.~Petronzio,
  Nucl.\ Phys.\  B {\bf 158}, 295 (1979).
  
\bibitem{Tobe:2002zj}
  K.~Tobe and J.~D.~Wells,
  Phys.\ Rev.\  D {\bf 66}, 013010 (2002)
  [arXiv:hep-ph/0204196].



\bibitem{Sher:1988mj}
  M.~Sher,
  Phys.\ Rept.\  {\bf 179}, 273 (1989).

\bibitem{Altarelli:1994rb}
  G.~Altarelli and G.~Isidori,
  Phys.\ Lett.\  B {\bf 337}, 141 (1994).

\bibitem{DiazCruz:1992uw}
  J.~L.~Diaz-Cruz and A.~Mendez,
  Nucl.\ Phys.\  B {\bf 380}, 39 (1992).

 
\bibitem{Veltman:1997nm}
  M.~J.~G.~Veltman,
  ``Reflections on the Higgs system,'' 
  CERN-YELLOW-97-05 (1997).
  
\bibitem{Atwood:1996vj}
  D.~Atwood, L.~Reina and A.~Soni,
  Phys.\ Rev.\  D {\bf 55}, 3156 (1997)
  [arXiv:hep-ph/9609279].
  

\bibitem{Lunghi:2007ak}
  E.~Lunghi and A.~Soni,
  JHEP {\bf 0709}, 053 (2007)
  [arXiv:0707.0212 [hep-ph]].
  
\bibitem{Lubicz:2007yu}
  V.~Lubicz {\it et al.}  [UTfit Collaboration],
  Nucl.\ Phys.\ Proc.\ Suppl.\  {\bf 163}, 43 (2007).

  \bibitem{Gupta:soon}
  S. Gupta, J. Wells, to appear.

   
  \bibitem{Higgs type models}
  Type I and type II Higgs doublet model ideas were developed in the following  papers with the terminology set in the last: \\
  S.~L.~Glashow and S.~Weinberg,
  Phys.\ Rev.\  D {\bf 15}, 1958 (1977). \\
  H.~E.~Haber, G.~L.~Kane and T.~Sterling,
  Nucl.\ Phys.\  B {\bf 161}, 493 (1979).\\
  L.~J.~Hall and M.~B.~Wise,
  Nucl.\ Phys.\  B {\bf 187}, 397 (1981).



\bibitem{Su:2009fz}
  S.~Su and B.~Thomas,
  arXiv:0903.0667 [hep-ph].


\bibitem{Ambroso:2008kb}
  M.~Ambroso, V.~Braun and B.~A.~Ovrut,
  JHEP {\bf 0810}, 046 (2008)
  [arXiv:0807.3319 [hep-th]].

\bibitem{Misiak:2006zs}
For a recent discussion, see
  M.~Misiak {\it et al.},
  Phys.\ Rev.\ Lett.\  {\bf 98}, 022002 (2007)
  [arXiv:hep-ph/0609232].

\bibitem{Wells:2003tf}
  J.~D.~Wells,
  arXiv:hep-ph/0306127.
  J.~D.~Wells,
  Phys.\ Rev.\  D {\bf 71}, 015013 (2005)
  [arXiv:hep-ph/0411041].

\bibitem{split susy}
  N.~Arkani-Hamed and S.~Dimopoulos,
  JHEP {\bf 0506}, 073 (2005)
  [arXiv:hep-th/0405159].
  G.~F.~Giudice and A.~Romanino,
  Nucl.\ Phys.\  B {\bf 699}, 65 (2004)
  [Erratum-ibid.\  B {\bf 706}, 65 (2005)]
  [arXiv:hep-ph/0406088].
  N.~Arkani-Hamed, S.~Dimopoulos, G.~F.~Giudice and A.~Romanino,
  Nucl.\ Phys.\  B {\bf 709}, 3 (2005)
  [arXiv:hep-ph/0409232].


\bibitem{Polchinski:1992ed}
  J.~Polchinski,
  arXiv:hep-th/9210046.

\bibitem{Giudice:2008bi}
  G.~F.~Giudice,
  arXiv:0801.2562 [hep-ph].

\bibitem{Martin:1997ns}
See, e.g., S.~P.~Martin, ``A supersymmetry primer,''
  hep-ph/9709356.

\bibitem{little higgs}
For a review, see
  M.~Schmaltz and D.~Tucker-Smith,
  Ann.\ Rev.\ Nucl.\ Part.\ Sci.\  {\bf 55}, 229 (2005)
  [arXiv:hep-ph/0502182].

\bibitem{conformal}
For an exploratory vision of possibilities, see
  P.~H.~Frampton and C.~Vafa,
  arXiv:hep-th/9903226.


\bibitem{xdim reviews}
For theory reviews of extra dimensions, see
  R.~Sundrum,
  arXiv:hep-th/0508134.
  R.~Rattazzi,
  arXiv:hep-ph/0607055.

\bibitem{Lane:2009ct}
For a recent approach to technicolor, see
  K.~Lane and A.~Martin,
  arXiv:0907.3737 [hep-ph].


\bibitem{top condensate}
  C.~T.~Hill,
  Phys.\ Lett.\  B {\bf 266}, 419 (1991).
  S.~P.~Martin,
  Phys.\ Rev.\  D {\bf 46}, 2197 (1992)
  [arXiv:hep-ph/9204204].
  C.~T.~Hill,
  Phys.\ Lett.\  B {\bf 345}, 483 (1995)
  [arXiv:hep-ph/9411426].
  R.~S.~Chivukula, B.~A.~Dobrescu, H.~Georgi and C.~T.~Hill,
  Phys.\ Rev.\  D {\bf 59}, 075003 (1999)
  [arXiv:hep-ph/9809470].

\bibitem{higgsless}
  C.~Csaki, C.~Grojean, L.~Pilo and J.~Terning,
  Phys.\ Rev.\ Lett.\  {\bf 92}, 101802 (2004)
  [arXiv:hep-ph/0308038].
  C.~Csaki, C.~Grojean, H.~Murayama, L.~Pilo and J.~Terning,
  Phys.\ Rev.\  D {\bf 69}, 055006 (2004)
  [arXiv:hep-ph/0305237].
  Y.~Cui, T.~Gherghetta and J.~D.~Wells,
  arXiv:0907.0906 [hep-ph].


\bibitem{Douglas:2006es}
For example, see sec.\ II.F.3 in
  M.~R.~Douglas and S.~Kachru,
  Rev.\ Mod.\ Phys.\  {\bf 79}, 733 (2007)
  [arXiv:hep-th/0610102],
and sec.\ 3.4 in
  J.~Kumar,
  Int.\ J.\ Mod.\ Phys.\  A {\bf 21}, 3441 (2006)
  [arXiv:hep-th/0601053].

\bibitem{Pilaftsis:1999qt}
  A.~Pilaftsis and C.~E.~M.~Wagner,
  Nucl.\ Phys.\  B {\bf 553}, 3 (1999)
  [arXiv:hep-ph/9902371].

\bibitem{Carena:2002bb}
  M.~S.~Carena, J.~R.~Ellis, S.~Mrenna, A.~Pilaftsis and C.~E.~M.~Wagner,
  Nucl.\ Phys.\  B {\bf 659}, 145 (2003)
  [arXiv:hep-ph/0211467].
  
\bibitem{Polonsky:1999qd}
  N.~Polonsky,
  arXiv:hep-ph/9911329.
  

\bibitem{Haber:1990aw}
  H.~E.~Haber and R.~Hempfling,
  Phys.\ Rev.\ Lett.\  {\bf 66}, 1815 (1991).

\bibitem{Okada:1990vk}
  Y.~Okada, M.~Yamaguchi and T.~Yanagida,
  Prog.\ Theor.\ Phys.\  {\bf 85}, 1 (1991).



\bibitem{Ellis:1990nz}
  J.~R.~Ellis, G.~Ridolfi and F.~Zwirner,
  Phys.\ Lett.\  B {\bf 257}, 83 (1991).

\bibitem{Carena:2002es}
  M.~S.~Carena and H.~E.~Haber,
  Prog.\ Part.\ Nucl.\ Phys.\  {\bf 50}, 63 (2003)
  [arXiv:hep-ph/0208209].

\bibitem{Martin:2002wn}
  S.~P.~Martin,
  Phys.\ Rev.\  D {\bf 67}, 095012 (2003)
  [arXiv:hep-ph/0211366];
  Phys.\ Rev.\  D {\bf 75}, 055005 (2007)
  [arXiv:hep-ph/0701051].


\bibitem{Loinaz:1998ph}
  W.~Loinaz and J.~D.~Wells,
  Phys.\ Lett.\  B {\bf 445}, 178 (1998)
  [arXiv:hep-ph/9808287].
  
\bibitem{Carena:1998gk}
  M.~S.~Carena, S.~Mrenna and C.~E.~M.~Wagner,
  Phys.\ Rev.\  D {\bf 60}, 075010 (1999)
  [arXiv:hep-ph/9808312].
  
\bibitem{Mrenna:2000qh}
  S.~Mrenna and J.~D.~Wells,
  Phys.\ Rev.\  D {\bf 63}, 015006 (2001)
  [arXiv:hep-ph/0001226].
  
\bibitem{Heinemeyer:2000fa}
  S.~Heinemeyer, W.~Hollik and G.~Weiglein,
  Eur.\ Phys.\ J.\  C {\bf 16}, 139 (2000)
  [arXiv:hep-ph/0003022].
  
\bibitem{Haber:2000kq}
  H.~E.~Haber, M.~J.~Herrero, H.~E.~Logan, S.~Penaranda, S.~Rigolin and D.~Temes,
  Phys.\ Rev.\  D {\bf 63}, 055004 (2001)
  [arXiv:hep-ph/0007006].
  
\bibitem{Gunion:2002zf}
  J.~F.~Gunion and H.~E.~Haber,
  Phys.\ Rev.\  D {\bf 67}, 075019 (2003)
  [arXiv:hep-ph/0207010].
  


\bibitem{ArkaniHamed:1998rs}
  N.~Arkani-Hamed, S.~Dimopoulos and G.~R.~Dvali,
  Phys.\ Lett.\  B {\bf 429}, 263 (1998)
  [arXiv:hep-ph/9803315].

\bibitem{Antoniadis:1998ig}
  I.~Antoniadis, N.~Arkani-Hamed, S.~Dimopoulos and G.~R.~Dvali,
  Phys.\ Lett.\  B {\bf 436}, 257 (1998)
  [arXiv:hep-ph/9804398].

\bibitem{ArkaniHamed:1998nn}
  N.~Arkani-Hamed, S.~Dimopoulos and G.~R.~Dvali,
  Phys.\ Rev.\  D {\bf 59}, 086004 (1999)
  [arXiv:hep-ph/9807344].

\bibitem{Giudice:1998ck}
  G.~F.~Giudice, R.~Rattazzi and J.~D.~Wells,
  Nucl.\ Phys.\  B {\bf 544}, 3 (1999)
  [arXiv:hep-ph/9811291].



\bibitem{Hewett:2002hv}
  J.~L.~Hewett and M.~Spiropulu,
  Ann.\ Rev.\ Nucl.\ Part.\ Sci.\  {\bf 52}, 397 (2002)
  [arXiv:hep-ph/0205106].



\bibitem{Kribs:2006mq}
  G.~D.~Kribs,
  arXiv:hep-ph/0605325.

\bibitem{Randall:1999ee}
  L.~Randall and R.~Sundrum,
  Phys.\ Rev.\ Lett.\  {\bf 83}, 3370 (1999)
  [arXiv:hep-ph/9905221].
  
\bibitem{Csaki:2004ay}
  C.~Csaki,
  arXiv:hep-ph/0404096.
  


\bibitem{Goldberger:1999uk}
  W.~D.~Goldberger and M.~B.~Wise,
  Phys.\ Rev.\ Lett.\  {\bf 83}, 4922 (1999)
  [arXiv:hep-ph/9907447].

\bibitem{Giudice:2000av}
  G.~F.~Giudice, R.~Rattazzi and J.~D.~Wells,
  Nucl.\ Phys.\  B {\bf 595}, 250 (2001)
  [arXiv:hep-ph/0002178].

\bibitem{Csaki:2000zn}
  C.~Csaki, M.~L.~Graesser and G.~D.~Kribs,
  Phys.\ Rev.\  D {\bf 63}, 065002 (2001)
  [arXiv:hep-th/0008151].

\bibitem{Hewett:2002nk}
  J.~L.~Hewett and T.~G.~Rizzo,
  JHEP {\bf 0308}, 028 (2003)
  [arXiv:hep-ph/0202155].

\bibitem{Gunion:2003px}
  J.~F.~Gunion, M.~Toharia and J.~D.~Wells,
  Phys.\ Lett.\  B {\bf 585}, 295 (2004)
  [arXiv:hep-ph/0311219].


\bibitem{Dominici:2009pq}
  D.~Dominici and J.~F.~Gunion,
  arXiv:0902.1512 [hep-ph].

\bibitem{Azatov:2008vm}
  A.~Azatov, M.~Toharia and L.~Zhu,
  Phys.\ Rev.\  D {\bf 80}, 031701 (2009)
  [arXiv:0812.2489 [hep-ph]].



  
\bibitem{Burgess:2000yq}
  C.~P.~Burgess, M.~Pospelov and T.~ter Veldhuis,
  Nucl.\ Phys.\  B {\bf 619}, 709 (2001)
  [arXiv:hep-ph/0011335].

\bibitem{Davoudiasl:2004be}
  H.~Davoudiasl, R.~Kitano, T.~Li and H.~Murayama,
  Phys.\ Lett.\  B {\bf 609}, 117 (2005)
  [arXiv:hep-ph/0405097].

  
\bibitem{Barger:2007im}
  V.~Barger, P.~Langacker, M.~McCaskey, M.~J.~Ramsey-Musolf and G.~Shaughnessy,
  Phys.\ Rev.\  D {\bf 77}, 035005 (2008)
  [arXiv:0706.4311 [hep-ph]].


\bibitem{Joshipura:1992ua}
  A.~S.~Joshipura and S.~D.~Rindani,
  Phys.\ Rev.\ Lett.\  {\bf 69}, 3269 (1992).

\bibitem{Choudhury:1993hv}
  D.~Choudhury and D.~P.~Roy,
  Phys.\ Lett.\  B {\bf 322}, 368 (1994)
  [arXiv:hep-ph/9312347].

\bibitem{Gunion:1993jf}
  J.~F.~Gunion,
  Phys.\ Rev.\ Lett.\  {\bf 72}, 199 (1994)
  [arXiv:hep-ph/9309216].

\bibitem{Frederiksen:1994me}
  S.~G.~Frederiksen, N.~Johnson, G.~L.~Kane and J.~Reid,
  Phys.\ Rev.\  D {\bf 50}, 4244 (1994).

\bibitem{Binoth:1996au}
  T.~Binoth and J.~J.~van der Bij,
  Z.\ Phys.\  C {\bf 75}, 17 (1997)
  [arXiv:hep-ph/9608245].

\bibitem{Martin:1999qf}
  S.~P.~Martin and J.~D.~Wells,
  Phys.\ Rev.\  D {\bf 60}, 035006 (1999)
  [arXiv:hep-ph/9903259].

\bibitem{Godbole:2003it}
  R.~M.~Godbole et al.,
  Phys.\ Lett.\  B {\bf 571}, 184 (2003)
  [arXiv:hep-ph/0304137].

\bibitem{Davoudiasl:2004aj}
  H.~Davoudiasl, T.~Han and H.~E.~Logan,
  Phys.\ Rev.\  D {\bf 71}, 115007 (2005)
  [arXiv:hep-ph/0412269].


\bibitem{kinetic mixing}
  B.~Holdom,
  Phys.\ Lett.\  B {\bf 166}, 196 (1986).
  F.~del Aguila, M.~Masip and M.~Perez-Victoria,
  Nucl.\ Phys.\  B {\bf 456}, 531 (1995)
  [hep-ph/9507455].
  K.~R.~Dienes, C.~F.~Kolda and J.~March-Russell,
  Nucl.\ Phys.\  B {\bf 492}, 104 (1997)
  [hep-ph/9610479].
  K.~S.~Babu, C.~F.~Kolda and J.~March-Russell,
  Phys.\ Lett.\  B {\bf 408}, 261 (1997)
  [hep-ph/9705414].
  T.~G.~Rizzo,
  Phys.\ Rev.\  D {\bf 59}, 015020 (1999)
  [hep-ph/9806397].
  S.~A.~Abel, J.~Jaeckel, V.~V.~Khoze and A.~Ringwald,
  Phys.\ Lett.\  B {\bf 666}, 66 (2008)
  [arXiv:hep-ph/0608248].
  D.~Feldman, Z.~Liu and P.~Nath,
  Phys.\ Rev.\  D {\bf 75}, 115001 (2007)
  [arXiv:hep-ph/0702123].

\bibitem{Schabinger:2005ei}
  R.~Schabinger and J.~D.~Wells,
  Phys.\ Rev.\  D {\bf 72}, 093007 (2005)
  [arXiv:hep-ph/0509209].
 
\bibitem{Bowen:2007ia}
  M.~Bowen, Y.~Cui and J.~D.~Wells,
  JHEP {\bf 0703}, 036 (2007)
  [arXiv:hep-ph/0701035].

  
\bibitem{Wells:2008xg}
  J.~D.~Wells,
  arXiv:0803.1243 [hep-ph].



\bibitem{Kumar:2006gm}
  J.~Kumar and J.~D.~Wells,
  Phys.\ Rev.\  D {\bf 74}, 115017 (2006)
  [arXiv:hep-ph/0606183].

\bibitem{Gopalakrishna:2008dv}
  S.~Gopalakrishna, S.~Jung and J.~D.~Wells,
  Phys.\ Rev.\  D {\bf 78}, 055002 (2008)
  [arXiv:0801.3456 [hep-ph]].

   
\bibitem{Strassler:2006im}
  M.~J.~Strassler and K.~M.~Zurek,
  Phys.\ Lett.\  B {\bf 651}, 374 (2007)
  [arXiv:hep-ph/0604261].
  
\bibitem{Strassler:2006ri}
  M.~J.~Strassler and K.~M.~Zurek,
  Phys.\ Lett.\  B {\bf 661}, 263 (2008)
  [arXiv:hep-ph/0605193].
  

\bibitem{Bhattacharyya:2007pb}
  G.~Bhattacharyya, G.~C.~Branco and S.~Nandi,
  Phys.\ Rev.\  D {\bf 77}, 117701 (2008)
  [arXiv:0712.2693 [hep-ph]].


\bibitem{MarchRussell:2008yu}
  J.~March-Russell, S.~M.~West, D.~Cumberbatch and D.~Hooper,
  JHEP {\bf 0807}, 058 (2008)
  [arXiv:0801.3440 [hep-ph]].

\bibitem{Espinosa:2008kw}
  J.~R.~Espinosa, T.~Konstandin, J.~M.~No and M.~Quiros,
  Phys.\ Rev.\  D {\bf 78}, 123528 (2008)
  [arXiv:0809.3215 [hep-ph]].

\bibitem{Ahlers:2008qc}
  M.~Ahlers, J.~Jaeckel, J.~Redondo and A.~Ringwald,
  Phys.\ Rev.\  D {\bf 78}, 075005 (2008)
  [arXiv:0807.4143 [hep-ph]].



\bibitem{Gopalakrishna:2009yz}
  S.~Gopalakrishna, S.~J.~Lee and J.~D.~Wells,
  arXiv:0904.2007 [hep-ph].



  
\end{thebibliography}
\end{document}